\def\etal{et al.\ }
\def\eg{{\em eg.\ }}

\def\spose#1{\hbox to 0pt{#1\hss}}
\def\approxlt{\mathrel{\spose{\lower 3pt\hbox{$\sim$}}
	\raise 2.0pt\hbox{$<$}}}
\def\approxgt{\mathrel{\spose{\lower 3pt\hbox{$\sim$}}
	\raise 2.0pt\hbox{$>$}}}

\def\degmark{$^\circ$}
\def\<{\thinspace}
\def\s{\hbox{\phantom{5}}}	

\def\cm{{\rm\thinspace cm}}

\def\erg{{\rm\thinspace erg}}

\def\K{{\rm\thinspace K}}
\def\keV{{\rm\thinspace keV}}

\def\km{{\rm\thinspace km}}
\def\kpc{{\rm\thinspace kpc}}

\def\Mpc{{\rm\thinspace Mpc}}
\def\Msun{\hbox{$\rm\thinspace M_{\odot}$}}

\def\s{{\rm\thinspace s}}
\def\yr{{\rm\thinspace yr}}

\def\ergpcmsqps{\hbox{$\erg\cm^{-2}\s^{-1}\,$}}
\def\ergpspcmsq{\hbox{$\erg\cm^{-2}\s^{-1}\,$}}
\def\ergpspkpcsq{\hbox{$\erg\s^{-1}\kpc^{-2}\,$}}
\def\ergpspA{\hbox{$\erg\s^{-1}\AA^{-1}\,$}}

\def\ergps{\hbox{$\erg\s^{-1}\,$}}

\def\kmps{\hbox{$\km\s^{-1}\,$}}

\def\Msunpyr{\hbox{$\Msun\yr^{-1}\,$}}

\def\kmpspMpc{\hbox{$\kmps\Mpc^{-1}$}}

\documentstyle[psfig]{mn}  
\begin{document}

\def\Hb{H$\beta$\ }
\def\Hbn{H$\beta$}
\def\Ha{H$\alpha$\ }
\def\Han{H$\alpha$}
\def\otw{[OII]$\lambda$3727}
\def\oth{[OIII]$\lambda$5007}
\def\niib{[NII]$\lambda\lambda$6548,6584}
\def\D{${\rm D}_{4000}$\ } 
\def\Dn{${\rm D}_{4000}$}
	
\title[Optical spectra of the central cluster galaxies of the ROSAT
Brightest Cluster Sample ]{The ROSAT Brightest Cluster Sample (BCS): -- III. 
Optical spectra of the central cluster galaxies}

\author[C.S. Crawford et al ]
{\parbox[]{6.in} {C.S. Crawford$^{1}$, 
S.W. Allen$^{1}$, 
H. Ebeling$^{2}$, A.C. Edge$^{1,3}$ and  A.C. Fabian$^{1}$ %
\\
\footnotesize
1. Institute of Astronomy, Madingley Road, Cambridge CB3 0HA \\
2. Institute for Astronomy, 2680 Woodlawn Drive, Honolulu HI~96822, USA \\
3. Department of Physics, University of Durham, South Road, Durham DH1
3LE \\}}
\maketitle
\begin{abstract}

\noindent
We present new spectra of dominant galaxies in X-ray selected clusters of
galaxies, which combine with our previously published spectra to form a
sample of 256 dominant galaxies in 215 clusters. 177 of the clusters are
members of the ROSAT Brightest Cluster Sample (BCS; Ebeling \etal 1998), and
18  have no previous measured redshift.  This is the first paper in a
series correlating the properties of brightest cluster galaxies and their
host clusters in the radio, optical and X-ray wavebands.

27 per cent of the central dominant galaxies have emission-line spectra, all
but five with line intensity ratios typical of cooling flow nebulae. A
further 6 per cent show only [NII]$\lambda\lambda6548,6584$ with \Ha in
absorption. We find no  evidence for an increase in the frequency of
line emission with X-ray
luminosity. Purely X-ray-selected clusters at low redshift have a higher
probability of containing line emission.
 The projected separation between the optical position
of the dominant galaxy and its host cluster X-ray centroid is less for the
line-emitting galaxies than for those without line emission, consistent with
a closer association of the central galaxy and the gravitational centre in
cooling flow clusters.

The more \Han-luminous galaxies have
larger emission-line regions and show a higher ratio of Balmer 
 to forbidden line emission, although there is a continuous trend of
ionization behaviour across four decades in \Ha luminosity. Galaxies with the  more
luminous line emission  (L(\Han)$>10^{41}$\ergps) show a
significantly bluer continuum, whereas lower-luminosity and [NII]-only line
emitters have continua that differ little from those of non-line emitting
dominant galaxies. Values of the Balmer decrement in the more luminous
systems commonly imply intrinsic reddening of E(B-V)$\sim0.3$, and when this
is corrected for, the excess blue light can be characterized by a population
of massive young stars. Several of the galaxies require a large 
population of O stars, which also provide sufficient photoionization to
produce the observed \Ha luminosity. The large number of lower-mass stars relative to the O star
population suggests that this anomalous population is due to a series of
starbursts in the central galaxy. 

The lower \Han-luminosity systems show a higher ionization state and few
massive stars, requiring instead the introduction of a harder source of
photoionization, such as turbulent mixing layers, or low-level nuclear
activity. The line emission from the systems showing only [NII] is very
similar to low-level LINER activity commonly found in many normal
elliptical galaxies.

\end{abstract}

\begin{keywords} 
surveys -- 
galaxies: clusters: general -- cooling flows --
galaxies: elliptical and lenticular, cD -- 
galaxies: stellar content --
X-rays:galaxies. 
\end{keywords}

\section{Introduction}
\label{sec:intro}

\noindent We have compiled the ROSAT Brightest Cluster Sample (BCS;
Ebeling etal 1998, hereafter Paper~I), which is a 90 per cent flux-complete sample of the 201
X-ray brightest clusters of galaxies in the
northern extragalactic hemisphere ($\delta\ge0$\degmark and
$\left|b\right|\ge20$\degmark) within $z=$0.3. The clusters were selected
from their X-ray emission in the ROSAT All-Sky Survey (RASS; Voges 1992),
and include both clusters listed in optical catalogues and systems newly
discovered through their X-ray properties (Crawford \etal 1995; Paper~I).
 The BCS is now redshift-complete at unabsorbed fluxes greater than
$4.4\times10^{-12}$\ergpcmsqps in the $0.1-2.4$\keV\ ROSAT band (Paper~I).
Paper~I also tabulates a further two clusters with $z>0.3$
that fulfil all the other BCS criteria (RXJ1532.9+3021 and Z1953). The
BCS is the largest statistically complete X-ray selected sample of
clusters to date, and has been used to derive the cluster log~N-log~S
distribution (Paper~I), the cluster-cluster correlation function
(Edge \etal 1999a), and also the X-ray luminosity function out to a redshift
of 0.3 (Ebeling \etal 1997, Paper~II). The BCS forms a unique statistical sample to
define the characteristics of clusters at other wavebands and their
dependence on the global cluster properties, without the biases inherent in
optically- or radio-selected samples.

As part of the selection and compilation of the BCS, we have published
optical spectra of central dominant galaxies in both Abell and Zwicky
clusters without a previously known redshift, and also of the new
ROSAT-discovered clusters (Allen \etal 1992, Crawford \etal 1995; hereafter
A92 and C95 respectively). Preliminary analyses of the spectra of these
galaxies show that around one-third of them display anomalies such as a
spectrum featuring luminous emission lines and an excess ultraviolet/blue
continuum. Such features are well-known to be associated with galaxies at
the centres of cooling flows (Hu, Cowie \& Wang 1985; Johnstone, Fabian \&
Nulsen 1987 (hereafter JFN); Heckman \etal 1989; Romanishin 1987; Donahue,
Stocke \& Gioia 1992; McNamara \& O'Connell 1989, 1992, 1993), and
preferentially with those that contain a radio source (Heckman \etal 1989,
C95).  The blue continuum is directly attributable to a modest amount of
recent massive star formation, which can also power some of the luminous
line emission (Allen 1995).

This paper is the first in a series in which we compile and correlate the
properties of the brightest cluster galaxies (BCG) and their host clusters
in three wavebands. We present optical spectra of the dominant galaxies of
87 per cent of the BCS; subsequent papers will compile results from optical
and radio images (Edge \etal 1999b, 1999c) and the cluster X-ray
properties (Crawford \etal 1999; Ebeling \etal 1999) in order to perform a full
statistical analysis of the sample. A value of the Hubble constant of
$H_0=50$\kmpspMpc\ and a cosmological deceleration parameter of $q_0=0.5$
have been used throughout this paper; where cluster properties are quoted
from the literature, we convert them to this assumed cosmology.

\section{Observations and Data reduction}

\subsection{Target selection and samples}
\label{sec:target}

\noindent
The BCS is comprised of clusters culled from a positional cross-correlation
of X-ray sources detected in the RASS with the optical cluster catalogues of
Abell and Zwicky (Abell 1958; Abell, Corwin \& Olowin 1989; Zwicky \etal
1961-1968), as well as systems detected as extended X-ray sources in the
RASS and confirmed as clusters using the Palomar Observatory Sky Survey
(POSS) and deep optical CCD imaging (Paper~I). The latter, purely
X-ray selected systems are referred 
to in this paper by the ROSAT X-ray source name, eg RXJ0107.4+3227. (Any
more commonly used names are referred to in the 
notes accompanying the tables). 
We identified the
dominant galaxies from comparison of the X-ray emission centroid and extent
(using ROSAT pointed observations where possible rather than the limited
information available from the RASS) against the POSS and maps provided by
the Automated Plate Measuring Machine at Cambridge. Our previous work
(A92 and C95) showed that the BCG is usually found within 1-2 arcmin of
the centroid of the X-ray emission from the cluster. Where there is no
simple identification of a single dominant galaxy near the X-ray centroid,
we aimed to observe all possible candidates, and in cases where the cluster
is known or suspected to be part of a multiple cluster system (eg A2249,
A2256), to observe both (or more) dominant galaxies. For detailed
information about the best candidate dominant galaxies within a cluster, and
for the positions of good candidates remaining unobserved due to lack of
time, see the notes appended to the log of observations presented in
Tables~\ref{tab:bcslog} and \ref{tab:notbcslog}. In the course of compiling
the BCS we have taken spectra of the BCG in several clusters that were not
included in the final sample. These extra clusters were selected from the
RASS using exactly the same (original) criteria, but were eventually
discarded due to their faint X-ray flux, low Galactic latitude or negative
declination. As they are a natural extension to the BCS, we include their
BCG spectra in this paper to enlarge our sample, but list them separately
from the `true' BCS in the log of observations (Table~\ref{tab:notbcslog}).

\subsection{Previous data}
\label{sec:previous}
Optical spectra of 88 dominant galaxies in 79 clusters have already been
presented by us in A92  and C95. The majority of these observations were
taken using the Faint Object Spectrograph (FOS) on the 2.5m Isaac Newton
Telescope (INT) on La Palma, with the primary aim of determining the
redshift of BCS clusters. The FOS yields low-resolution spectra in two
orders, with dispersion of 5.4\AA/pixel over 3500--5500\AA\ and
10.7\AA/pixel over 5000--10500\AA, taken from a 6 arcsec-long slit.
Observations were usually made at the parallactic angle to minimize the
effects of atmospheric dispersion, and the slit width was matched to the
seeing of typically around 1.2 arcsec FWHM. The FOS spectra were reduced
using an automated optimal extraction package, and thus {\em spatially
extended} spectral information is not available for these galaxies. The
short slit length may also mean that the sky-subtraction is less accurate
for the lower-redshift galaxies. The final spectra were corrected for
atmospheric extinction and Galactic reddening, and flux calibrated. C95 
also includes ten spectra at  moderate resolution (5\AA/pixel) taken using
the Intermediate dispersion Spectroscopic and Imaging System on the William
Herschel Telescope (WHT). All the previous spectra are included in the log
of observations given in Tables~\ref{tab:bcslog} and \ref{tab:notbcslog},
although the reader is referred to A92 and C95 for a more detailed
account of the observations and data reduction. The reference for the paper
containing the first publication of the spectrum is listed in the tenth
column of Tables~\ref{tab:bcslog} and \ref{tab:notbcslog} (A92,  C95 and W94 for White \etal 1994).

A92 contains an observation of what was thought to be the central
cluster galaxy in the `sub'-BCS cluster A2552. This $z=0.128$ galaxy (at RA
23 11 31.1, Dec 03 37 55; J2000) had a peculiar spectrum, with narrow
emission lines,  curious line ratios and an exceptionally strong blue
continuum with no obvious stellar absorption features. 
Subsequent optical imaging has since shown that this galaxy has a
peculiar morphology, and is too bright to be the dominant galaxy for the
surrounding cluster (Edge \etal 1999c). 
The best candidate for the dominant cluster galaxy is the double system 
at RA  23 11 33.3, Dec 03 38 05 (J2000), at a redshift of 0.3025 (Ebeling,
private communication). 
Reconsideration of our noisy
spectrum of another nearby galaxy at RA 23 11 33.3 and Dec 03 38 05 (J2000)
gives a  tentative redshift of $z=0.306$; this galaxy was originally assigned a redshift of
0.137 in C95 based on an erroneous identification of [NII] emission.

\subsection{New observations}
\label{sec:new}

The majority of the optical spectra presented in this paper were acquired
during two week-long runs during 1995 May 03 -- 09 and 1996 Dec 29 -- 1997 Jan
03. A further six galaxies were observed during service time on 1998 Mar 16
and 29. The observations were made using the Intermediate Dispersion
Spectrograph (IDS) on the INT at La Palma, with the R150V grating on the
235mm camera to yield an intermediate resolution spectrum with dispersion of
6.5\AA/pixel. In combination with the TEK3 CCD detector the spectra covered
the observed wavelength range 3300--9500\AA\ with a dispersion of 0.7 arcsec/pixel
in the spatial direction. The weather was clear for nine of the fourteen
nights, with seeing conditions of typically 1.2 arcsec, varying between
1--1.5 arcsec. The long slit was set at a width of 1.3 arcsec and usually
oriented at the parallactic angle, except where a more favourable
orientation would include a second dominant galaxy in the same observation
(those galaxies known to be observed with the slit {\sl not} at parallactic
angle are marked by a dagger in column 9 of Tables~\ref{tab:bcslog}  and
\ref{tab:notbcslog}). Exposures were
routinely of 1000s, and three or four observations of flux standards were
made during each night with the slit opened up to 5.5 arcsec.

The data were reduced using standard procedures in the software package
IRAF. After subtraction of a bias frame, the data were flat-fielded using a
normalised flat created from long exposures of a tungsten lamp. Any spatial
dispersion of the spectrum due to atmospheric refraction was corrected at
this stage so that the continuum intensity peaked in the same row along the
detector. The data were wavelength-calibrated using exposures of
copper-argon and copper-neon arc-lamps, flux-calibrated, and corrected for
atmospheric extinction. The Galactic reddening expected toward each galaxy
was calculated from neutral hydrogen columns (Stark \etal 1992) using the
formula of Bohlin, Savage \& Drake (1978), and the spectra corrected using
the law of Cardelli, Clayton \& Mathis (1989) by the amount listed in column
8 of Tables~\ref{tab:bcslog} and \ref{tab:notbcslog}. The data were
sky-subtracted using background regions selected to be well outside the
galaxy extent, as 
determined from fitting spatial profiles of the galaxy continuum. The slit
was sufficiently long to achieve good sky subtraction for all but M87, the
dominant galaxy of the most nearby cluster in our sample, 
 RXJ1230.7+1220 (Virgo). Although we also took
short exposures of a nearby F8 star after each galaxy observation with the
aim of removing atmospheric absorption features from the galaxy spectra, we
eventually decided that such correction might introduce more uncertainty in
the final emission line fluxes. Thus we have not applied this correction, but
throughout the paper we note where detection of an emission line will
be affected by atmospheric absorption; the galaxies most affected are those
in the redshift ranges 0.044$<z<$0.048 and 0.158$<z<$0.170.

\subsection{Overview of observations}
\label{sec:overview}
A full log of observations is presented in Tables~\ref{tab:bcslog}  and
\ref{tab:notbcslog}. Observations of
dominant galaxies in clusters from the BCS are presented first in Table~\ref{tab:bcslog}, with the
remainder listed in Table~\ref{tab:notbcslog}. `Sub'-BCS clusters that fulfil all the BCS
criteria {\sl except} that the RASS flux falls below the threshold for
inclusion (and lies in the range $3.7-4.4\times10^{-12}$\ergpspcmsq) are
marked as such in the final column of Table~\ref{tab:notbcslog}, as are the clusters included
in an X-ray selected sample of Abell clusters (marked XBACS; Ebeling \etal
1996). Galaxies are identified by the name of their host cluster in column
1, with their individual catalogued names listed in the final column and notes to
the table. Where more than one
galaxy in a cluster is observed, each is labelled a, b with an indication of
their relative position (North, South, etc) following in brackets. Further
elaboration on the properties of the individual galaxies (including if one
is a better candidate for the dominant cluster galaxy) and their host
clusters are given in the notes to the tables. 
The coordinates of the galaxy observed are given in columns 2 and 3, as
measured to better than one arcsec 
from the Space Telescope Science Institute Digitized Sky Survey (DSS), and the
galaxies are presented in RA order within each table. The redshift is given in
column 4, and is in bold type if it is newly determined (see
section~\ref{sec:newz}), and in brackets if
uncertain due to insufficient signal-to-noise in the spectrum. Column 5
shows  whether or not the galaxy has optical \Han-line emission ($\times$ or
$\surd$), or is marked by an \lq N' if there is only
\niib\ line emission (see section~\ref{sec:occurrence}). Clusters marked by [$\times$] show no obvious line
emission, but the expected H$\alpha$ and [NII] complex falls within a
region of atmospheric absorption, and we cannot completely rule out very low level
\niib\ emission in these cases. Columns 6--9 show the total 
exposure, airmass, magnitude of expected Galactic reddening
and  slit position angle respectively for the new observations. 
A $\dag$ in column 9 marks an observation where the position
angle of the slit is known not to be at the parallactic angle. Column 10
gives either the run of the new IDS observations (M95 for May 1995, D96 for
December 1996, M98 for March 1998) or the reference for the first publication of the spectrum
(A92,  C95 or  W94; see section~\ref{sec:previous}).

\subsection{Supplementary data from the literature}
\label{sec:supple}
Twenty-seven clusters remain for which we have not observed the BCG, and these
are listed in Table~\ref{tab:bcsnotlog}. The identification of the possible
BCG for each cluster has been made by us from comparison of the X-ray
properties with the digitized POSS in the same way as for the observed
galaxies; the notes accompanying the table indicate which 
of the galaxies listed  could be the
BCG. 
Eight clusters have galaxies observed by Ebeling, Henry \& Mullis (1999),
and we have used NED to search the literature for  basic information (ie
whether the BCG has optical emission lines or not) on the remainder. 
 Only five clusters remain for which we do not have any spectral
information for the BCG  (A75, A84, A104, A1235 and A2108).
In Table~\ref{tab:bcsnotlog} we list the cluster name, galaxy coordinates
and redshift in columns 1--3 respectively. Where possible we list the
redshift of the individual galaxy if known from the literature; otherwise we
list the cluster redshift and mark it as such in the final column. Whether
or not the galaxy is known to have \Han-line emission is marked in column 4, with
a reference to the source for the redshift (and emission-line properties) in
column 5. Individual names are given in the final column, with more detailed
information in the notes to the table.

\onecolumn
\begin{table}
\caption{Log of the observations -- Clusters in the BCS \label{tab:bcslog}}
\begin{tabular}{llccccrccrll}
             &   & & &       & & & & & & & \\
Cluster      &   & RA         & DEC      & Redshift & Lines? & Exp &
Airmass & A(V) & p.a. & Ref & Notes\\
             &   & (J2000)    & (J2000)  &       &  & (sec)  & & (mag) & (deg) & \\
             &   & & &       & & & & & & & \\
 A7           &   & 00 11 45.2 & 32 24 57 & 0.1017 & $\times$ & 1000 & 1.096 & 0.374 & 272 & D96 & \\
 A21 &a (SE)      & 00 20 37.5 & 28 39 30 & 0.0967 & $\times$ & 1000 & 1.529 & 0.293 & 155$\dag$ & D96 &  \\
     &b (NW)      & 00 20 37.2 & 28 39 36 & 0.0966 & $\times$ & $''$ & $''$  & $''$  & $''$ & $''$ &  \\
 A76  &           & 00 39 55.9 & 06 50 55 & 0.0407 & $\times$ & 1000 & 2.194 & 0.258 &  62 & D96 & IC1568 \\
 Z235           & & 00 43 52.1 & 24 24 22 & 0.083 & $\surd$ & & & & & C95 \\
 A115 &           & 00 55 50.5 & 26 24 39 & 0.1970 & $\surd$ & 1000 & 2.065 & 0.363 &  70 & D96 & \\
 RXJ0107.4+3227 & & 01 07 24.9 & 32 24 46 & 0.0175 & $\surd$ & 1000 & 1.004 & 0.359 & 316 & D96 & NGC383 \\
 A168           & & 01 14 57.5 & 00 25 52 & 0.0443 & [$\times$] & 1000 & 1.904 & 0.226 & 234 & D96 & UGC797 \\
 RXJ0123.2+3327 & & 01 23 11.6 & 33 27 36 & 0.0141 & N & 1000 & 1.512 & 0.349 & 259 & D96 & NGC499 \\
 RXJ0123.6+3315 & & 01 23 39.8 & 33 15 22 & 0.0169 & $\times$ & 1000 & 1.945 & 0.349 &  70 & D96 & NGC507 \\
 A193           & & 01 25 07.6 & 08 41 59 & 0.0484 & [$\times$] & 1000 & 1.488 & 0.307 & 238 & D96 & UGC977 \\
 A267         &   & 01 52 41.9 & 01 00 27 & 0.230 & $\times$ & & & & & C95 \\
 A262           & & 01 52 46.5 & 36 09 08 & 0.0166 & $\surd$ & 1000 & 1.043 & 0.358 & 245 & D96 & NGC708 \\
 A272           & & 01 55 10.5 & 33 53 50 & 0.0898 & $\times$ & 1000 & 1.347 & 0.373 &  85 & D96 & \\
 RXJ0228.2+2811 & & 02 28 03.6 & 28 10 34 & 0.0351 & $\times$ & 1000 & 1.608 & 0.522 &  70 & D96 & IC227\\
 A376           & & 02 46 04.0 & 36 54 21 & 0.0482 & [$\times$] & 1000 & 1.960 & 0.449 & 255 & D96 & UGC2232 \\
 A400 &a (N)      & 02 57 41.5 & 06 01 37 & 0.0238 & N & 1000 & 1.210 & 0.630 & 355$\dag$ & D96 & MCG+01-08-027\\
      &b (S)      & 02 57 41.6 & 06 01 22 & 0.0253 & N & $''$ & $''$  & $''$  & $''$      & $''$ &  $''$ \ \ \ \ $''$ \\
 A399 &           & 02 57 53.1 & 13 01 52 & 0.0699 & $\times$ & 1000 & 1.296 & 0.727 & 259 & D96 & UGC2438 \\
 A401 &           & 02 58 57.8 & 13 34 59 & 0.0737 & N& 1000 & 1.382 & 0.700 & 243 & D96 & UGC2450 \\
 Z808         &   & 03 01 38.2 & 01 55 15  & 0.169 & $\surd$ &  & & & & C95 & \\
 A407 &           & 03 01 51.5 & 35 50 30 & 0.0484 & [$\times$] & 1000 & 1.668 & 0.820 &  79 & D96 & UGC2489 \\
 A409         &   &  03 03 21.1 & 01 55 34   &  0.153 & $\times$ & & & & & C95 \\
 RXJ0338.7+0958&  & 03 38 40.5 & 09 58 12 & 0.0338 & $\surd$ & 1000 & 1.387 & 1.187 & 331$\dag$  & D96 \\
 RXJ0341.3+1524   & & 03 41 17.5 & 15 23 49 & 0.029 & $\times$ & & & & & C95 & \\
 RXJ0352.9$+$1941 & &  03 52 58.9 & 19 41 00 & 0.109 & $\surd$ & & & & $\dag$  & C95 & \\
 A478             & & 04 13 25.3 & 10 27 56 & 0.086 & $\surd$ & 1000 & 1.058 & 0.670 & 315 & W94 &\\
 RXJ0419.6+0225&  & 04 19 37.9 & 02 24 35 & 0.0133 & $\times$ & 1000 & 1.389 & 0.767 &  49 & D96 & NGC1550 \\
 RXJ0439.0+0715&  & 04 39 00.6 & 07 16 05 & {\bf 0.2452} & $\times$ & 1000 & 1.265 & 0.747 &  70 & D96 & \\
 RXJ0439.0$+$0520 & & 04 39 02.2   & 05 20 45  & 0.208 & $\surd$ & & & & & C95 & \\
 A520          &  & 04 54 03.8 & 02 53 33 & (0.2024) & $\times$ & 1000 & 2.161 &  0.520 & 240 & D96 & \\
 A523          &  & 04 59 12.9 & 08 49 42 & {\bf 0.1036} & $\times$ & 1000 & 1.343 & 0.800 &  57 & D96 & \\
 RXJ0503.1$+$0608 & & 05 03 07.0    & 06 07 58  & 0.088 &$\times$  & & & & & C95 & \\
 A566          &  & 07 04 28.6 & 63 18 41 & 0.0945 & $\times$ & 1000 & 1.224 & 0.348 & 344 & D96 & MCG+11-09-031\\
 A576  &a         & 07 21 21.5 & 55 48 40 & 0.0368 & $\times$ & 1000 & 1.123 & 0.381 &   7 & D96 & \\
       &b         & 07 21 32.5 & 55 45 28 & 0.0408 & $\times$ & 1000 & 1.123 & 0.381 &   7 & D96 & MCG+09-12-061\\
 A586  &          & 07 32 20.4 & 31 38 02 & 0.1702 & $\times$ & 1000 & 1.027 & 0.343 & 333 & D96 & \\
 RXJ0740.9+5526&  & 07 40 58.2 & 55 25 39 & 0.0340 & $\times$ & 1000 & 1.298 & 0.309 & 296 & M95 & UGC3957 \\
 RXJ0751.3+5012 &a (S) & 07 51 17.7 & 50 10 46 & 0.0218 & $\times$ & 1000 & 1.265 & 0.339 & 113 & M95 & UGC4051 \\
                &b (N:E) & 07 51 21.1 & 50 14 10 & 0.0236 & $\surd$ & 1000 & 1.268 & 0.339 & 265$\dag$  & M95 & UGC4052\\
                &c (N:W) & 07 51 18.8 & 50 14 07 & 0.0228 & $\times$ & $''$ & $''$  & $''$  & $''$ & $''$ & \\
 A602 &a (NW)     & 07 53 16.5 & 29 24 06 & 0.0601 & $\times$ & 1000 & 1.078 & 0.290 & 267 & D96 & \\
      &b (SE)     & 07 53 26.4 & 29 21 35 & 0.0600 & $\times$ & 1000 & 1.048 & 0.290 & 267 & D96 & \\
 RXJ0819.6+6336 & & 08 19 25.7 & 63 37 28 & {\bf 0.1186} & $\times$ & 1000 & 1.371 & 0.272 & 305 & M95 & \\
 A646           & & 08 22 09.6 & 47 05 54 & 0.1268 & $\surd$ & 1000 & 1.218 & 0.269 & 290 & D96 & \\
 Z1665          & & 08 23 21.7 & 04 22 22 & {\bf 0.0311} & N & 1000 & 1.100 & 0.182 & 351 & D96 & IC505 \\
 A655           & & 08 25 28.9 & 47 08 01 & 0.129 & $\times$ & 1000 & 1.153 & 0.266 & 300 & M95 & \\
 A667         &  & 08 28 05.8   & 44 46 03   &  0.145 & $\times$ & & & & & C95 & \\
 A671           & & 08 28 31.5 & 30 25 54 & 0.051 & N & 1000 & 1.293 & 0.262 &  66 & M95 & IC2378\\
 A665           & & 08 30 57.3 & 65 50 32 & 0.1824 & $\times$ & 1000 & 1.385 & 0.282 & 310 & M95 & \\
 A689         &  & 08 37 24.6   & 14 58 22 &  0.2793 & $\times$ & & & & & C95 & \\
 A697         &  &  08 42 57.5  & 36 22 01   &  0.282 & $\times$ & & & & & C95 & \\
 A750           & & 09 09 12.6 & 10 58 29 & (0.177) & N & 1000 & 1.184 & 0.253 &  66 & M95 & \\
 A763           & & 09 12 35.3 & 16 00 03 & 0.0892 & $\times$ & 1000 & 1.080 & 0.233 & 225 & D96 & \\
 A757           & & 09 13 07.8 & 47 42 32 & 0.0520 & $\times$ & 1000 & 1.058 & 0.109 & 175 & D96 & \\
 A773        & a (S) & 09 17 53.4   & 51 43 39   &  0.216 & $\times$ & & & & & C95 & \\
            & b (N)&  09 17 53.5   & 51 44 03  &  0.224 & $\times$ & & & & & C95 & \\
 A795           & & 09 24 05.3 & 14 10 22 & 0.1355 & $\surd$ & 1000 & 1.042 & 0.234 & 210 & D96 & \\
 Z2701        &  & 09 52 49.2   & 51 53 06  &  0.215 & $\surd$ & & & & & C95 & \\
 Z2844          & & 10 02 36.6 & 32 42 26 & 0.0502 & $\times$ & 1000 & 1.138 & 0.101 & 270 & D96 & \\
 A961           & & 10 16 23.0 & 33 38 19 & 0.124 & $\times$ & 1000 & 1.190 & 0.113 & 90 & M98  & \\
 A963           & & 10 17 03.6 & 39 02 51 & 0.2059 & $\times$ & 1000 & 1.038 & 0.093 & 232 & D96 & \\
\end{tabular}
\end{table}
\vfill\eject

\addtocounter{table}{-1}
\begin{table}
\caption{Log of the observations -- Clusters in the BCS continued }
\begin{tabular}{llccccrccrll}
             &   & & &      & & & & & & & \\
Cluster      &   & RA         & DEC      & Redshift & Lines? & Exp &
Airmass & A(V) & p.a. & Ref &\\
             &   & (J2000)    & (J2000)  &       & & (sec)  & & (mag) & (deg) & & \\
             &   & & &       & & & & & & & \\
 A980           & & 10 22 28.4 & 50 06 21 & 0.158 & [$\times$] & 1000 & 1.144 & 0.070 & 128 & M95 & \\
 Z3146          & & 10 23 39.6 & 04 11 12 & 0.2906 & $\surd$ & & & & & A92 & \\
 A990   & & 10 23 39.8 & 49 08 39 & 0.142 & $\times$ & & & & $\dag$ & C95 & \\
 Z3179  & &  10 25 58.0 & 12 41 09 &  0.1432  & $\surd$ &  & & & &A92 & \\
 A1033  & & 10 31 44.3  & 35 02 30  &  0.1259   &$\times$ &   & & & &A92 & \\
 A1068   & & 10 40 44.4 & 39 57 12  &  0.1386 & $\surd$ & & & && A92 &\\
 A1132          & & 10 58 23.5 & 56 47 43 & 0.1365 & $\times$ & 1000 & 1.141 & 0.042 & 166 & M95 & \\
 A1177          & & 11 09 44.4 & 21 45 34 & 0.0323 &  $\times$ & 1000 & 1.008 & 0.077 & 113 & M95 & NGC3551 \\
 A1185 &a (N)     & 11 10 38.3 & 28 46 03 & 0.0348 & $\times$ & 1000 & 1.005 & 0.115 &  45 & M95 & NGC3550\\
       &b  (S,NW) & 11 10 42.9 & 28 41 36 & 0.0331 & $\times$ & 1000 & 1.027 & 0.115 & 332$\dag$  & M95 & NGC3552 \\
       &c  (S,SE) & 11 10 47.9 & 28 39 37 & 0.0288 & $\times$ & $''$ &$''$  &$''$ & $''$ & $''$ & NGC3554 \\
 A1190 &          & 11 11 43.6 & 40 49 15 & 0.0772 & $\times$ & 1000 & 1.693 & 0.112 & 260 & M95 & \\
 A1201 &          & 11 12 54.4 & 13 26 10 & 0.1679 &  [$\times$] & 1000 & 1.040 & 0.107 &  10 & D96 & \\
 A1204   & & 11 13 20.3 & 17 35 41  &  0.1706 & $\surd$ &  & & & & A92 & \\
 A1246 &          & 11 23 58.8 & 21 28 48 & 0.1904 & $\times$ &      &       &       & & A92 & \\
 A1302           & & 11 33 14.6 & 66 22 46 & 0.1148 & $\times$ & 1000 & 1.264 & 0.071 &   5 & D96 & \\
 A1314          & & 11 34 49.2 & 49 04 41 & 0.0337 & $\times$ & 1000 & 1.095 & 0.112 & 144 & M95 & IC712 \\
 A1361   & & 11 43 39.5 & 46 21 22  &  0.1167  & $\surd$ &  & & & & A92 & \\
 A1366          & & 11 44 36.7 & 67 24 20 & 0.1155 & $\times$ & 1000 & 1.281 & 0.090 &   0 & M95 & \\
 A1413          & & 11 55 17.8 & 23 24 20 & 0.1427 & $\times$ & 1000 & 1.005 & 0.146 & 180 & M95 & \\
 A1423      &  &  11 57 17.2  & 33 36 41   & 0.213 & $\times$ & & & & $\dag$ & C95 & \\
 A1437 &a (N)     & 12 00 25.3 & 03 20 50 & 0.1336 & $\times$ & 1000 & 1.112 & 0.122 &  10 & M95 & \\
       &b (S)     & $''$       & $''$     & 0.1315 & $\times$ & $''$ & $''$ & $''$ & $''$  & $''$ &\\
 Z4803 &          & 12 04 26.8 & 01 53 45 & 0.0194 & $\times$ & 1000 & 1.166 & 0.125 & 210 & M95 & NGC4073 \\
 RXJ1205.1$+$3920 & & 12 05 10.2   & 39 20 49  & 0.037 & $\times$ & & & & & C95 & \\
 RXJ1206.5+2810 & & 12 06 38.8 & 28 10 27 & 0.0281 & $\surd$ & 1000 & 1.009 & 0.113 & 275$\dag$  & D96 & NGC4104 \\
 Z4905         &  & 12 10 16.8 & 05 23 11 & {\bf 0.0766} & N & 1000 & 1.108 & 0.099 &  20 & M95 & \\
 Z5029         &   & 12 17 41.0 & 03 39 23 & 0.0767 & $\times$ & 1000 & 1.110 & 0.115 & 190 & M95 & \\
 RXJ1223.0+1037 &  & 12 23 06.6 & 10 37 17 & 0.0259 & $\surd$ & 1000 & 1.625 & 0.149 & 300 & D96 & NGC4325 \\
 A1553 &a (NW)    & 12 30 43.3 & 10 34 44 & 0.1634 & [$\times$] & 1000 & 1.272 & 0.133 & 303 & D96 & MCG+02-32-109 \\
       &b (SE)    & 12 30 48.9 & 10 32 48 & 0.1715 & [$\times$] & 1000 & 1.355 & 0.133 & 303 & D96 & \\
 RXJ1230.7+1220 & & 12 30 49.5 & 12 23 22 & 0.0036 & $\surd$ &  500 & 1.271 & 0.163 & 240 & M95 & M87\\
 Z5247          & & 12 34 17.5 & 09 46 00 & 0.229 & $\times$ & 1000 & 1.140& 0.125 & 220 & M98 &  \\ 
 A1589          & & 12 41 17.4 & 18 34 30 & 0.0709 & $\times$ & 1000 & 1.018 & 0.127 & 180 & M95 & MCG+03-32-083\\
 A1656          & & 12 59 35.6 & 27 57 35 & 0.0237 & $\times$ & 1000 & 1.010 & 0.061 & 262 & M95 & NGC4874 \\
 A1668          & & 13 03 46.6 & 19 16 18 & 0.0640 & $\surd$ & 1000 & 1.014 & 0.147 & 180 & M95 &  IC4130\\
 A1677 &   & 13 05 50.7 & 30 54 20 & 0.1845  & $\times$   & & & &  & A92 & MCG+05-31-128\\
 A1682 & a & 13 06 45.7 & 46 33 31 &  0.2190  &  $\times$ & & & & & A92 & \\
       & b & 13 06 49.8 & 46 33 35  &  0.2330  & $\times$ &  & & & & A92 & \\
 RXJ1320.1+3308 &a (W) & 13 20 14.6 & 33 08 39 & 0.0360 & $\surd$ & 1000 & 1.016 & 0.070 & 250$\dag$  & D96 & NGC5096 \\
                &b (E) & 13 20 17.6 & 33 08 44 & 0.0377 & $\times$ & $''$ & $''$ & $''$ & $''$ & $''$ & NGC5098 \\
 RXJ1326.3+0013 & & 13 26 17.6 & 00 13 17 & {\bf 0.0821} & $\times$ & 1000 & 1.403 & 0.122 & 230 & M95 & \\
 A1758a         & & 13 32 38.4  & 50 33 38 & 0.2792 & $\times$ & & & &  & A92 & \\
 A1763          & & 13 35 20.1 & 41 00 05 & 0.2280 & $\times$ & 1000 & 1.023 & 0.061 &   3 & M95 & \\
 A1767          & & 13 36 08.1 & 59 12 24 & 0.0715 & N & 1000 & 1.160 & 0.119 & 180 & M95 & MCG+10-19-096\\
 A1775 &a (SE)    & 13 41 50.5 & 26 22 15 & 0.0700 & $\times$ & 1000 & 1.004 & 0.070 & 123$\dag$  & M95 & UGC8669 \\
       &b (NW)    & 13 41 49.1 & 26 22 27  & 0.0758 & $\times$ & $''$ & $''$ & $''$ & $''$ & $''$ & \\
 A1773          & & 13 42 09.6 & 02 13 39 & 0.0763 & $\times$ & 1000 & 1.140 & 0.118 & 200 & M95 & \\
 A1795          & & 13 48 52.5 & 26 35 37 & 0.062 & $\surd$ & 1000 & 1.011 & 0.077 & 165$\dag$  & M95 & MCG+05-33-005\\
 A1800          & & 13 49 23.5 & 28 06 27 & 0.0750 & $\times$ & 1000 & 1.000 & 0.080 & 330 & M95 & UGC8738 \\
 A1809          & & 13 53 06.3 & 05 08 59 & 0.0796 & $\times$ & 1000 & 1.704 & 0.136 & 300 & D96 & \\
 A1831          & & 13 59 15.0 & 27 58 34 & 0.0760 & $\times$ & 1000 & 1.007 & 0.093 &  80 & M95 & MCG+05-33-033\\
 A1835          & & 14 01 02.0 & 02 52 45 & 0.2523 & $\surd$ & 1000 & 1.189 & 0.150 & 214 & M95 & \\
   & & $''$  & $''$ & $''$ & $\surd$ &  & & & & A92 & \\
   & & $''$ & $''$  & $''$ & $\surd$ & 1000 & 1.111 & 0.155 & 180 & M95 & \\
 A1885        &  &  14 13 43.6  & 43 39 45  & 0.090 & $\surd$ & & & & $\dag$  & C95&  \\
 Z6718        &  & 14 21 35.8   & 49 33 05  & 0.071 & $\times$ & & & & & C95 & \\
 A1902        &  &  14 21 40.6  & 37 17 31  & 0.160 & [$\times$] & & & & & C95 & \\
 A1914          & & 14 25 56.8   & 37 49 00 & 0.170 & [$\times$] & 1000 & 1.023 & 0.064 & 238 & M95 & \\
 A1918         &  & 14 25 22.5 & 63 11 55 & 0.139 & $\times$ & 1000 & 1.212 & 0.117 & 183 & M98 &  \\
 A1927          & & 14 31 06.6 & 25 38 02 & 0.0967 & $\times$ & 1000 & 1.253 & 0.147 &  67 & D96 & \\
 A1930          & & 14 32 37.9 & 31 38 49 & 0.1316 & $\surd$ & 1000 & 1.011 & 0.077 & 290 & M95 & \\
\end{tabular}
\end{table}
\vfill\eject

\addtocounter{table}{-1}
\begin{table}
\caption{Log of the observations -- Clusters in the BCS continued }
\begin{tabular}{llccccrccrll}
             &   & & &       & & & & & & & \\
Cluster      &   & RA         & DEC      & Redshift & Lines? & Exp &
Airmass & A(V) & p.a. & Ref & Notes\\
             &   & (J2000)    & (J2000)  &       & & (sec)  & & (mag) & (deg) & & \\
             &   & & &       & & & & & & & \\
 RXJ1440.6+0327 &a (SE) & 14 40 42.9 & 03 27 58 & 0.0274 & $\times$ & 1000 & 1.108 & 0.189 & 285$\dag$  & M95 & NGC5718 \\
                &b (NW) & 14 40 39.0 & 03 28 13 & 0.0268 & $\times$ & $''$ & $''$ & $''$ & $''$ & $''$ & IC1042 \\
 RXJ1442.2+2218 & & 14 42 19.4 & 22 18 13 & {\bf 0.0970} & $\surd$ & 1000 & 1.563 & 0.173 & 289 & D96 & UGC9480 \\
 A1978          & & 14 51 09.5 & 14 36 45 & {\bf 0.1468} & $\times$ & 1000 & 1.031 & 0.118 &   2 & M95&  \\
 A1983 &a (S)     & 14 52 55.4 & 16 42 11 & 0.0442 & [$\times$] & 1000 & 1.357 & 0.136 & 194$\dag$  & D96 & MCG+03-38-044\\
       &b (N)     & 14 52 57.0 & 16 43 40 & 0.0459 & [$\times$] & $''$ & $''$ & $''$ & $''$ & $''$ & \\
 A1991 &          & 14 54 31.4 & 18 38 34 & 0.0595 & $\surd$ & 1000 & 1.020 & 0.163 & 148 & M95 & NGC5778 \\
 Z7160 &          & 14 57 15.1 & 22 20 35 & 0.2578 & $\surd$ & & & & & A92 & \\
 A2009 &          & 15 00 19.6 & 21 22 11 & 0.1532 & $\surd$ & 1000 & 1.023 & 0.221 &  46 & M95 & \\
 A2034        & a &   15 10 10.2 &  33 34 03 & 0.115 & $\times$ & & & & & C95 & \\
              & b &   15 10 11.6 &  33 29 13 & 0.111 &$\times$ & & & & & C95 & \\
 A2029 &          & 15 10 56.2 & 05 44 43 & 0.0786 & $\times$ & 1000 & 1.088 & 0.204 & 170 & M95 & IC1101 \\
 A2033  &         & 15 11 26.6 & 06 20 58 & 0.078 & N & 1000 & 1.115 & 0.198 & 243 & M95 & UGC9756 \\
 A2050 &  &         15 16 18.0 & 00 05 22 & 0.118 & $\times$ & 1000 & 1.140 &0.316 & 165  & M98 & \\
 A2052  &         & 15 16 44.6 & 07 01 18 & 0.0351 & $\surd$ & 1000 & 1.103 & 0.195 &  30 & M95 & UGC9799 \\
 A2055 &a (SE)    & 15 18 45.8 & 06 13 57 & 0.1019 & N & 1000 & 1.087 & 0.214 & 286$\dag$  & M95 &  \\
       &b (SW)    & 15 18 40.7 & 06 14 19 & 0.1031 & $\times$ & $''$ & $''$ & $''$ & $''$ & $''$ & \\
       &c (N)     & 15 18 45.0 & 06 16 28 & 0.1056 & $\times$ & 1000 & 1.112 & 0.214 & 210 & M95 & \\
 A2064          & & 15 20 52.2 & 48 39 40 & 0.0741 & $\times$ & 1000 & 1.538 & 0.123 & 269 & D96 & \\
 A2061 &a (S)     & 15 21 11.1 & 30 35 03 & 0.0753 & $\times$ & 1000 & 1.048 & 0.128 & 272 & M95 & MCG+08-28-020 \\
       &b (N)     & 15 21 20.5 & 30 40 16 & 0.077 & $\times$ & 1000 & 1.002 & 0.128 & 242 & M95 & \\
 RXJ1522.0+0741 & & 15 21 51.9 & 07 42 32 & 0.0451 & [$\times$] & 1000 & 1.115 & 0.202 & 115 & M95 & NGC5920 \\
 A2065 &a (W)     & 15 22 24.1 & 27 42 53 & 0.0698 & $\times$ & 1000 & 1.046 & 0.195 & 110$\dag$  & M95 & MCG+05-36-020\\
       &b         & 15 22 27.7 & 27 42 36 & 0.0710 & $\times$ & $''$ & $''$ & $''$ & $''$ & $''$ & \\
       &c (E)     & 15 22 29.2 & 27 42 28 & 0.0749 & $\times$ & $''$ & $''$ & $''$ & $''$ & $''$ & \\
 A2063 &          & 15 23 05.4 & 08 36 35 & 0.0342 & $\times$ & 1000 & 1.066 & 0.199 & 180 & M95 & MCG+02-39-020\\
 A2069 &a  (SE)   & 15 24 08.4 & 29 52 56 & 0.1138 & $\times$ & 1000 & 1.148 & 0.134 & 331$\dag$  & M95 & \\
       &b  (NW)   & 15 24 07.4 & 29 53 20 & 0.1098 & $\times$ & $''$ & $''$ & $''$ & $''$ & $''$ & \\
       &c  (NW)   & $''$       & $''$     & 0.1135 & $\times$ & $''$ & $''$ & $''$ & $''$ & $''$ & \\
 A2072 &          & 15 25 48.7 & 18 14 11 & {\bf 0.127} & $\surd$ & 1000 & 1.025 & 0.262 & 210 & M95 & \\
 RXJ1532.9+3021 & & 15 32 53.8 & 30 21 00 & {\bf 0.3615} & $\surd$ & 1000 & 1.036 & 0.147 & 270 & M95 & \\
 A2107        &   & 15 39 39.1 & 21 46 57 & 0.042 & $\times$ & 1000 & 1.021 & 0.298 & 230 & M95 & UGC9958 \\
 A2111 & a  (NW)   & 15 39 40.4 & 34 25 28 & 0.2317 & $\times$ & 1000 & 1.334 & 0.129 & 340 & D96 & \\
       &b  (SE)   & 15 39 41.8 & 34 24 44  & 0.2300 & $\times$ & $''$ & $''$ & $''$ & $''$ & $''$ & \\
 A2110 &          & 15 39 50.8 & 30 43 05 & 0.0976 & $\times$ & 1000 & 1.004 & 0.159 & 180 & M95 & \\
 A2124 &          & 15 44 59.1 & 36 06 35 & 0.0663 & $\times$ & 1000 & 1.035 & 0.118 & 297 & M95 & UGC10012 \\
 A2142 &a  (SE)   & 15 58 20.0 & 27 14 02 & 0.089 & $\times$ & 1000 & 1.044 & 0.262 & 301$\dag$  & M95 & \\
       &b  (NW)   & 15 58 13.3 & 27 14 55 & 0.095 & $\times$ & $''$ & $''$ & $''$ & $''$ & $''$ & \\
 A2147 &          & 16 02 17.0 & 15 58 30 & 0.0357 & $\times$ & 1000 & 1.030 & 0.226 & 200 & M95 & UGC10143 \\
 A2151a &         & 16 04 35.7 & 17 43 19 & 0.035 & $\times$ & 1000 & 1.115 & 0.227 & 240 & M95 & NGC6041 \\
 RXJ1604.9+2356 & & 16 04 56.6 & 23 55 58 & 0.0324 & $\times$ & 1000 & 1.024 & 0.330 &  63 & M95 & NGC6051 \\
 A2175          & & 16 20 31.2 & 29 53 29 & 0.0961 & $\times$ & 1000 & 1.041 & 0.166 & 270 & M95 & \\
 A2199          & & 16 28 38.5 & 39 33 05 & 0.031 & $\surd$ & 1000 & 1.096 & 0.058 & 289 & M95 & NGC6166\\
 A2204          & & 16 32 46.9 & 05 34 33 & 0.1514 & $\surd$ & 1000 & 1.093 & 0.378 & 317$\dag$  & M95 & \\
 A2219 & a & 16 40 12.9 & 46 42 43  &  0.2250  &  $\times$ & & & & & A92 & \\
      & b &  16 40 19.6 & 46 42 43 &  0.2248  &  $\times$ & & & & & A92 & \\
      & c &  16 40 21.9  & 46 42 48 &  0.2344  &  $\times$ & & & & & A92 & \\
 RXJ1657.8+2751 &    & 16 57 58.1 & 27 51 16 & 0.035 & $\times$ & 500 &      1.526 & 0.284 & 105 & M98 & NGC6269 \\
 A2244          & & 17 02 42.5 & 34 03 39 & 0.0980 &  $\times$ & 500 & 1.012 & 0.134 & 310 & M95 & \\
 A2256   &a  (SE) & 17 03 35.6 & 78 37 46 & 0.0541 &  $\times$ & 500 & 1.565 & 0.275 & 300$\dag$  & M95 & NGC6331 \\
         &b  (SE) & 17 03 33.7 & 78 37 49 & 0.0565 &  $\times$ & $''$ & $''$ & $''$ & $''$ & $''$ & \ \ \  $''$ \\
         &c  (NW) & 17 03 29.0 & 78 37 57   & 0.0599 &  $\times$ & $''$ & $''$ & $''$ & $''$ & $''$ &  \\
         &d  (E)  & 17 04 26.9 & 78 38 27 & 0.0596 & $\times$ & 1000 & 1.573 & 0.275 & 155 & M95 & MCG+13-12-018 \\
 A2249   &        & 17 09 48.6 & 34 27 34 & 0.0873 & $\times$ & 1000 & 1.005 & 0.166 &   0 & M95 & \\
 A2255  &a (NE)   & 17 12 34.8 & 64 04 16 & 0.0845 & $\times$ & 1000 & 1.235 & 0.170 & 229$\dag$  & M95 & \\
        &b (SW)   & 17 12 28.7 & 64 03 40  & 0.0750 & $\times$ & $''$ & $''$ & $''$ & $''$ & $''$ & \\
 RXJ1715.3+5725&  & 17 15 22.6 & 57 24 43 & 0.0282 & $\surd$ & 1000 & 1.150 & 0.169 & 160 & M95 & NGC6338 \\
 A2254       &  & 17 17 45.8  & 19 40 51  & 0.178 & $\times$ & & & & & C95 & \\
 Z8197         &  & 17 18 11.8 & 56 39 56 & {\bf 0.1140} & $\surd$ & 1000 & 1.132 & 0.176 &   0 & M95 & \\
 A2259       &  &  17 20 09.6 &  27 40 09 & 0.164 & [$\times$] & & & & & C95 & \\
 RXJ1720.1+2638&  & 17 20 10.1 & 26 37 32 & 0.1611 & $\surd$ & 1000 & 1.001 & 0.262 &   0 & M95 & \\
 A2261     &  & 17 22 27.2  & 32 07 58  & 0.224 & $\times$ & & & & & C95 & \\
\end{tabular}
\end{table}
\vfill\eject

\addtocounter{table}{-1}
\begin{table}
\caption{Log of the observations -- Clusters in the BCS continued }
\begin{tabular}{llccccrccrll}
             &   & & &      & & & & & & & \\
Cluster      &   & RA         & DEC      & Redshift & Lines? & Exp &
Airmass & A(V) & p.a. & Ref & Notes\\
             &   & (J2000)    & (J2000)  &       & & (sec)  & & (mag) & (deg) & & \\
             &   & & &       & & & & & & & \\
 A2294    &  & 17 24 11.3  & 85 53 13  & 0.178 & $\surd$ & & & & & C95 & \\
 RXJ1733.0+4345&  & 17 33 02.1 & 43 45 35 & {\bf 0.0331} & $\surd$ &1000 & 1.036 & 0.164 & 190 & M95 & IC1262\\
 RXJ1740.5+3539& a (NE) & 17 40 34.4 & 35 39 14 & {\bf 0.0416} & $\times$ & 1000 & 1.015 & 0.176 & 226$\dag$  & M95 & MCG+06-39-010\\
        & b  (SW)       & 17 40 32.1  & 35 38 47 & {\bf 0.0448} & [$\times$] & $''$ & $''$  & $''$  & $''$  & $''$ & MCG+06-39-009\\
 Z8276  &         & 17 44 14.5 & 32 59 30 & 0.075 & $\surd$ & 1000 & 1.010 & 0.240 & 297 & M95 & \\
 RXJ1750.2+3505 & & 17 50 16.9 & 35 04 59 & {\bf 0.1712} & $\surd$ &1000 & 1.014 & 0.205 & 310 & M95 & \\
 Z8338          & & 18 11 05.2 & 49 54 34 & 0.047 & $\times$ & 600 & 1.629 & 0.284 & 90 & M98 & NGC6582 \\
A2318          & & 19 05 11.0 & 78 05 38 &  0.1405      & $\times$ &      &       &       &     &  A92 & \\
 RXJ2114.1+0234 & & 21 13 56.0 & 02 33 56 & {\bf 0.0497} & $\times$ & 700  & 2.579 & 0.437 &  60 & D96 & IC1365 \\
 RXJ2129.6+0005 & & 21 29 39.9 & 00 05 23 & {\bf 0.2346} & $\surd$ & 1000 & 1.593 & 0.272 & 310 & M95 &  \\
 A2390          & & 21 53 36.7 & 17 41 45 & 0.2328 & $\surd$ & 1000 & 1.754 & 0.453 &  68 & D96 & \\
 A2409          & & 22 00 52.6 & 20 58 10 &  0.1470  & $\times$ &  & & & & A92 & \\
 A2443          & & 22 26 08.0 & 17 21 25 & 0.1105 & $\times$ & 1000 & 2.157 & 0.349 &  65 & D96 & \\
 A2457          & & 22 35 40.9 & 01 29 07 & 0.0592 & $\times$ & 1000 & 1.913 & 0.383 &  60 & D96 & \\
 A2495          & & 22 50 19.6 & 10 54 13 & 0.0808 & $\surd$ & 1000 & 1.759 & 0.353 &  60 & D96 & MCG+02-58-021\\
 Z8852          & & 23 10 22.3 & 07 34 52 & 0.0399 &  $\times$ & 600 & 2.510 & 0.325 &  60 & D96 & NGC7499\\
 A2572a  &a (NE)  & 23 17 13.5 & 18 42 30 & 0.0400 & $\times$ & 1000 & 2.050 & 0.316 & 222 & D96 & NGC7578B\\
         &b (SW)  & 23 17 11.9  & 18 42 05 & 0.0400 & $\times$ & $''$ & $''$ & $''$ & $''$ & $''$ & NGC7578A \\
 A2572b  &        & 23 18 30.3 & 18 41 21 & 0.0370 & $\times$ & 1000 & 1.774 & 0.313 & 247 & D96 & NGC7597 \\
 A2589          & & 23 23 57.4 & 16 46 39 & 0.0407 & $\times$ & 1000 & 1.154 & 0.277 & 240 & D96 & NGC7647 \\
 A2593          & & 23 24 20.1 & 14 38 50 & 0.0423 & $\times$ &  600 & 2.409 & 0.279 &  65 & D96 & NGC7649 \\
 A2626 &a (NE)    & 23 36 30.7 & 21 08 49 & 0.0556 & $\times$ & 1000 & 1.383 & 0.289 & 212$\dag$  & D96 & IC5338 \\
       &b (SW)    & $''$       & $''$     & 0.0552 & $\surd$ & $''$ & $''$ & $''$ & $''$ & $''$ & \\
 A2627       & a (N) & 23 36 42.1   & 23 55 30  & 0.127 & N & & & & & C95 & \\
             & b (S) & 23 36 42.5   & 23 54 46  & 0.122 & N & & & & & C95 & \\
 A2634  &         & 23 38 29.3 & 27 01 53 & 0.0298 & $\surd$ & 1000 & 1.253 & 0.338 & 257 & D96 & NGC7720 \\
 A2657  &         & 23 44 57.3 & 09 11 36 & 0.0401 & $\times$ & 1000 & 1.123 & 0.377 & 220 & D96 & MCG+01-60-030\\
 A2665  &         & 23 50 50.6 & 06 09 00 & 0.0567 & $\surd$ & 1000 & 1.712 & 0.409 &  60 & D96 & MCG+01-60-039\\
 A2675  &         & 23 55 42.6 & 11 20 35 & 0.0746 & $\times$ & 1000 & 2.211 & 0.321 &  65 & D96 & \\

\end{tabular}
\end{table}
\vfill\eject

\newpage

\begin{table}
\caption{Log of the observations -- Clusters not included in either the BCS  
due to their faint X-ray flux, declination $<0$, or low
Galactic latitude. \label{tab:notbcslog}}
\begin{tabular}{llccccrccrll}
             &   & & &       & & & & & & & \\
Cluster      &   & RA         & DEC      & Redshift & Lines? & Exp &
Airmass & A(V) & p.a. & Ref & Notes\\
             &   & (J2000)    & (J2000)  &       & & (sec)  & & (mag) & (deg) & & \\
             &   & & &       & & & & & & & \\
 Z353         &   & 01 07 40.7 & 54 06 33   & (0.109) &  $\times$  & & & &  &C95 & low $|b|$\\
 A291 $\dag$   &   & 02 01 43.1 & --02 11 47  & 0.196 & $\surd$  & & & & &C95 & \\
 A531         &   &  05 01 16.3 & --03 33 47 & 0.094 & $\times$ & & & & &  C95 & \\
 RXJ0510.7$-$0801 & & 05 10 47.8  & --08 01 44 & 0.217 & $\times$ & & & &  & C95 & \\
 Z1121        &  &   06 31 22.7  & 25 01 07   &  0.083 & N  & & & &  & C95  & low $|b|$, 3C162 \\
 Z1133        &  &  06 38 04.0  & 47 47 56   &  0.174 & $\times$ & & & &  & C95  & low $|b|$\\
 A611         &  &   08 00 56.7 & 36 03 25   &  0.288 & N & & & & & C95 & sub-BCS\\
 A621         &  &  08 11 12.2  & 70 02 30   &  0.223 & $\times$ & & & & & C95 & sub-BCS\\
 RXJ0821.0$+$0752 & & 08 21 02.4   & 07 51 47   & 0.110 & $\surd$ & & & & $\dag$ & C95 & sub-BCS\\
 A761         &  & 09 10 35.9  & --10 35 00&  0.091 & $\times$ & & & &  & C95 & XBACS\\
 RXJ1000.5$+$4409 & a & 10 00 25.1   & 44 09 14  & 0.155 & $\surd$ & & & &  & C95 & \\
                  & b & 10 00 31.1   & 44 08 44  & 0.153 & $\surd$ & & & &  & C95 & \\
 A971         &  &  10 19 52.0  & 40 59 19  & 0.093 & $\times$ & & & &$\dag$ & C95 & sub-BCS \\
 A1023  & &10 27 58.5 & --06 47 56 &  0.1165  & N & & & &  & A92 & \\
 A1035          & & 10 32 14.0 & 40 16 17 & 0.078 & $\times$ & 1000 & 1.120 & 0.093 & 110 & M95 & sub-BCS\\
 A1045  & &10 35 00.1 & 30 41 39  &  0.1381  &  $\times$ & & & & & A92 & sub-BCS\\
 A1084   & & 10 44 32.9 & --07 04 08 &  0.1329   & $\surd$  & & & & & A92 & XBACS \\
 A1173          & & 11 09 15.2 & 41 33 43 & {\bf 0.0767} & $\times$ & 1000 & 1.028 & 0.102 & 342 & M95 & sub-BCS\\
 Z3916 $\dag$       &  &  11 14 21.9  & 58 23 20   & 0.204 & $\surd$ & & & &  & C95 & \\
 Z4673        &  & 11 56 55.6 & 24 15 37     &  0.1419  & $\times$  & & & & & A92 & sub-BCS\\
 Z5604        &  & 12 57 21.6  & 69 30 20  &  0.227 & $\times$ & & & &  & C95 & \\
 A1651        & & 12 59 22.4 & --04 11 44 & 0.0860   & $\times$ & & & &  & A92 & XBACS\\
 A1664  & & 13 03 42.5 & --24 14 41 &  0.1276 & $\surd$ & & & &  & A92 & XBACS\\
 A1672  & & 13 04 27.1 & 33 35 16  &  0.1882  & $\times$ & & & &  & A92 & \\
 A1703 & a & 13 15 05.1 & 51 49 04  &  0.2836  &$\times$ & & & &  & A92 & \\
      & b & 13 15 11.0 & 51 49 04  &  0.2336  & $\times$ & & & &  & A92 & \\
 RXJ1449.5+2746 & & 14 49 27.9 & 27 46 51 & 0.0311 & N & 1000 & 1.723 & 0.133 & 287 & D96 & sub-BCS, MCG+05-35-018 \\
 A2104  & &15 40 08.0  &--03 18 16  &  0.1554  & $\times$ & & & &  & A92 & XBACS\\
 A2146  & &  15 56 13.8 & 66 20 55  &  0.2343  &  $\surd$ & & & & & A92 & sub-BCS\\
 Z7833  & & 16 10 00.7 & 67 10 15  &  0.2136  & $\times$ & & & &  & A92 & \\
 A2187  & & 16 24 13.9 & 41 14 38 & 0.1839 & $\times$ & & & & & A92 & sub-BCS\\
 A2201 & & 16 26 58.8 & 55 28 30  & 0.130 & $\times$ & & & &  & C95 & \\ 
 A2208  & & 16 29 38.8 & 58 31 52  &  0.1329  & $\times$ & & & &  & A92 & \\
 A2228  & &16 48 00.8  & 29 56 59   &  0.1013  &  $\times$ & & & &  & A92 & Also AGN at same redshift \\ 
 Z8193  & & 17 17 19.1 & 42 26 59  &  0.1754  & $\surd$ & & & &  & A92 & \\
 A2292      &  & 17 57 06.7  & 53 51 38  & 0.119 & $\surd$ & & & &  & C95 & \\
 Z8451  & & 19 57 14.0 & 57 51 27  &  0.0884  & $\times$ & & & &  & A92  & low $|b|$\\
 Z8503  & &   21 22 19.8 & 23 10 33 &  0.1430  & $\times$ & & & &  & A92 & low $|b|$\\
 A2426  & &  22 14 31.5 & --10 22 26 &  0.0990  & $\times$ & & & &  & A92 & \\ 
 A2428  & & 22 16 15.6 & --09 20 00 &  0.0846  & $\times$ & & & &  & A92 & XBACS\\
 A2631        &  & 23 37 41.1   & 00 17 06 & 0.278 & $\times$ & & & & & C95 & sub-BCS\\

\end{tabular}
\end{table}

\newpage

\begin{table}
\noindent
{\bf Notes on individual entries in Tables~\ref{tab:bcslog}  and
\ref{tab:notbcslog}:}\\
{\bf A21}  Two galaxies form a double system in the slit, separated in projection by 6.8 arcsec (16kpc).
     Galaxy a (to the SE) is the brighter, but galaxy b has an asymmetric diffuse envelope (see Porter,
     Schneider \& Hoessel 1991). \\
{\bf A115} The central cluster galaxy contains the  radio source
   {\bf3C28}.
     The ROSAT HRI image of this cluster shows two components to the X-ray
     morphology, suggesting it to be a double cluster. The smaller
     component to the south is centred on a galaxy at RA 00 56 00, Dec
     26 20 37 (J2000)  but we have taken a spectrum
     for 3C28 at the centre of the main X-ray component.  \\
{\bf RXJ0107.4+3227}  This cluster is also known as the group {\bf IV Zw 038}, and the 
galaxy observed (NGC383) is associated with the
radio source {\bf3C31}. Colina \& Perez-Fournon (1990) note a central dust lane in this galaxy.\\
{\bf A168} The observed galaxy is offset by 3--4$'$ from the X-ray centroid,
 but the X-ray image of the cluster is broad with no tight core, and the
galaxy observed  is clearly the brightest cluster galaxy. \\
{\bf RXJ0123.2+3327} (NGC499) and {\bf RXJ0123.6+3315} (NGC507) form a double system.\\
{\bf A262} The central cluster galaxy is known from ground-based images
     to be bifurcated by a central dust lane (Colina \& Perez-Fournon 1990),which splits the galaxy 
     continuum light in our slit into  two components
     separated in projection by 1.6 arcsec (0.8kpc). The
     line emission is spatially centred exactly on the dust lane separating
these two components.\\
{\bf A272} We have taken a spectrum of the brightest dominant galaxy, although
a second candidate dominant galaxy lies 2.4 arcmin to the N. \\
{\bf A400} The dominant galaxies are a dumbell system separated by 16.2 arcsec
     ($\sim11$kpc) in projection, and are associated with the twin radio
source {\bf 3C75}.  \\
{\bf Z808} The observed galaxy is associated with the radio source {\bf 4C$+$01.06}. \\
{\bf A407} Several bright galaxies appear to be embedded in a diffuse optical
halo within a region of diameter $\sim$60kpc.  We present the spectrum only of the brightest galaxy. \\
{\bf RXJ0338.7+0958} is the cluster 2A~0335+096.\\
{\bf RXJ0341.3+1524} The observed galaxy is catalogued as  {\bf III~Zw~054}. \\
{\bf A520} This is a bimodal cluster, and we have observed the 
     brightest (to the SW of the cluster) of three 
     dominant galaxies. This cluster was detected in the Einstein Medium
Sensitivity Survey      (EMSS; Stocke et al. 1991). \\
{\bf A566} is associated with the radio source {\bf 4C$+$63.10}. \\
{\bf A576} Galaxy b, the more southerly of the two  dominant galaxies observed
is the more probable central cluster galaxy. \\
{\bf A586} Two galaxies lie on our slit separated only by 5.1 arcsec (20kpc) in
projection. We present only the spectrum of the main galaxy 
     to the NW  of the two. \\
{\bf RXJ0751.3+5012} This cluster has a disturbed X-ray morphology. The three galaxies 
     for which we have spectra are the bright more southerly galaxy (a)
     and both components of a dumbell system to the north 
     (galaxy b to the E and c to the W of the pair; separated in
     projection  by 22.1 arcsec (15kpc)).\\
{\bf A602} We have taken the spectrum of each of the two equally bright dominant
galaxies in this double cluster.\\
{\bf A671} We have taken the spectrum only of the larger, brighter galaxy
surrounded by diffuse optical emission, which we assume is the BCG. A second very bright dominant
galaxy lies 3.1  arcmin to the SE. \\
{\bf A665} There is a second, less bright galaxy also within the slit, 
     10.6 arcsec (42kpc) to the SE of the dominant galaxy. We present
only the spectrum of the dominant galaxy.  \\
{\bf A689} This galaxy was observed in C95,  but the redshift given
in that paper is wrong. The spectrum is very noisy, and shows a 
 very blue galaxy with uncertain stellar absorption features. The new redshift has been
determined by Ebeling, Henry \& Mullis (1999), and is confirmed as the
redshift of the cluster from spectra of two other galaxies. 
The X-ray source
is coincident with a moderately bright radio source, which  could contribute significantly to the X-ray flux of
the total cluster emission. \\
{\bf A750} The cluster contains an AGN at the same redshift. There are
two equally bright dominant galaxies, separated by 94 arcsec, but we
only present a spectrum of the galaxy to the SE; the other lies at RA 09 09
07, Dec 10 67 51 (J2000). This cluster and the
AGN were detected in the EMSS (Stocke \etal 1991). \\
{\bf A763} The identification of this source is confused. Our optical
identification of the point-like object at the X-ray centroid 
(RA 09 12 31.0, Dec 15 55 25; J2000) is of a
star, despite it being associated with a moderate, spatially extended
radio source. The spectrum presented in this paper is of the 
 dominant galaxy 4.7 arcmin to the N of this star, which despite being the
nearest obvious BCG, is probably too far  to be associated with the X-ray source. \\
{\bf A773} The two galaxies appear to be equally dominant. \\
{\bf RXJ1000.5+4409} Galaxy a is the more likely central dominant galaxy. \\
{\bf Z3179} There is an IRAS-detected, foreground ($z=0.032$) spiral galaxy 110$''$ to the NW
of the observed galaxy. \\
{\bf A1023} Position of observed galaxy revised from A92. \\
{\bf A1035} The cluster is contaminated by a foreground group at $z=0.067$
(Maurogordato \etal 1997). \\
{\bf A1185} This X-ray source is very extended. Galaxy a is the brightest of
the three observed, and agrees with the X-ray centroid. \\
{\bf A1190} We have taken a spectrum only of the brighter of two nearby dominant
galaxies, that to the SE of the pair; the radio
source {\bf 4C$+$41.23}. is associated with the other galaxy. \\
{\bf A1201} We have taken a spectrum of the galaxy closest to the X-ray centroid. 
A brighter galaxy to the SE probably belongs to a foreground cluster. \\
{\bf A1204} Note that the declination given in A92 is incorrectly
given as $+11$. \\
{\bf A1361} The observed galaxy is associated with the radio source {\bf 4C$+$46.23}.\\
{\bf A1413} A second, smaller galaxy also lies on the slit, at a separation of only 
9.5 arcsec (31kpc) to the N. We present the spectrum only of the main galaxy. 
Maurogordato \etal (1997) note the presence of a small foreground group of
galaxies at $z=0.1$. \\
{\bf A1437} The main galaxy is close dumbell, of separation 3.2 arcsec (10 kpc) \\
{\bf Z4803} This source is also catalogued as the poor group {\bf MKW4}. \\ 
{\bf RXJ1206.5+2810}  The observed galaxy (NGC4104) is a double galaxy in
the group {\bf MKW4s}, the two components separated by 
      2.8 arcsec (2.2kpc). The emission line system
      spans the two systems. The cluster is an EMSS source (Stocke \etal
1991).  \\
{\bf Z4905} The observed galaxy has a very large and asymmetric halo on the POSS.\\
{\bf Z5029} The observed galaxy is not associated with the radio source {\bf
4C$+$04.41} which lies in this cluster.  \\
\end{table}
\vfill\eject

\begin{table}
\noindent
{\bf  Notes on individual entries in Tables~\ref{tab:bcslog} and
\ref{tab:notbcslog} continued:}  \\
{\bf RXJ1223.0+1037} The observed galaxy (NGC4325) may have nuclear activity as the X-ray source
     is barely extended in archival PSPC data, although the line emission is
quite muted.  Alternatively it could be an example of a
     highly focussed cooling flow in a low luminosity group.\\
{\bf A1553} We have taken the spectrum of each of the two equally dominant
bright galaxies in
this double cluster.\\
{\bf RXJ1230.7+1220} is M87 in the Virgo cluster, and associated with the
radio source {\bf3C274}. \\
{\bf A1656} This is the Coma cluster, and we have taken the spectrum only of
the more Western of the two dominant galaxies, NGC4874 .\\
{\bf A1668} The measured redshift is significantly different from that in
Rhee \& Katgert (1988). \\
{\bf A1672} Position of observed galaxy revised from A92. \\
{\bf A1677} Previously noted in A92 as Z5694. Position revised from original.\\
{\bf A1682} The two galaxies are equally dominant. \\
{\bf A1703} Galaxy a is the more likely central dominant galaxy. \\
{\bf RXJ1320.1+3308} Two equally bright galaxies lie along our slit position,
     with a projected separation of 38.1 arcsec (38kpc).\\
{\bf A1758} This double cluster has its two components listed separately in Paper~I. 
We list here the dominant galaxy associated only with
A1758a, which produces approximately 40 per cent of the combined luminosity; there is a
bright less dominant galaxy also at RA 13 32 51.9, Dec 50 31 36 (J2000). \\
{\bf A1763} The observed galaxy is associated with the radio source {\bf 4C$+$41.26}.\\
{\bf A1775} The two dominant galaxies form a double system, separated in projection by 21.1
     arcsec ($\sim40$kpc); they are associated with the radio source
     {\bf 4C$+$26.41}. \\
{\bf A1795} The observed galaxy is associated with the radio source {\bf 4C$+$26.42}. \\
{\bf A1831} The cluster shows a foreground cluster at $z=0.062$
(Maurogordato \etal
1997). \\
{\bf A1835} We present three different observations of the same central cluster galaxy, one
from A92, and two from our May observing run at slightly different position angles.  \\
{\bf A1914} This is a complex binary cluster, and we have only taken a
spectrum of the brighter 
    (to the SW) of the two dominant galaxies. 
    The radio source {\bf 4C$+$38.39} is probably a radio halo and hence not associated with the
observed galaxy. \\
{\bf RXJ1440.6+0327} The source is also catalogued as the poor group  {\bf MKW8} and the two galaxies observed are
    also catalogued as Arp171.    The galaxy (a) to the SE is probably the dominant
    galaxy; the two galaxies on the slit are separated in projection by
    by 59.2 arcsec (44kpc). \\
{\bf RXJ1449.5+2746}  The observed galaxy is offset 
from the centroid of this relatively
compact X-ray source,  indicating that it may well be unrelated to the X-ray
emission. \\ 
{\bf A1983} Two equally bright dominant galaxies lie along the slit position, separated by 92.6 arcsec (112kpc)\\
{\bf Z7160 } This cluster is the  EMSS cluster MS1455+223 (Stocke \etal 1991).\\
{\bf A2009} The cluster contains the radio source {\bf 4C$+$21.44} which
is, however, is unlikely
to be associated with the observed galaxy.\\
{\bf A2034} Galaxy b is the more likely central dominant galaxy. \\
{\bf A2029} The observed galaxy is associated with the radio source {\bf 4C$+$06.53}. \\
{\bf A2033} There are two galaxies along the slit; we take the spectrum of the
     main galaxy to the NE as the central dominant galaxy. The galaxy 38
     arcsec to the SW is a  foreground AGN at a redshift of
0.038. \\
{\bf A2052} The observed galaxy is associated with {\bf3C317}. \\
{\bf A2055} There are three equally bright  dominant galaxies;
     a and b lie along the same slit position, separated in
     projection by 80 arcsec (200kpc). None of these redshifts agrees with the
     previously published redshift of 0.053 in Struble \& Rood (1991).\\
{\bf A2064} The redshift of the observed galaxy is inconsistent with the
previously published redshift of 0.1074 from Owen, White \& Thronson (1988)
from which the cluster redshift has been derived. We find no evidence
to support the previous redshift in our spectrum.\\ 
{\bf A2061} This bimodal cluster is highly extended.
We have observed  the two equally bright dominant galaxies either side of the
X-ray centroid.\\
{\bf RXJ1522.0+0741} The source is also catalogued as the poor group  {\bf MKW3s}.
The central galaxy (NGC5920) is associated with the radio source {\bf 3C318.1}. \\
{\bf A2065} There are three galaxies along our chosen slit position; galaxy a
    is 52 arcsec to the NW, and galaxy c lies 21 arcsec to the SE of
the much less bright galaxy b. Galaxies a and c are both dominant galaxies. 
\\
{\bf A2069} This cluster is particularly extended and
 the X-ray centroid lies between two bright galaxies; one the central dominant
   galaxy in A2069 (galaxy a). Our spectrum of the second bright galaxy
   at RA 15 24 22.7 and Dec 29 57 25 (J2000) shows it to be a foreground
   active galaxy     at a redshift of 0.076. This galaxy may be a weak
   contributor to the total X-ray emission from the cluster, and we note that
   its redshift suggests that it may be a outlying member of the nearby
   cluster A2061. Also along our slit position for galaxy a
   are two other cluster galaxies (b and c) 26 arcsec to the NW of galaxy a, and separated from each other by
   only 4.1 arcsec (11kpc). This cluster is an EMSS cluster. \\
{\bf A2072} The X-ray centroid is offset by $2.5'$ from the observed galaxy,
which is the nearest obvious BCG. \\
{\bf A2111} This cluster has two equally dominant central galaxies, separated by 47 arcsec (220kpc) in
   projection. The X-rays from this cluster are elongated (Wang, Ulmer \& Lavery 1997). \\
{\bf A2142} We  have taken the spectrum of each of the two equally bright dominant galaxies in
this binary  cluster. They lie along the same slit position, separated by
104 arcsec (238kpc). \\
{\bf A2146} The emission line ratios of the observed galaxy and a pointed
X-ray observation of the cluster show that this galaxy 
contains an AGN (Allen 1995), but the deep ROSAT HRI image shows that the
majority of the flux is from extended cluster gas. \\
{\bf A2151} This cluster is resolved into two components by the RASS because
of its low redshift. We list here the dominant galaxy in A2151a, which
produces 93 per cent of the combined luminosity of the system.\\
{\bf RXJ1604.9+2356} The source is also catalogued as the poor group {\bf AWM4}. The
observed galaxy is associated with the radio source {\bf 4C$+$24.36}.  \\
{\bf A2199} The observed galaxy (NGC6166) is associated with {\bf 3C338}. \\
{\bf A2204} Two galaxies lie along our slit position, separated by 4 arcsec
(14kpc). The galaxy to the SE has the strong extended line emission
associated with it, and we present the spectrum of only this galaxy. \\
\end{table}
\vfill\eject

\begin{table}
\noindent 
{\bf  Notes on individual entries in Tables~\ref{tab:bcslog} and
\ref{tab:notbcslog}  continued:}  \\
{\bf A2219} Galaxy b is the most likely central dominant galaxy. \\
{\bf A2228} The X-ray flux and centroid of the cluster source is likely to be
significantly affected by the presence of an AGN within, but not at the
centre of the cluster (see A92). \\
{\bf RXJ1657.8+2751} This source is also catalogued as the poor group AWM5. \\
{\bf A2256} This cluster is a well-known binary system. Galaxies a and b form a common system,
   separated by 6.6 arcsec ($\sim$10kpc), and galaxy c is a further 19 arcsec 
   to the NW along this slit. Galaxy d is a bright dominant galaxy to the E
   of galaxies a,b and c, and the best candidate for a single  central cluster
   galaxy. \\
{\bf A2249} This cluster is a binary system and relatively extended in the X-ray. We have only taken a spectrum of one of the
   two dominant giant elliptical galaxies (the more eastern) as it is in
better agreement with the X-ray centroid.  The cluster is associated with 
the radio source {\bf 4C$+$34.45}. \\
{\bf A2255} This cluster has two dominant galaxies separated by 53.7 arcsec (109kpc); each of these
   galaxies appears to be a very close dumbell in its own right. \\
{\bf RXJ1733.0+4345} The observed galaxy  (IC1262) is associated with {\bf 4C$+$43.46}.\\
{\bf RXJ1740.5+3539} Two dominant galaxies lie along this slit position,
separated by 38.9 arcsec ($\sim$45kpc); galaxy b looks from the POSS to
be the more likely candidate for the central cluster galaxy.  \\
{\bf Z8276} is an uncatalogued radio-loud galaxy. \\
{\bf Z8338} This cluster has two close central galaxies; we observed  the
brighter to the East of the pair. \\
{\bf A2318} The X-ray centroid is offset from the galaxy observed, but note
that the  galaxy position given in A92 is in error. 
There is a Seyfert~1 galaxy coincident with the X-ray centroid at the redshift of the cluster,
$z=0.14$, (Edge, private communication) so this source is significantly contaminated. \\
{\bf RXJ2114.1+0234} The observed galaxy (IC1365) has an asymmetric optical
halo  on the POSS image, and is contained in the poor group {\bf II Zw 108}. \\
{\bf A2443} The cluster contains two  dominant galaxies. We have only
taken a spectrum of the brighter galaxy which is a better candidate for the
BCG. ; the other lies 2 arcmin to the
NW. This fainter galaxy is associated with {\bf 4C$+$17.89}. \\
{\bf Z8852} This highly extended X-ray source is the Pegasus Group. The
centre is occupied by two equally dominant galaxies: we have observed
NGC7499, and the other central galaxy NGC7503 
(which we did not observe) is associated with the radio source {\bf 4C$+$07.61}.
NGC7499 is closer to the X-emission centroid. \\ 
{\bf A2572a} This source is also catalogued as the compact group Hickson 94. There are two 
      galaxies along our slit, separated in projection 33.7 arcsec (37kpc).
Galaxy a is the brighter, but 
      galaxy b is embedded in an asymmetric optical halo on the POSS. \\
{\bf A2572b} There is a second, higher-redshift cluster in the background of
A2572b that may contribute to the total X-ray flux measured from this
cluster (Ebeling, Mendes de Oliveira \& White 1995). \\ 
{\bf A2626} This cluster has two close galaxies  separated by 3.4 arcsec (5kpc), embedded in an
       asymmetric halo extended to the SW component.  Galaxy b (SW) has 
        line emission, and is associated with the radio source 
 {\bf3C464}. \\
{\bf A2627} The two galaxies observed appear equally dominant. \\
{\bf A2634} The observed galaxy (NGC7720) is associated with {\bf 3C465}. \\
\end{table}

\onecolumn
\begin{table}
\caption{Clusters in the BCS for which we 
do not have a spectrum of the central
cluster galaxy.\label{tab:bcsnotlog}}
\begin{tabular}{llccccll}
       &      &   &   & &          &        & \\
Cluster      &   &   RA         & DEC      & Redshift & Lines? & Ref & Notes \\
             &&      (J2000)    & (J2000)  &          &        & & \\
             &&      & &          &        & & \\
 RXJ0000.1+0816 & & 00 00 07.1 & 08  16  49 & 0.040 & $\surd$ & EHM99  & UGC12890 \\                   
 RXJ0021.6+2803 & (S)   & 00 21 36.9 & 28 03 04 &0.094  & $\times$  & EHM99 & IV Zw 015 \\
                &(E)    & 00 21 44.0 & 28 03 56 & 0.094  & $\times$ & EHM99 & \\
 A68  & a (SE) &   00 37  06.8 & 09 09 25 &  0.250  & $\times$ & EHM99 & \\
      & b (NW) & 00 37 04.9 & 09 09 47 & 0.259 & $\times$ & EHM99 & \\ 
 A75&  a (W)  & 00 39 28.5 & 21 13 48 &  0.062  &  --    & SR91  & cluster redshift \\
   &  b (E)  & 00 39 42.3 & 21 14 06 &  0.058  & $\times$ & OLK95 & MCG+03-02-021\\
 A77&     & 00 40 28.2 & 29 33 22 &  0.071  & $\times$ & OLK95 & UGC428, EMSS \\
 A84&  a (SE) & 00 41 54.9 & 21 22 37 & 0.103  & -- & SR91 &  cluster redshift \\
   &  b (NW) & 00 41 41.3 & 21 24 10 & 0.102  & $\times$ & OWT88 & 4C+21.05 \\
 A104  & & 00 49 49.8 & 24 27 03 & 0.082 & -- & SR91 & cluster redshift\\
 RXJ0058.9+2657 &   & 00 58 22.7 & 26 51 57 & 0.048  & $\times$ & DC95 & NGC326, 4C+26.03  \\ 
 A147  & & 01 08 12.0  & 02 11 39 & 0.042 & $\times$ & P78 & UGC701 \\ 
 A160  & & 01 12 59.6  & 15 29 28 &  0.044 & $\times$ & OLK95  & MCG+02-04-010 \\
 A189  & & 01 25 31.3  & 01 45 34 & 0.019  & $\times$ & S78 & NGC533, EMSS\\ 
 Z1261 &   & 07 16 41.1 & 53 23 10 & 0.064 & $\times$ & GB82 & 4C+53.16 \\
 Z1478 &a & 07 59 44.3 &  53 58 57 &  0.104  & $\times$  & EHM99 & \\
       &b &07 59 41.0  & 54 00 11 & 0.104 &  $\times$ &  EHM99 & \\
       &c &07 59 39.2  & 54 00 54 & 0.104 &  $\times$ &  EHM99 & \\
Z1953 &    & 08 50 07.2 & 36 04 13.1 & 0.374 & $\times$ & EHM99 & \\
 Z2089 & a & 09 00 36.8 & 20 53 43  & 0.235 &  $\surd$ & EHM99 & \\
       & b & 09 00 40.1 & 20 54 35 & 0.235 & $\times$ & EHM99 & \\
       & c & 09 00 36.6 & 20 53 43 & 0.235& $\times$ & EHM99 & \\
A781 & a &  09 20 25.1 & 30 31  33 & 0.304  & $\times$ & EHM99 &\\
      & b &  09 20 25.6 & 30  29  40 & 0.293  & $\times$ & EHM99 & \\
 RXJ1053.7+5450 &   & 10 53 36.6 & 54 52 06  & 0.070  & $\times$ & EHM99 & \\
 A1235 &  & 11 23 15.6 & 19 35 53 & 0.104  & --      & SR91 & cluster redshift \\
 A1367 &  & 11 45 05.0 & 19 36 23 & 0.021  & $\surd$ & OLK & \\
 A2108 & a & 15 40 15.9 & 17 52 30 & 0.092 & -- & SR91 & cluster redshift\\
       & b & 15 40 18.9 & 17 51 25 & 0.092 & -- & SR91 & cluster redshift\\
       & c & 15 40 17.9 & 17 53 05 & 0.092 & -- & SR91 & cluster redshift\\
 A2218 &  & 16 35 49.1 & 66 12 45 & 0.172 & $\times$ & LPS92 & \\  
RXJ1651.1+0459 &   & 16 51 08.1 & 04 59 35 & 0.154  & $\surd$ & T93 & Her-A, 3C348, MCG+01-43-006   \\ 
A2241 & & 16 59 43.9 & 32 36 56 & 0.0984 & $\times$ & U76 &  PGC59392 \\
 A2312 &  & 18 54 06.1 & 68 22 57 & 0.095 & $\times$ & M97 & \\ 
 RXJ2214.7+1350 &   & 22 14 47.0 & 13 50 28 & 0.026 & $\surd$ & LM89 &  NGC7237, 3C442\\ 
 A2622 &  & 23 35 01.4 & 27 22 22 & 0.061 & $\times$ & OLK95 & \\
 Z9077 & & 23 50 35.4 & 29 29 40 & 0.095 & $\surd$ & DSG92 & EMSS\\
      &        &   & &          &        & \\
\end{tabular}
\ \ \ \ \  \ \ \ \  \ \ \ \  \ \ \ 

{\bf References: }
DC95 Davoust \& Considere 1995;
DSG92 Donahue \etal 1992; 
EHM99 Ebeling, Henry \& Mullis 1999;
GB82 Gregory \& Burns 1982;
LM89 Laurikainen \& Moles 1989;
LPS92 Le Borgne, Pello \& Sanahuja 1992; 
M97 Maurogordato \etal 1997; 
OLK95 Owen \etal 1995; 
OWT88 Owen \etal 1988;
P78 Peterson 1978; 
S78 Sandage 1978;
SR91 Struble \& Rood 1991; 
T93 Tadhunter \etal 1993; 
U76 Ulrich 1976\\
\end{table}

\begin{table}
\noindent 
{\bf Notes on individual entries  in Table~\ref{tab:bcsnotlog}:}\\ 
There are five dominant galaxies with no redshift  in the literature. The
position of that galaxy is given but the redshift of the cluster is
tabulated.   \\
{\bf RXJ0001.6+0816} The central galaxy shows [NII] and [SII] line emission
(EHM99). \\
{\bf RXJ0021.6+2803} Galaxy a is the BCG. \\
{\bf RXJ0058.9+2657} The central galaxy (NGC326) is a dumbbell galaxy. \\
{\bf A68} Galaxy a is the brighter galaxy, nearer the RASS centroid. \\
{\bf A75 } Galaxy a is the more likely BCG, but is very close to a bright star, making a observation
difficult. Galaxy b is the radio galaxy 0037+209 (Owen, Ludlow \& Keel 1995).\\
{\bf A77} This galaxy contains the radio source 0037+292 (Owen \etal 1995). \\
{\bf A84} Galaxy a is the better candidate for the BCG. The redshift for
galaxy b quoted by O'Dea \& Owen (1985) is from Owen \& White 
(1984; in preparation) for which we can find no subsequent reference. \\
{\bf A104} Owen \etal 1995 showed that another  galaxy in the cluster (harbouring the
radio source 0047+241) and  west of the 
dominant one  listed in Table\ref{tab:bcsnotlog},  at 
RA 00 49 41.8, Dec +24 26 42 (J2000) has \Ha+[NII] emission, but no
\otw\ emission. \\
{\bf Z1478} None of the galaxies listed in Table~\ref{tab:bcsnotlog} is the central  dominant
galaxy, which lies almost exactly behind a star, at
RA 07 59 40.4, Dec  +54 00 22 (J2000).\\
{\bf Z1953} HRI observations of this cluster suggest that as much as two
thirds of the X-ray flux assigned to this system based on the RASS
observation are in fact due to two X-ray point sources wihtin 7 arcmin of
the cluster (Ebeling, private communication). \\
{\bf Z2089} Galaxy (a) at the X-ray centroid has strong line emission with the line ratios
([NII]/\Ha and [OIII]/\Hbn) suggestive of an AGN. The X-ray source is clearly
resolved in a recent ROSAT HRI image supporting its  inclusion as a cluster
in the BCS. \\
{\bf A781} A third galaxy at RA 09 20 22.3, Dec 30 30 53 (J2000) shows \Ha\ in emission, but     
is foreground at a redshift of 0.126 (Ebeling, Henry \& Mullis 1999). It is not clear whether galaxy a
or b is the BCG, although b is brighter. \\
{\bf RXJ1053.7+5450} The observed galaxy is to the NW of a very extended
X-ray source. There are several other slightly fainter galaxies the SE.\\
{\bf A1367} The BCG is NGC3862, which is associated with
the strong radio source {\bf 3C264}, and  detected as an unresolved point source
in ROSAT images contributing
approximately 5\% to the total X-ray emission from the
cluster (Edge \& R\"ottgering 1995).
NGC3862 is  significantly offset from the centre of the highly
extended X-ray emission, and is observed to have a weak emission line
spectrum ([NII]$>$\Han, weak [OII]) with evidence for a blue continuum.
 HST observations reveal that
3C264 contains a nonthermal core and jet (Baum \etal 1998) which could
contribute to this blue continuum.
We have observed the galaxy NGC3860 at RA 11 44 49.1, DEC 19 47 44 (J2000),
which is the brightest galaxy in the broad cluster core. NGC3860 has
has strong extended line emission, dominated by \Han. \\
{\bf A2108} There are three galaxies of equal rank in the core of this cluster.\\
{\bf RXJ1651.1+0459} Hercules-A has moderate \otw\ and weak \oth\
([OIII]/[OII]=0.2) emitted by one of
two diffuse continuum components separated by 3 arcsec (10.5 kpc;
Tadhunter \etal 1993). \\
{\bf A2241} This optical position and redshift of this cluster is confused
in the literature;  A2241 appears to be a 
a superposition of an X-ray bright cluster at $z\sim0.1$ and an X-ray
faint group at $z\sim0.06$. The X-ray bright cluster
detected by ROSAT is clearly centred on PGC59392 at a redshift $z=0.0984$,
which is also associated with the radio source {\bf4C32.52C}. 
The galaxies at $z\sim0.06$ are more than ten arcmin from the X-ray peak, 
and form a separate system not detected by the RASS. \\
{\bf RXJ2214.7+1350} The cluster has two central galaxies (NGC7236/7) sharing a disturbed halo.\\
{\bf Z9077} This source is an EMSS cluster, MS 2348.0+2913 (Stocke \etal 1991). The central galaxy has 
\Ha+[NII] emission at a flux of $2.6\pm2.6\times10^{-15}$\ergpspcmsq
discovered from narrow-band imaging by Donahue \etal 1992. \\
~\\
\end{table}
\twocolumn 

\section{Data Analysis and Results }

\subsection{New redshifts}
\label{sec:newz}
We have obtained a total of eighteen new cluster redshifts for the BCS,
which are listed with a redshift marked in bold font in
Tables~\ref{tab:bcslog} and \ref{tab:notbcslog} (the redshift of the cluster
is assumed the same as that of its BCG). The redshifts were determined using
the cross-correlation technique detailed in A92 and C95. We have
compared the redshifts we have derived for our whole sample to all those
available in the literature, and find that errors in our redshift are
typically less than $\pm0.0005$. The only exceptions are those spectra taken
on the fourth night in the May 1995 run which have a less precise wavelength
calibration; these redshifts are quoted to within $\pm0.001$. One new
X-ray discovered cluster (RXJ1532.9+3021) was discovered to have a redshift
of 0.3615, the second highest in the BCS.

\subsection{Completeness of sample}
\label{sec:complete}

\begin{figure}
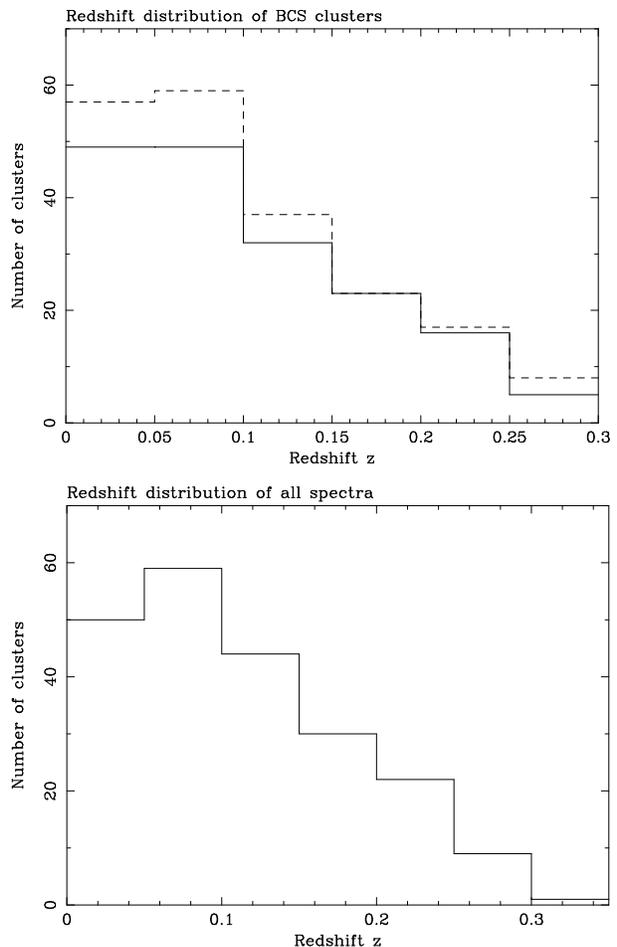

\vbox{
\psfig{figure=fig1top.ps,width=0.45\textwidth,angle=270}
\vspace{0.25cm}
\psfig{figure=fig1bot.ps,width=0.45\textwidth,angle=270}}
\caption{ 
\label{fig:zdist}
(Top) The redshift distribution of the 201 $z<0.3$ clusters in the
BCS (dashed line) and of those clusters in the BCS for which we have
spectra (solid line). 
(Bottom) The redshift distribution of all the clusters for which we have
spectra, ie all those in Tables~\ref{tab:bcslog}  and \ref{tab:notbcslog}.}
\end{figure}

Combining both the new observations presented here and those observations
from A92 and C95, we have compiled a total spectral sample of 256
dominant galaxies in 215 clusters. 213 of these spectra are of dominant
galaxies in 177 clusters of the BCS, leading to a completeness in coverage
of 87 per cent for this sample. With the sole exception of 3C264 in A1367,
we have spectra for all the BCG in the BCS down to an unabsorbed X-ray flux
limit of $7.9\times10^{-12}$\ergpspcmsq (98 clusters); this limit
corresponds to a luminosity of $3.6\times10^{44}$\ergps at $z$=0.1,
$1.5\times10^{45}$\ergps at $z$=0.2 and $3.5\times10^{45}$\ergps at $z$=0.3
(in the ROSAT 0.1-2.4\keV\ band).  [We have basic information on whether a
BCG has an emission-line spectrum or not down to an X-ray flux of
$7.1\times10^{-12}$\ergpspcmsq (109 clusters).]
Fig~\ref{fig:zdist} shows the redshift distribution for all clusters in the
BCS, and the distribution of those for which we have optical spectra; 
for redshift bins at $z<0.25$ we have spectra of over 
83 per cent of the BCS BCG. 

\subsection{Occurrence of line emission}
\label{sec:occurrence}

\begin{figure}
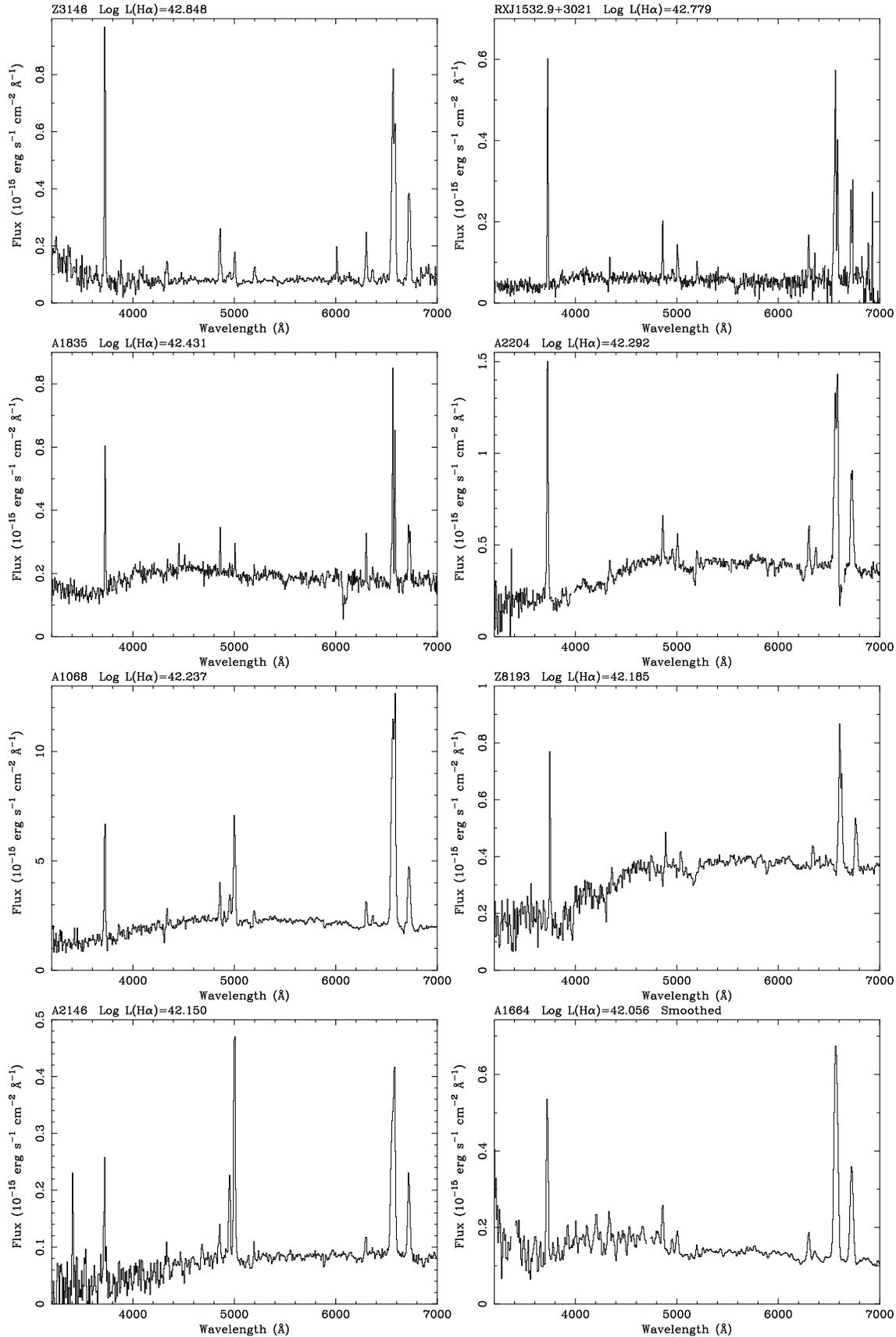

\centering
\vbox{\hbox{
\psfig{figure=fig2p1a.ps,width=0.425\textwidth,angle=270}
\psfig{figure=fig2p1b.ps,width=0.425\textwidth,angle=270}
}\hbox{
\psfig{figure=fig2p1c.ps,width=0.425\textwidth,angle=270}
\psfig{figure=fig2p1d.ps,width=0.425\textwidth,angle=270}
}\hbox{
\psfig{figure=fig2p1e.ps,width=0.425\textwidth,angle=270}
\psfig{figure=fig2p1f.ps,width=0.425\textwidth,angle=270}
}\hbox{
\psfig{figure=fig2p1g.ps,width=0.425\textwidth,angle=270}
\psfig{figure=fig2p1h.ps,width=0.425\textwidth,angle=270}
}
\caption{
\label{fig:emspectra}
The spectra of line-emitting central cluster galaxies in descending order of the observed \Ha luminosity; the
brightest line emitters with log L(\Han) from 42.9--42.05. Some of the
spectra have been smoothed (as noted in the title for each) and cosmic ray
hits have been removed. }
}\end{figure}

Many of the spectra show strong low-ionization emission lines commonly
observed in central cluster galaxies (eg: Heckman 1981, Hu \etal 1985, JFN,
Heckman \etal 1989, A92 and C95). The major coolants are hydrogen and
oxygen, and lines of nitrogen and sulphur are also strong. We have marked
which galaxies show such line emission by a tick ($\surd$) in column 5 of
Tables~\ref{tab:bcslog} and \ref{tab:notbcslog}.
We determine whether or not a galaxy has line emission if the lines of 
[NII]$\lambda\lambda$6548,6584 (the most common coolant) can be fit at the
redshift of the galaxy, at 
an intensity significantly ($3\sigma$) above the noise in the galaxy
continuum.  As we are sampling down to
lower values of X-ray luminosity than in A92 and C95, we also find a
population of galaxies that show only low-level \niib\ line emission with
\Ha in absorption. These galaxies are marked separately in column 5 of
Tables~\ref{tab:bcslog} and \ref{tab:notbcslog} by an \lq N'. We show the
slit spectrum of all the line emitters with L(\Han)$>10^{40}$\ergps in
Fig~\ref{fig:emspectra}, including for completeness those from A92 and C95.
The spectra are presented in descending order of the observed
\Ha luminosity (ie {\it not} corrected for any internal reddening as
detailed in section~\ref{sec:reddening}). Some of the galaxy spectra have
been smoothed with a box 3 pixels wide (these are noted in the title of each
plot).  Fig~\ref{fig:nemspectra} shows three examples of the galaxies that
only feature [NII] line emission. We determine a flux limit for the
detection of \Ha emission by investigating the redshift-dependence of the
fitted \Ha luminosity. We find that below a redshift of 0.05, a systematic
uncertainty in the level of stellar \Ha absorption expected from the
underlying galaxy continuum dominates our detection level, leading to an
upper limit to \Ha of $\sim4\times10^{39}$\ergps. Beyond this redshift, we find the
detection limit to rise according to the expected $L\propto z^2$ relation.
We fit this slope at what we estimate to be the least significant detection
of \Ha in emission in a galaxy at $z>0.05$, and extrapolate to obtain \Ha luminosity limits of
$1.9\times10^{40}$,
$7.8\times10^{40}$ and
 and $1.8\times10^{41}$\ergps at redshifts of 0.1, 0.2 and 0.3 respectively.

Taking the 203 members of the BCS sample (excluding the `b' components of the
A2572, A2151 and A1758 clusters and including the two clusters at $z>0.3$),
32$\pm$5 per cent show some level of line emission (ie including those with
only [NII] and no \Han), and 27$\pm$4 per cent show \Ha in emission; we
assume that the five clusters for which the emission-line properties of the
BCG are not known are {\em not} line-emitters, and obtain the errors
assuming that $\surd n$ statistics are applicable. The fractions of
line-emitters are similar if we include all the galaxies from
Table~\ref{tab:notbcslog} that are not {\em bona fide} members of the BCS,
at 34$\pm$5 per cent and 27$\pm$4 per cent for all emitters, and \Han-only
emitters respectively. 

Fig~\ref{fig:lxfreq} shows the frequency of line emitters as a function of
the X-ray luminosity ($L_X$ is taken from Paper~I) for all the
BCS sample, and then for all clusters included in this paper for which we
have an X-ray flux (ie including those in Table~\ref{tab:notbcslog} that are
in the sub-BCS and XBACS). The data have been grouped so as to contain an
equal number (45--50) of clusters per luminosity bin. In both samples there
is no compelling evidence for an increase in the frequency of line emission
with X-ray luminosity, the distribution in each case being consistent with a
constant fraction around 31.5$\pm$4 per cent (including 6 per cent with
[NII]-only line emission).

Fig~\ref{fig:zfreq} shows the frequency of line emitters as a function of
redshift, both for the BCS and for all the clusters in this paper. We have
grouped the sample into an equal number of clusters per redshift bin (50),
and again assume that the five clusters whose BCG optical properties are not
known are not line emitters. Both samples show the fraction to be high in
the lowest redshift bin;
assuming that the galaxy line luminosity is correlated with the cluster X-ray
luminosity, the threshold for detecting line-emitters will be lower at low
redshift. The frequency of line emission drops sharply at a redshift of
0.07, and then climbs back up to a fraction above 30 per cent above a
redshift $z>0.1$. The BCS X-ray flux limit pre-selects the most luminous
clusters at any epoch, and by a redshift of 0.2 will include only those
clusters whose X-ray luminosity is boosted by the presence of a massive
cooling flow (which would enhance the probability of line emission), or
because it is a binary cluster system with whose X-ray emission has been
enhanced because it is blended (and such systems usually do not have cooling
flows). Fig~\ref{fig:lxfreq}, however, indicates that the fraction of
line-emitters is not dependent on X-ray luminosity, so the rise in the
fraction of line emitters with redshift may reflect a real increase in the
number of sites promoting line emission at $z>0.1$. To test this, we looked
at the 10 clusters in the complete (ie $z<0.3$) BCS that have an X-ray
luminosity above $1.8\times10^{45}$\ergps, which corresponds to the flux
limit of $4.4\times10^{-12}$\ergpcmsqps at the redshift of the furthest
cluster in the sample ($z=0.29$). Only one of the five clusters with a
redshift $z<0.23$ has line emission in the central galaxy, whereas 4 out of
the five at $0.23<z<0.29$ do.
Dropping the X-ray luminosity cut-off to $1\times10^{45}$\ergps, however,
results in equal numbers of line-emitters in the 17 clusters below and 16
clusters above 0.23. Thus any evidence for evolution in the frequency of
occurrence of line emission (and thus presumably massive cooling flows) with redshift
(eg Donahue \etal 1992) is only very tentative, and can involve only the
most luminous clusters. 

Whilst a cooling flow is not the only possible cause of line emission around
a BCG, all but five of the 64 \Han-emitting central galaxies (those in
A1068, A2089, A2146, A2294, and RXJ0821.0+0752: see
section~\ref{sec:lineint} and notes to the tables) have line intensity ratios consistent with the
nebulae seen around the central cluster galaxies of known cooling flows. Removing
these five exceptions from consideration only decreases the overall
frequency of line emission by 2 per cent in either the BCS, or all clusters
for which we have spectra.

We find, however, a higher probability that a purely X-ray selected cluster
will contain an emission-line system around its BCG. Of the 39 `RXJ' BCS
clusters listed in Tables ~\ref{tab:bcslog} and \ref{tab:bcsnotlog}, 19 have
a BCG with line emission and 19 do not, and only 1 has [NII]-only emission. 
 Including also the `RXJ' clusters in
Table~\ref{tab:notbcslog}, 20 out of the 43 show \Ha in emission, with only
2 [NII]-only emitters. The fraction of line emitters is
significantly greater in the `RXJ' clusters than in the sample as a whole.
This is not simply a selection of the RXJ systems being at lower redshift,
and thus line emission being easier to detect in the central cluster galaxy.
Of the 29 Abell and 31 RXJ clusters at $z<0.05$ in Tables ~\ref{tab:bcslog}
and \ref{tab:bcsnotlog}, only 5 (and a sole
[NII]-only emitter) of the central galaxies in the Abell clusters show line
emission; 16 (plus 2 [NII]-only emitters) of the RXJ systems have line
emission. Assuming that the emission line nebulae are (nearly) all tracing
the presence of a cooling flow in these clusters then
this suggests that the X-ray selected clusters contain a higher fraction of
cooling flows. This is most likely due to not only the fact that the RASS X-ray
detection algorithm used to construct the BCS is most sensitive to 
point or peaked X-ray sources (Paper~I), but also that cooling
flows may enhance the X-ray luminosity of a cluster (Allen \& Fabian 1998).
Thus the X-ray selected samples  will
preferentially contain cooling flow clusters (see also Pesce \etal 1990). 

Only 36 of the 59 \Han-line emitters show significant [OII]$\lambda$3727
line emission, although this fraction of 61$\pm$13 per cent should be
regarded as a lower limit to the true fraction. Detectability of [OII] will
be affected both by any dust intrinsic to the galaxy (see
section~\ref{sec:reddening}), and by atmospheric refraction, given that the
slit placement is decided on the red galaxy image.

\onecolumn
\vfill\eject

\addtocounter{figure}{-1}
\begin{figure}
\centering
\vbox{\hbox{
\psfig{figure=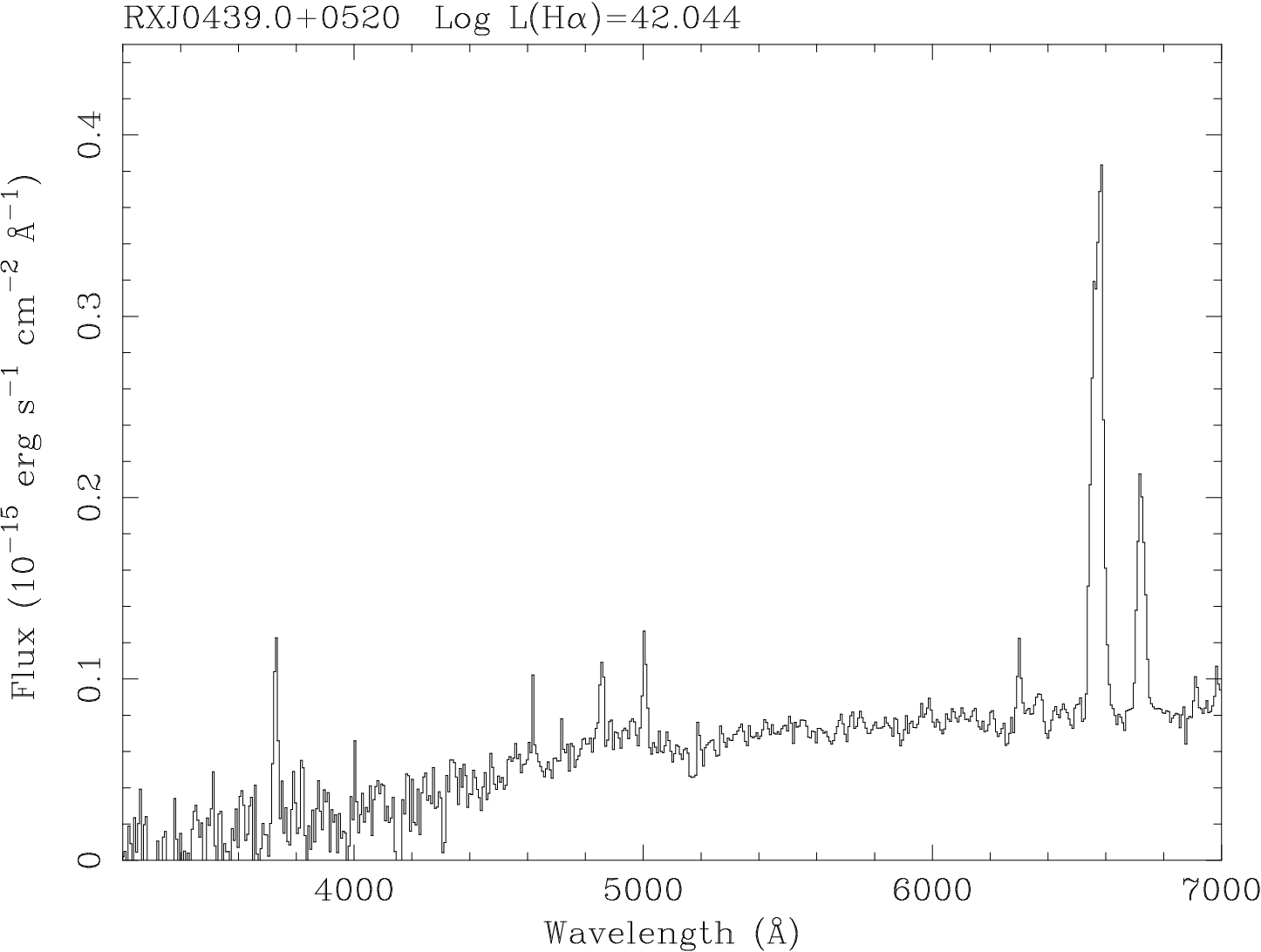,width=0.425\textwidth,angle=270}
\psfig{figure=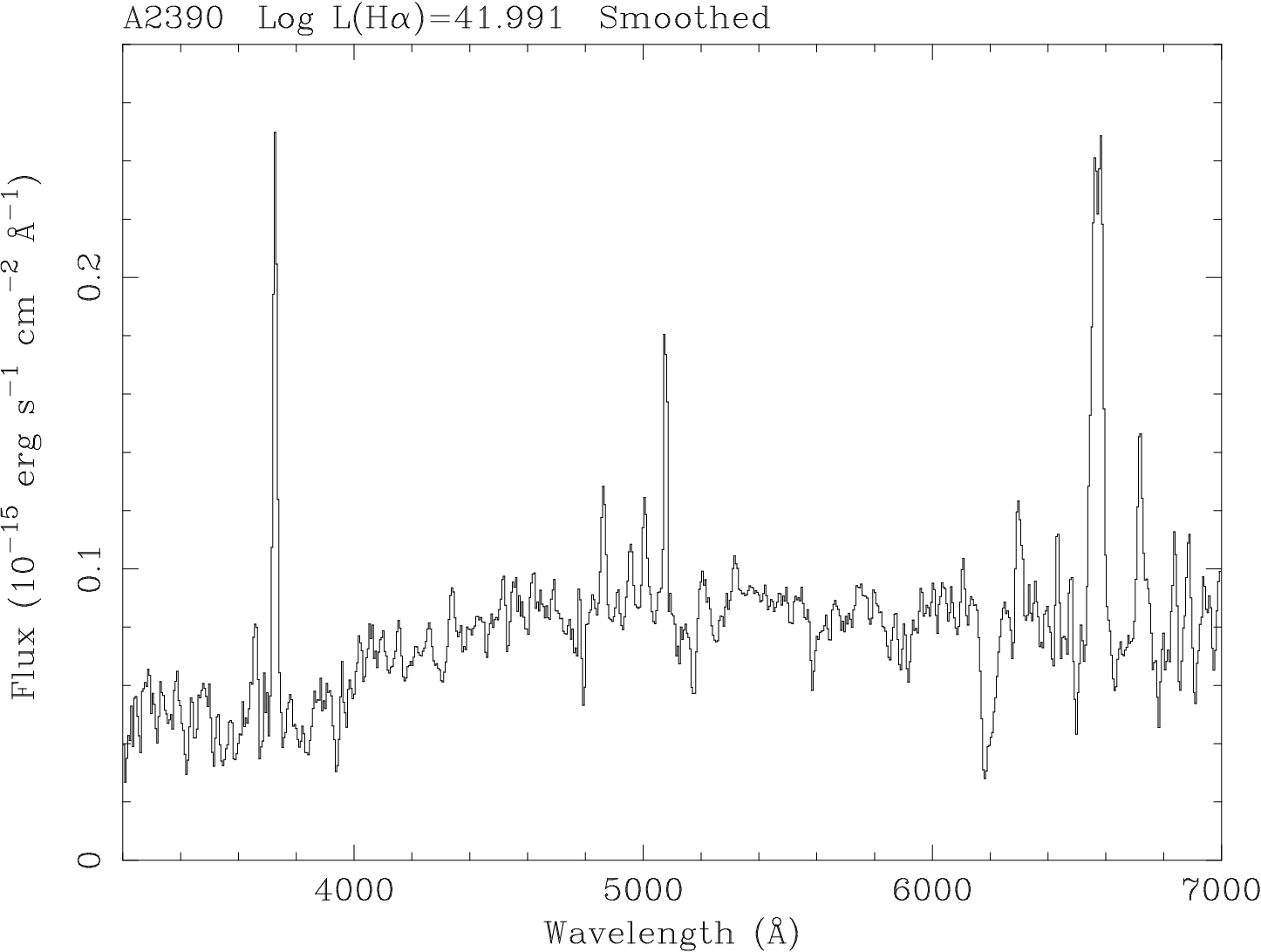,width=0.425\textwidth,angle=270}
}\hbox{
\psfig{figure=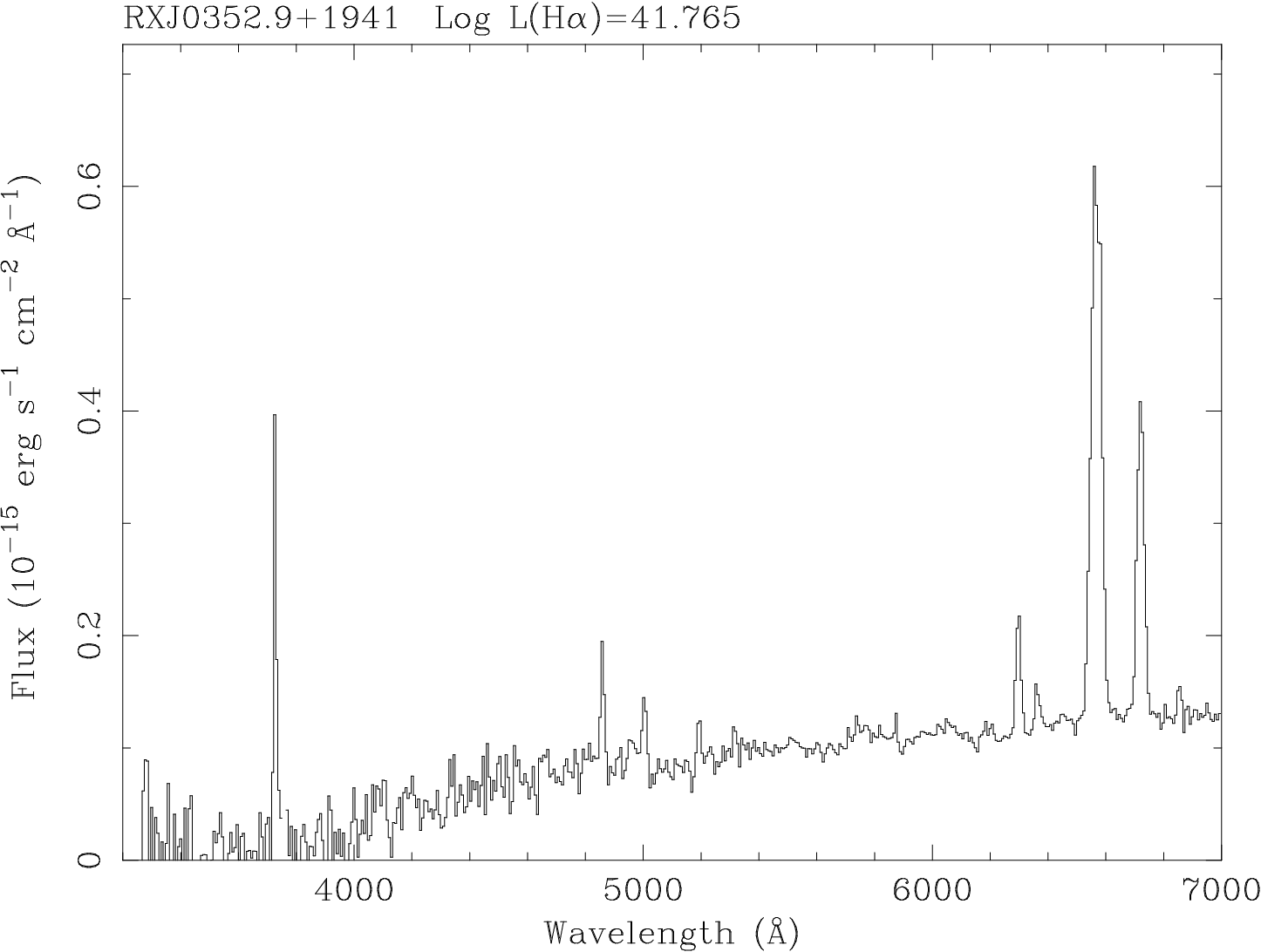,width=0.425\textwidth,angle=270}
\psfig{figure=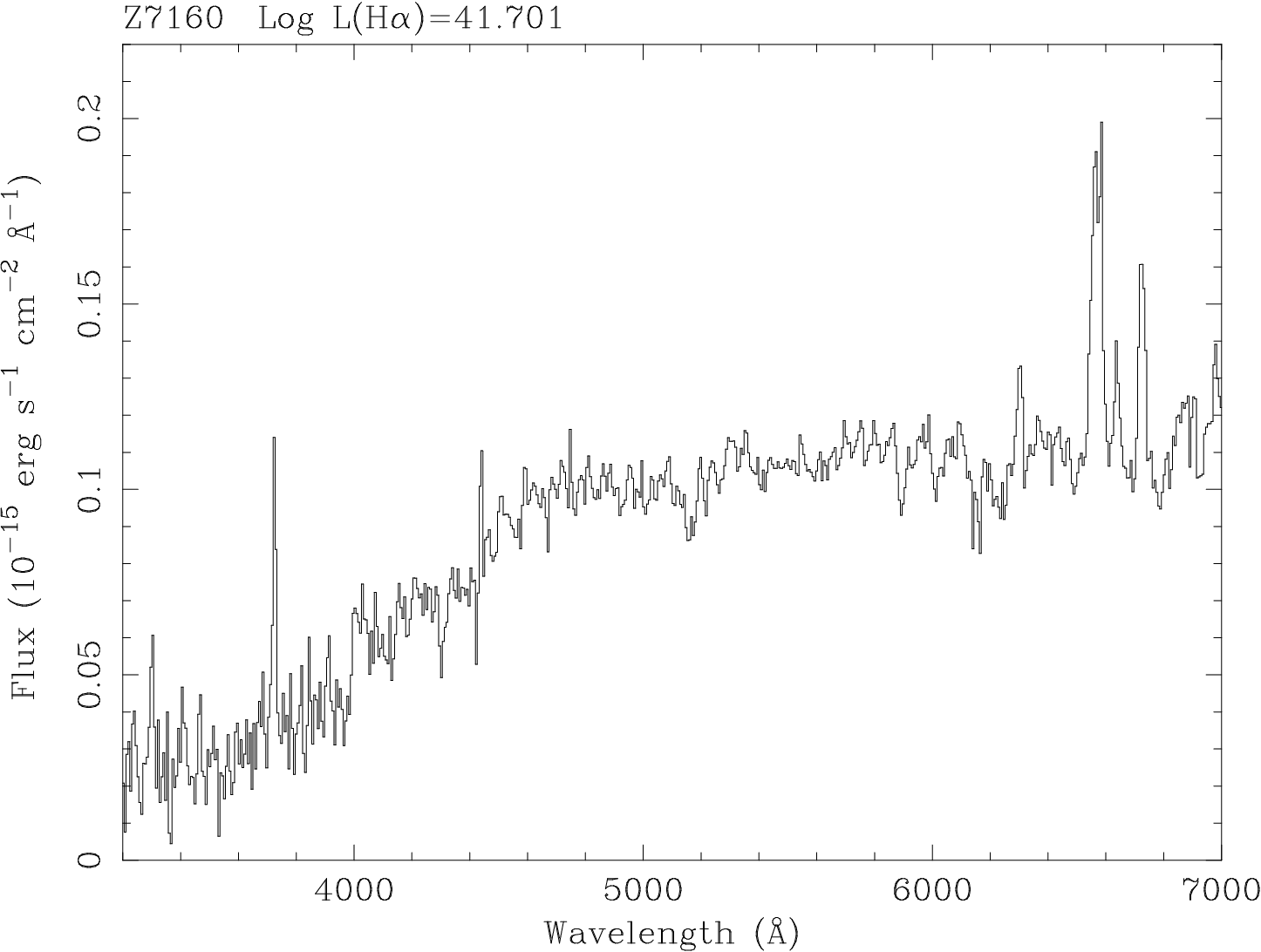,width=0.425\textwidth,angle=270}
}\hbox{
\psfig{figure=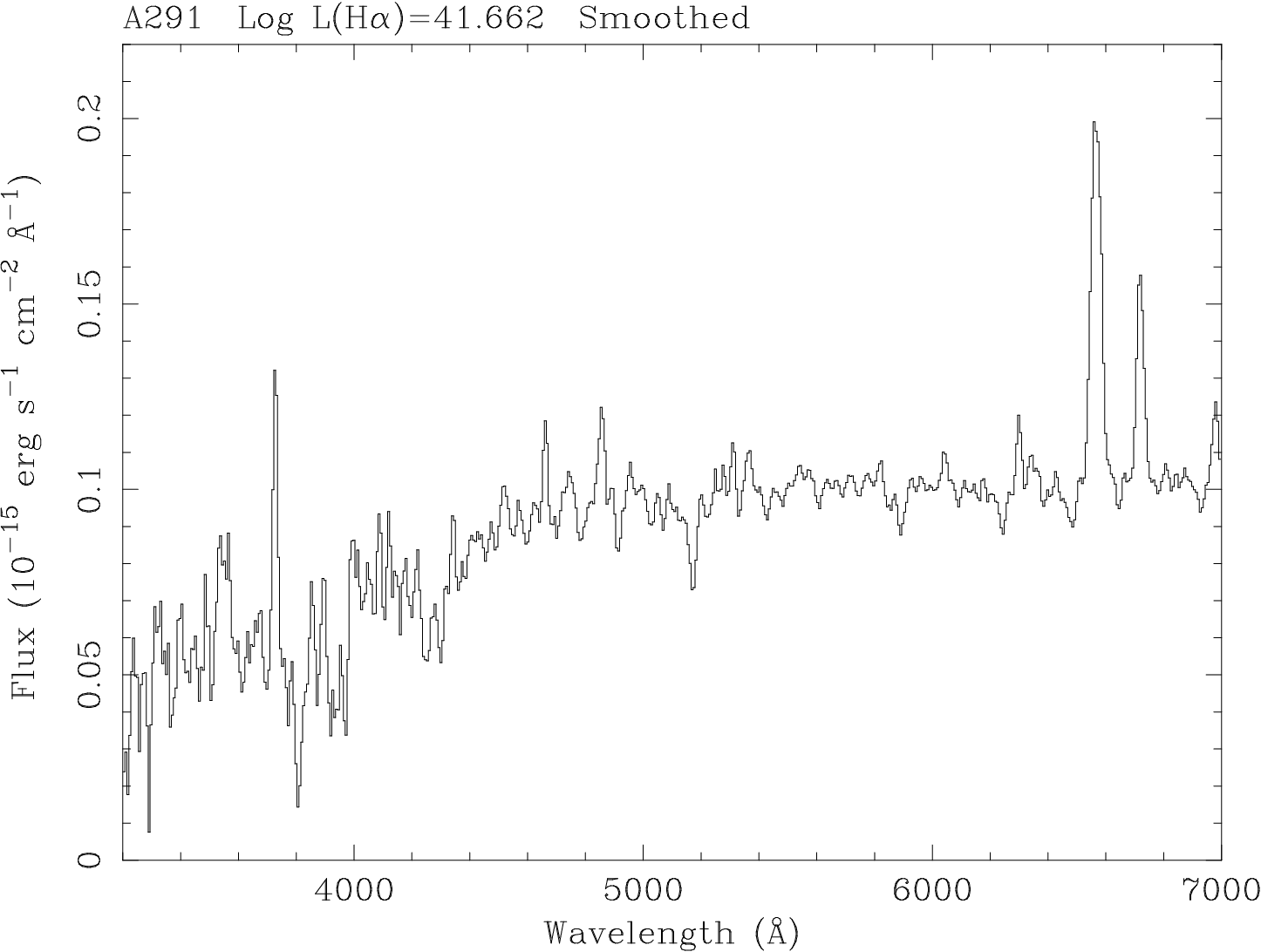,width=0.425\textwidth,angle=270}
\psfig{figure=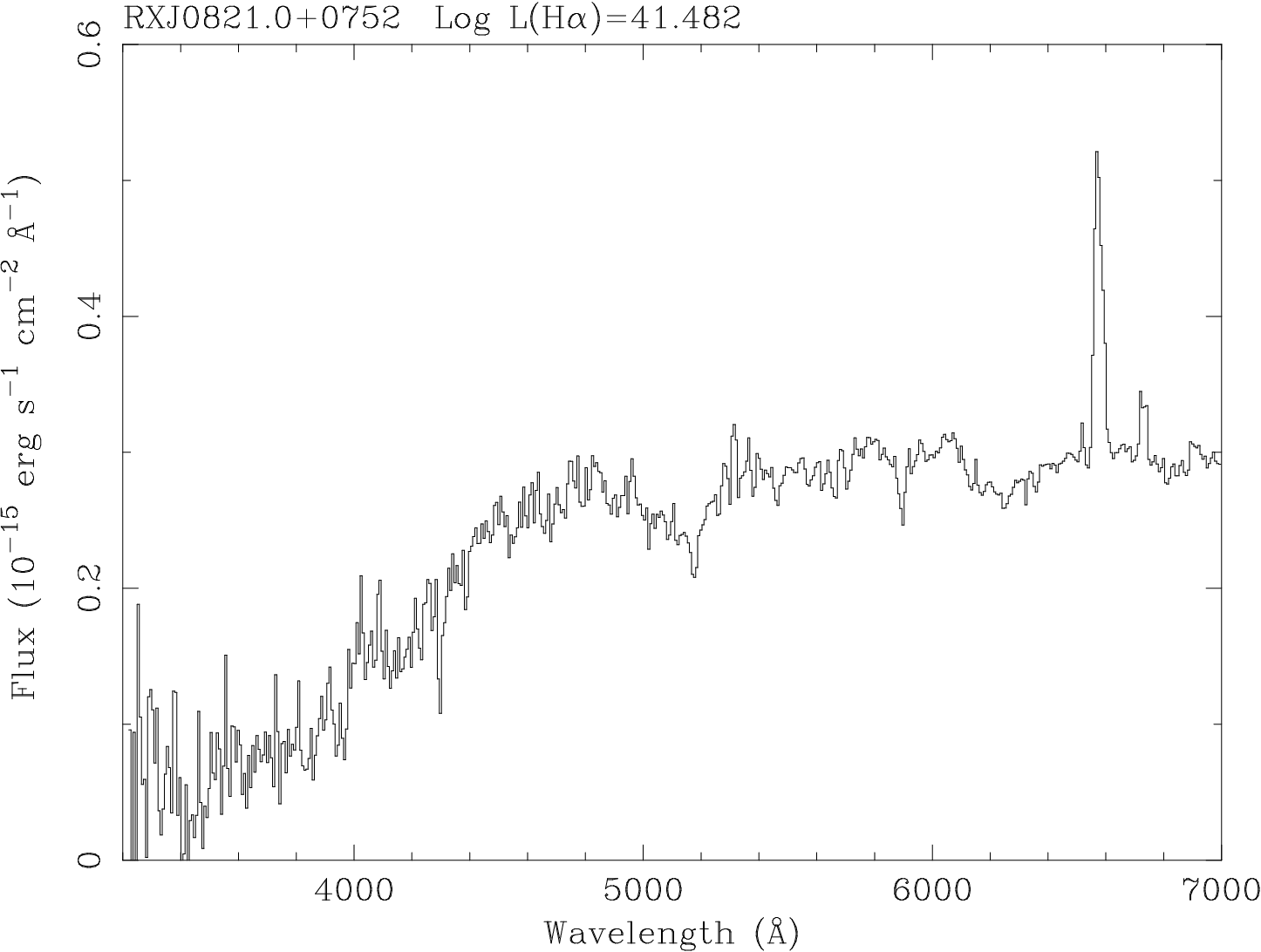,width=0.425\textwidth,angle=270}
}\hbox{
\psfig{figure=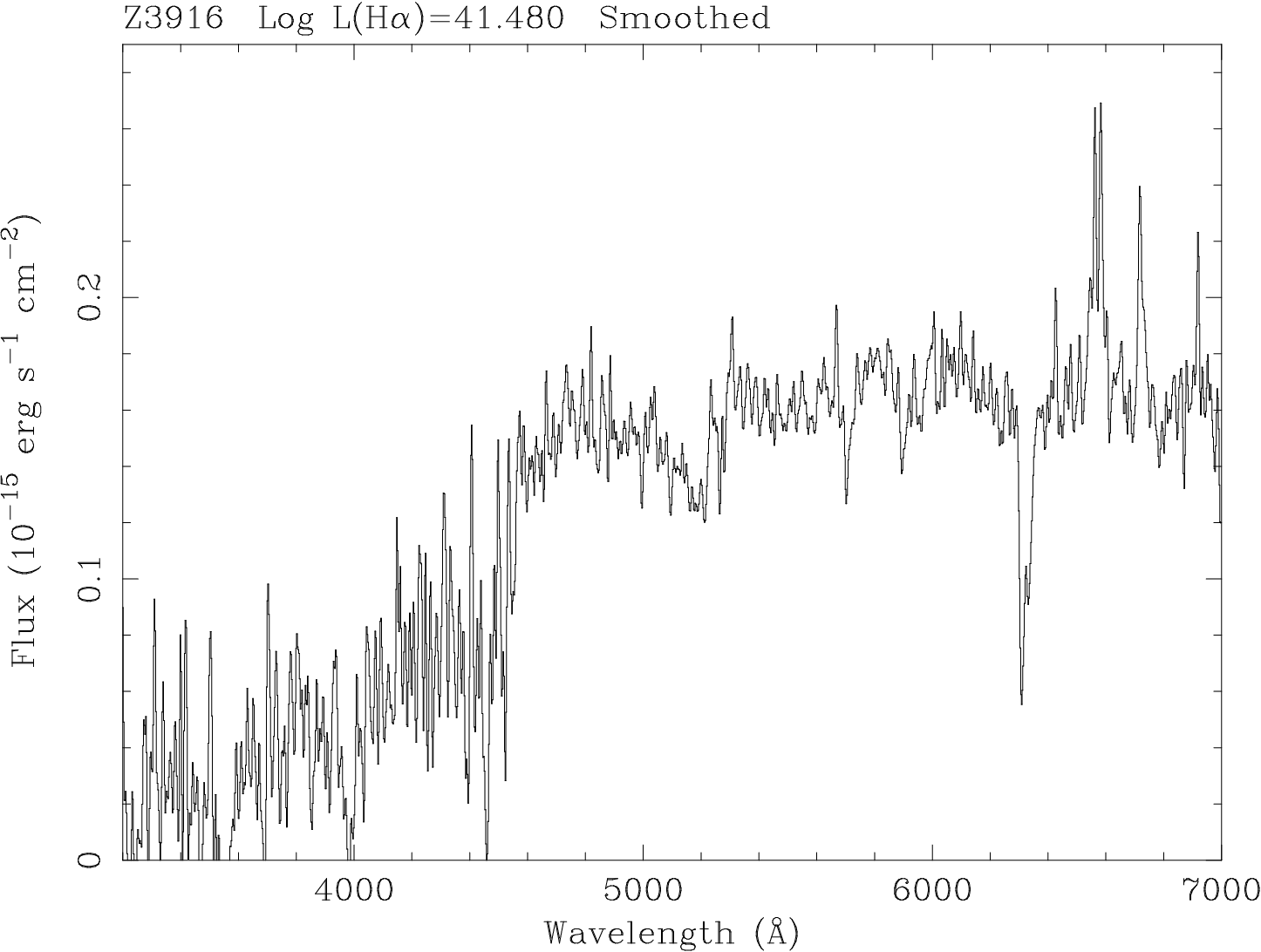,width=0.425\textwidth,angle=270}
\psfig{figure=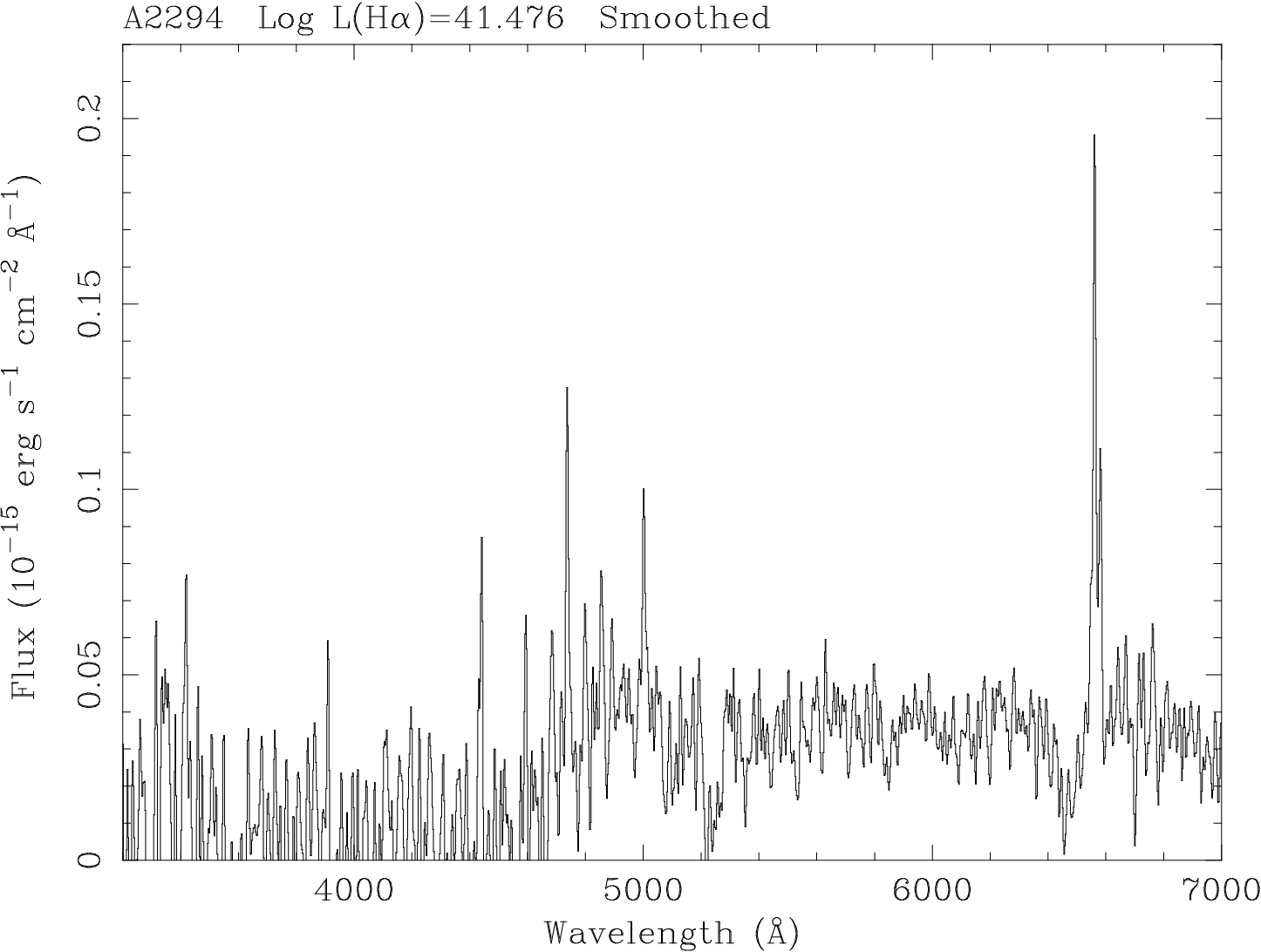,width=0.425\textwidth,angle=270}
}\caption{The spectra of line-emitting central 
cluster galaxies  continued; those with log L(\Han) in the range  42.05--41.47. }
}\end{figure}\vfill\eject

\addtocounter{figure}{-1}
\begin{figure}
\centering
\vbox{\hbox{
\psfig{figure=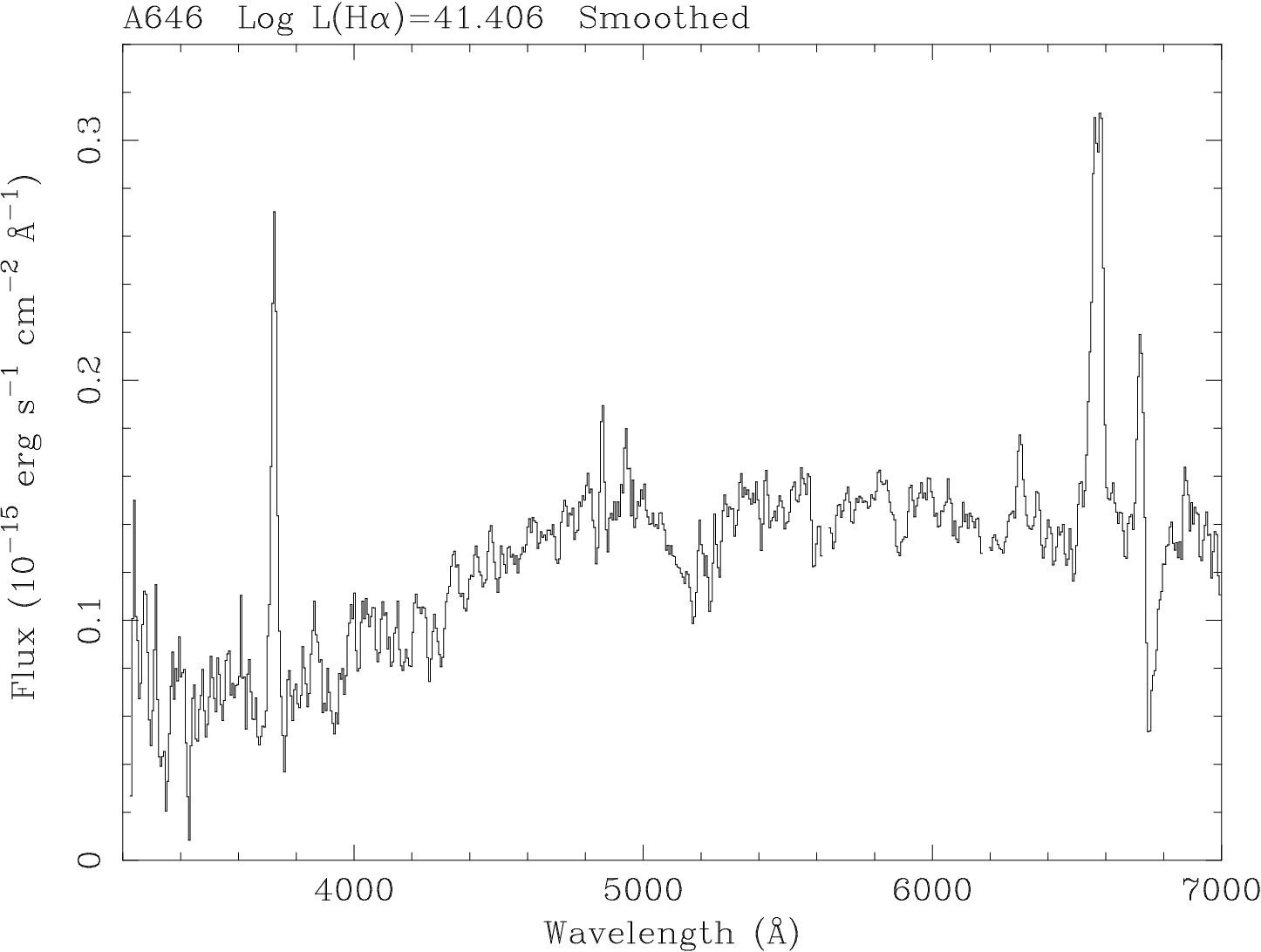,width=0.425\textwidth,angle=270}
\psfig{figure=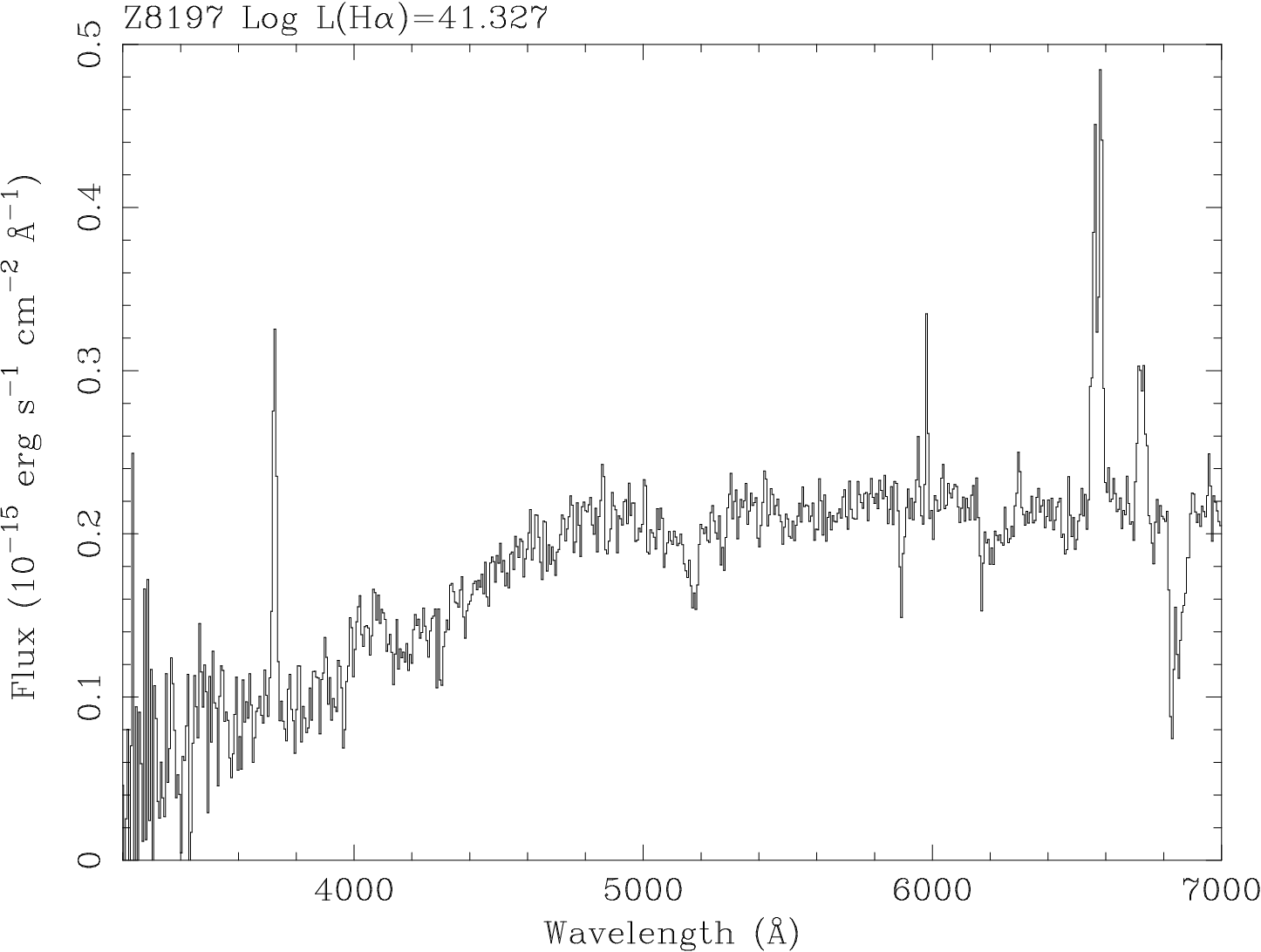,width=0.425\textwidth,angle=270}
}\hbox{
\psfig{figure=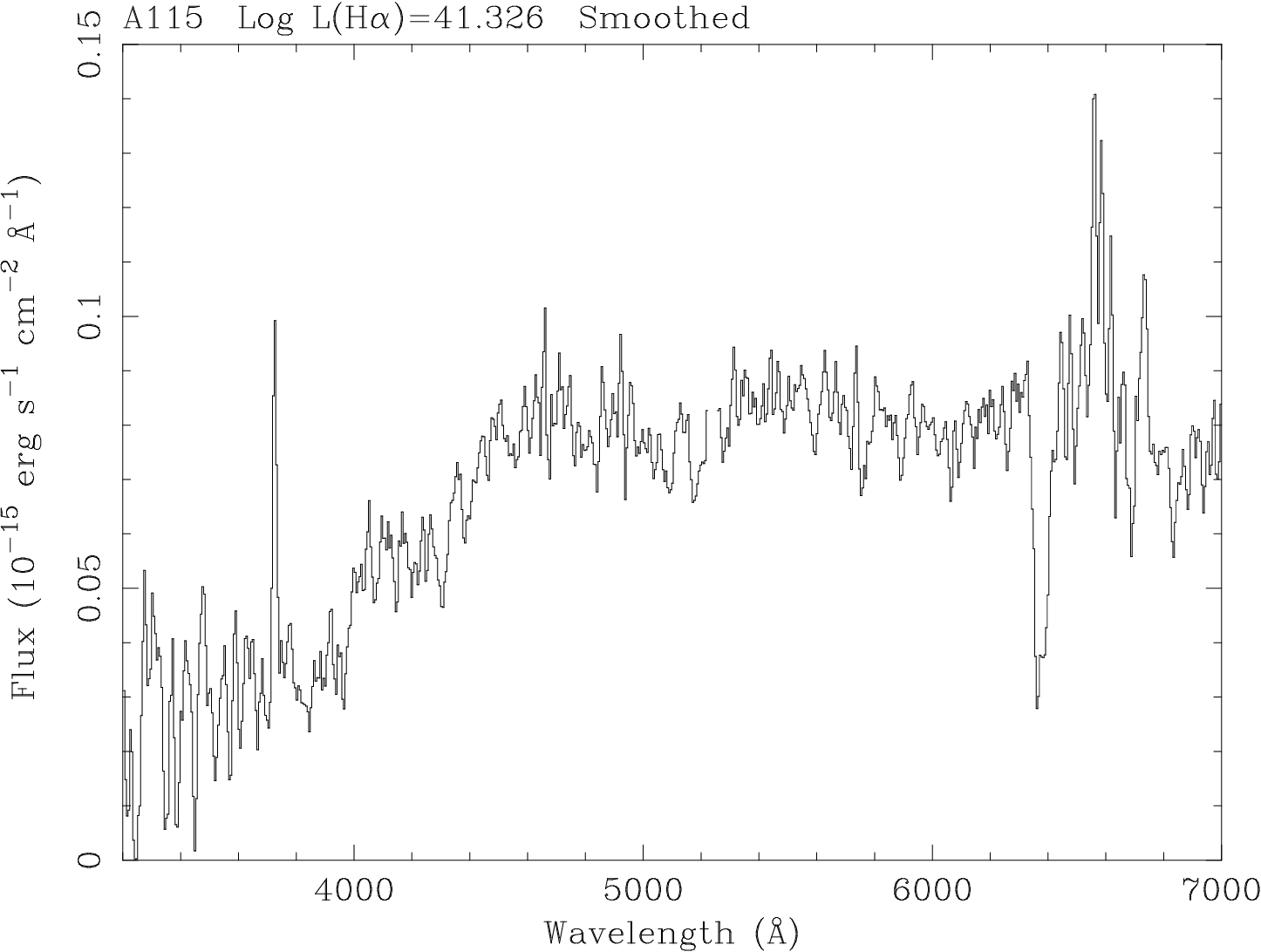,width=0.425\textwidth,angle=270}
\psfig{figure=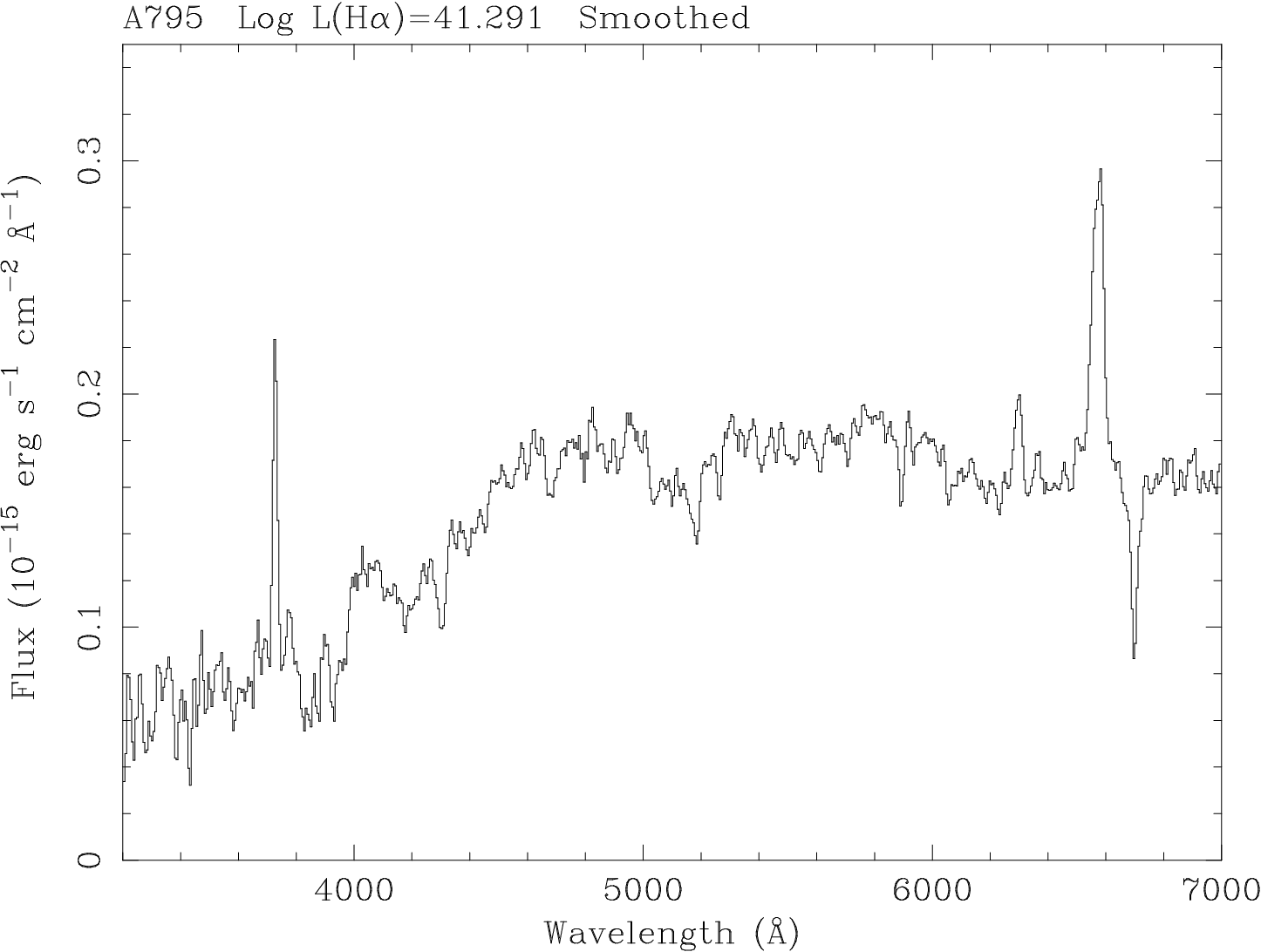,width=0.425\textwidth,angle=270}
}\hbox{
\psfig{figure=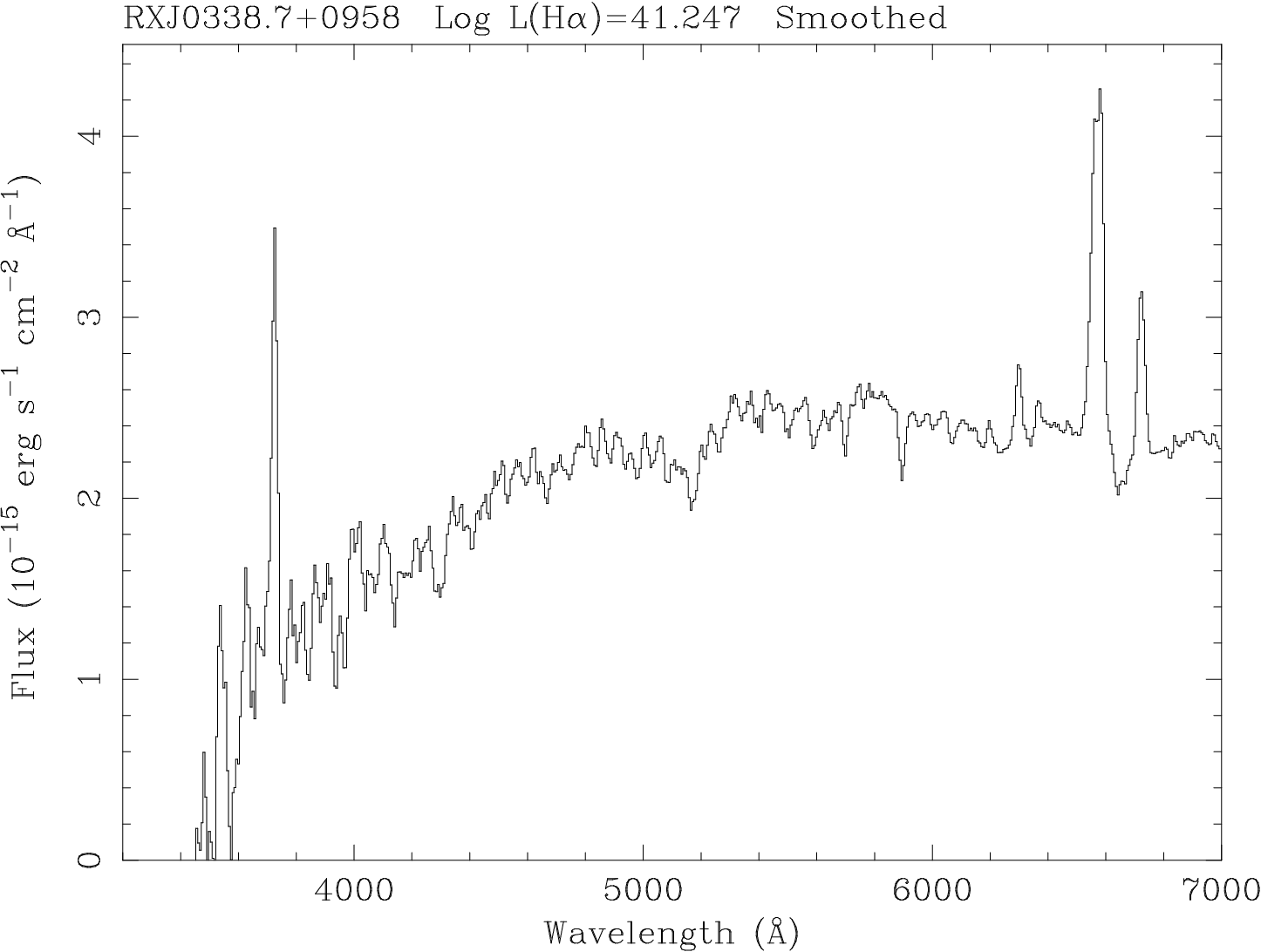,width=0.425\textwidth,angle=270}
\psfig{figure=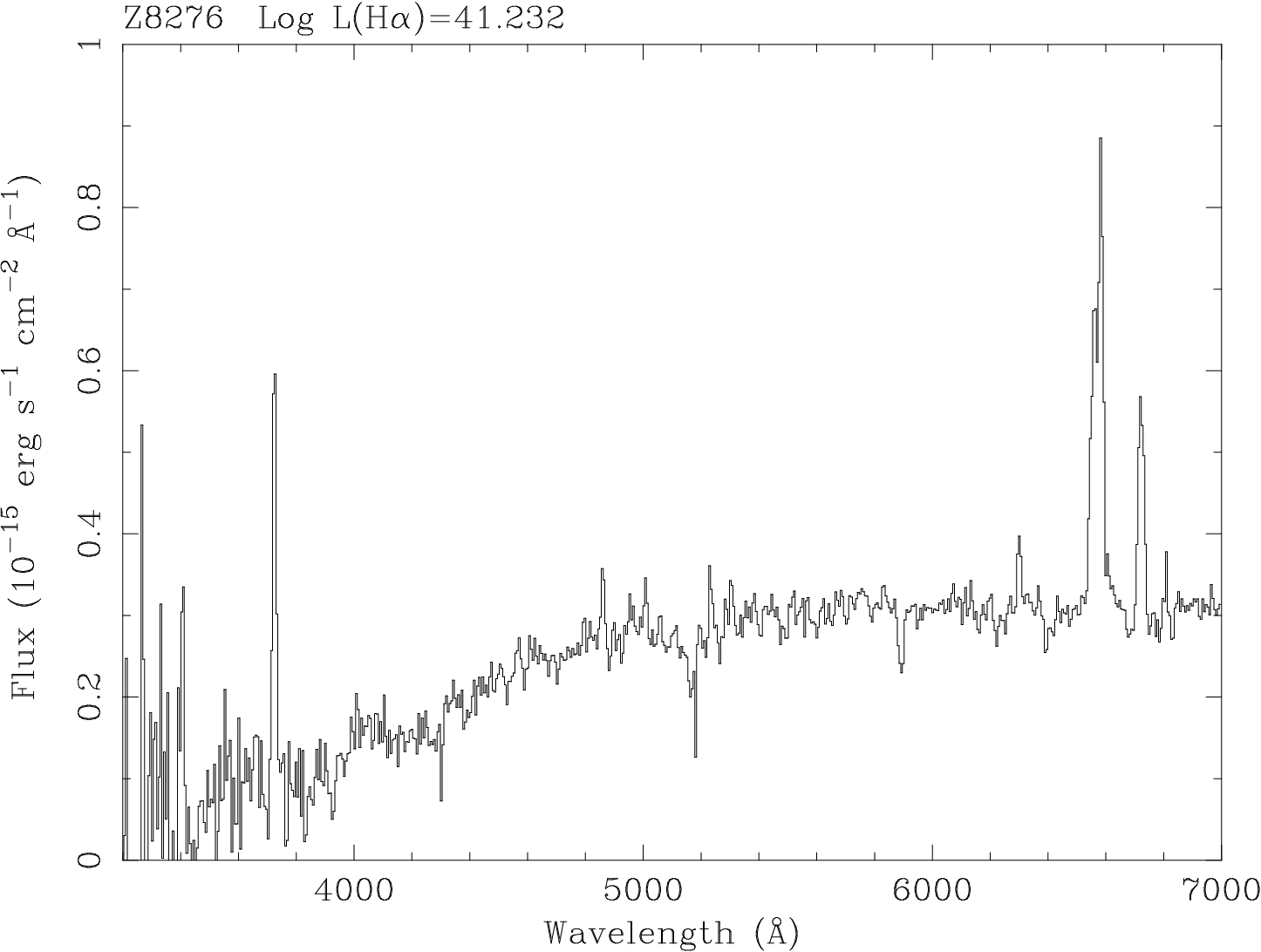,width=0.425\textwidth,angle=270}
}\hbox{
\psfig{figure=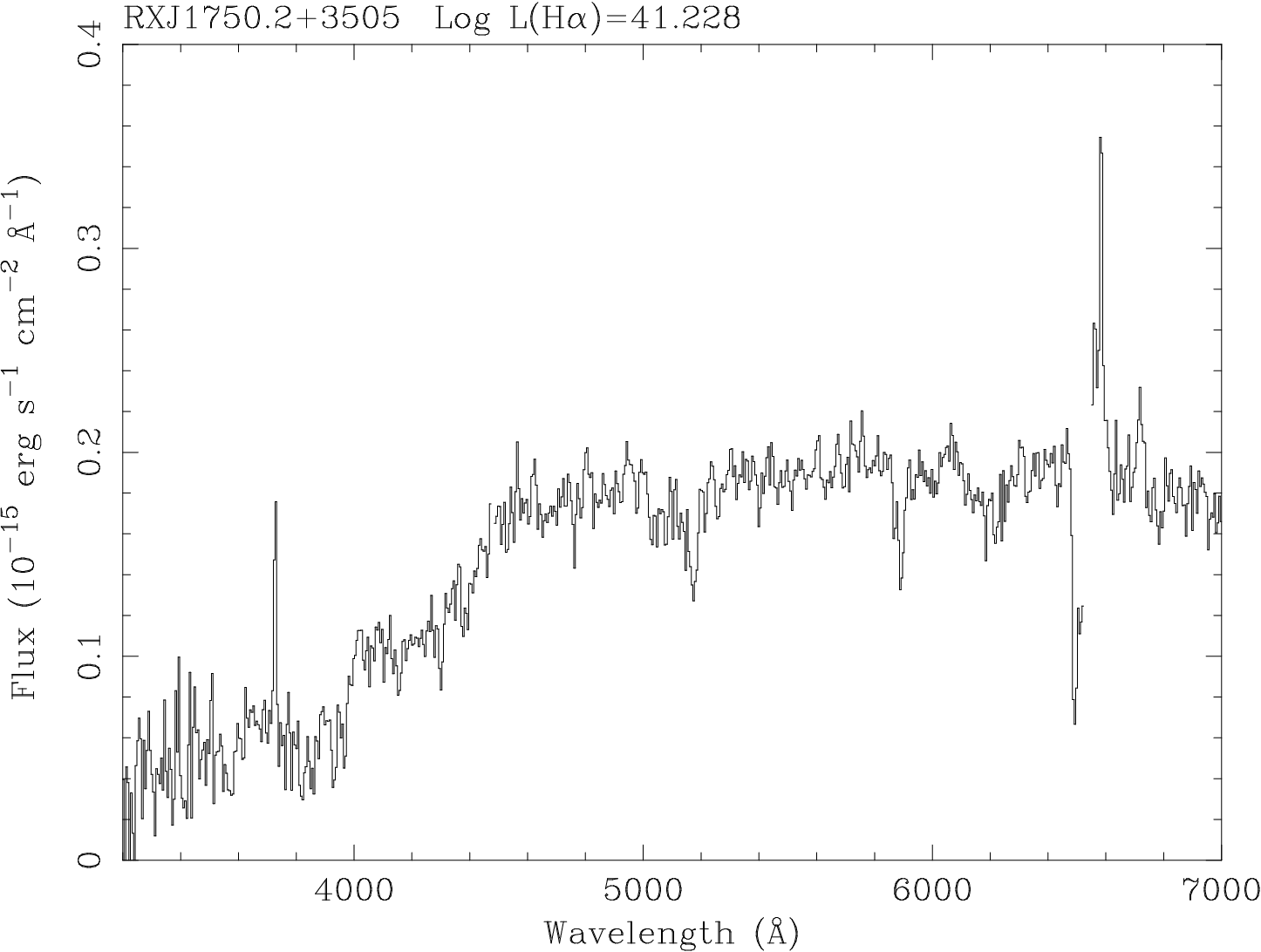,width=0.425\textwidth,angle=270}
\psfig{figure=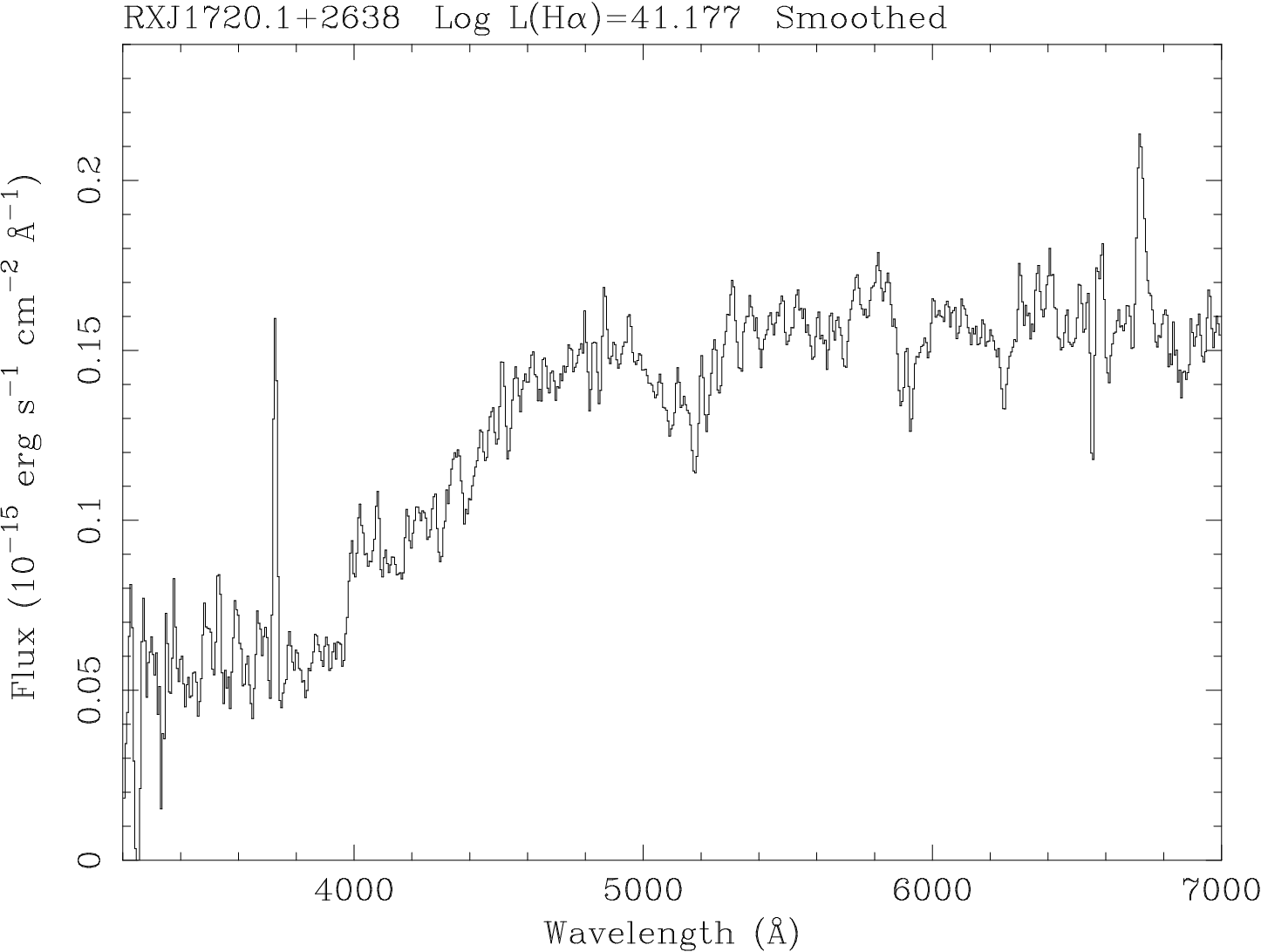,width=0.425\textwidth,angle=270}
}\caption{The spectra of line-emitting central cluster galaxies 
 continued; those with log L(\Han) in the range  41.41--41.17. }
}
\end{figure}\vfill\eject

\addtocounter{figure}{-1}
\begin{figure}
\centering
\vbox{
\hbox{
\psfig{figure=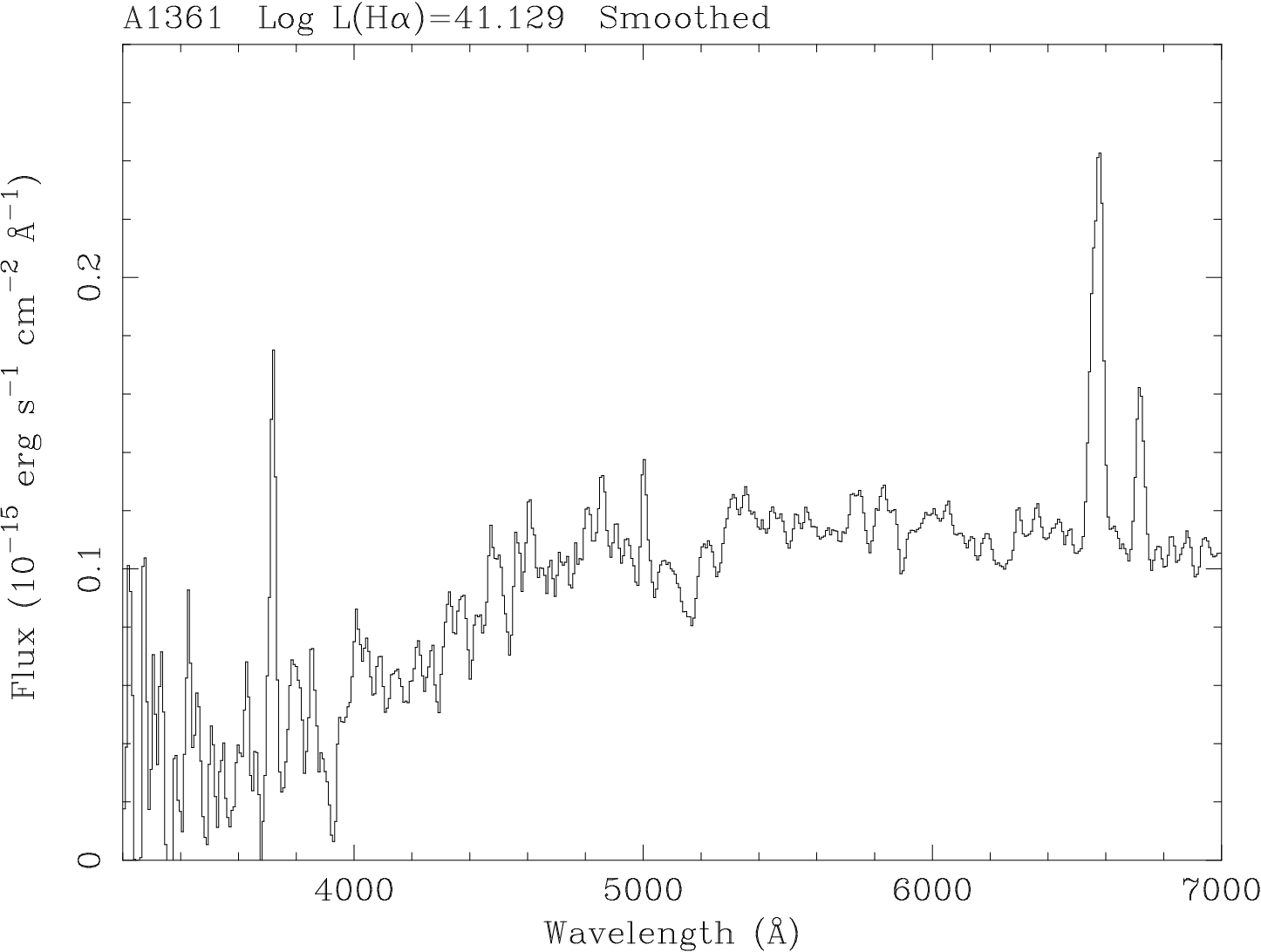,width=0.425\textwidth,angle=270}
\psfig{figure=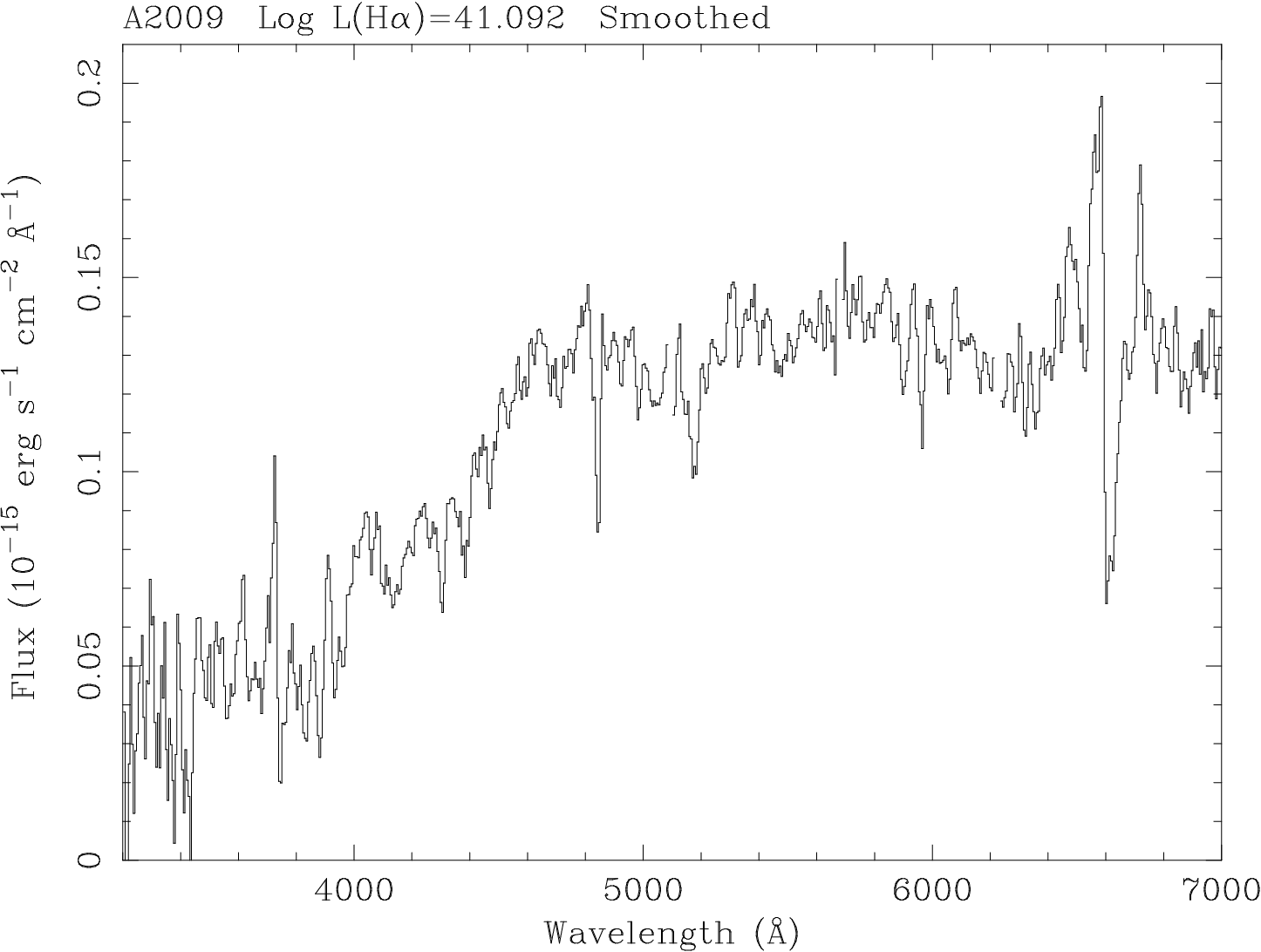,width=0.425\textwidth,angle=270}
}\hbox{
\psfig{figure=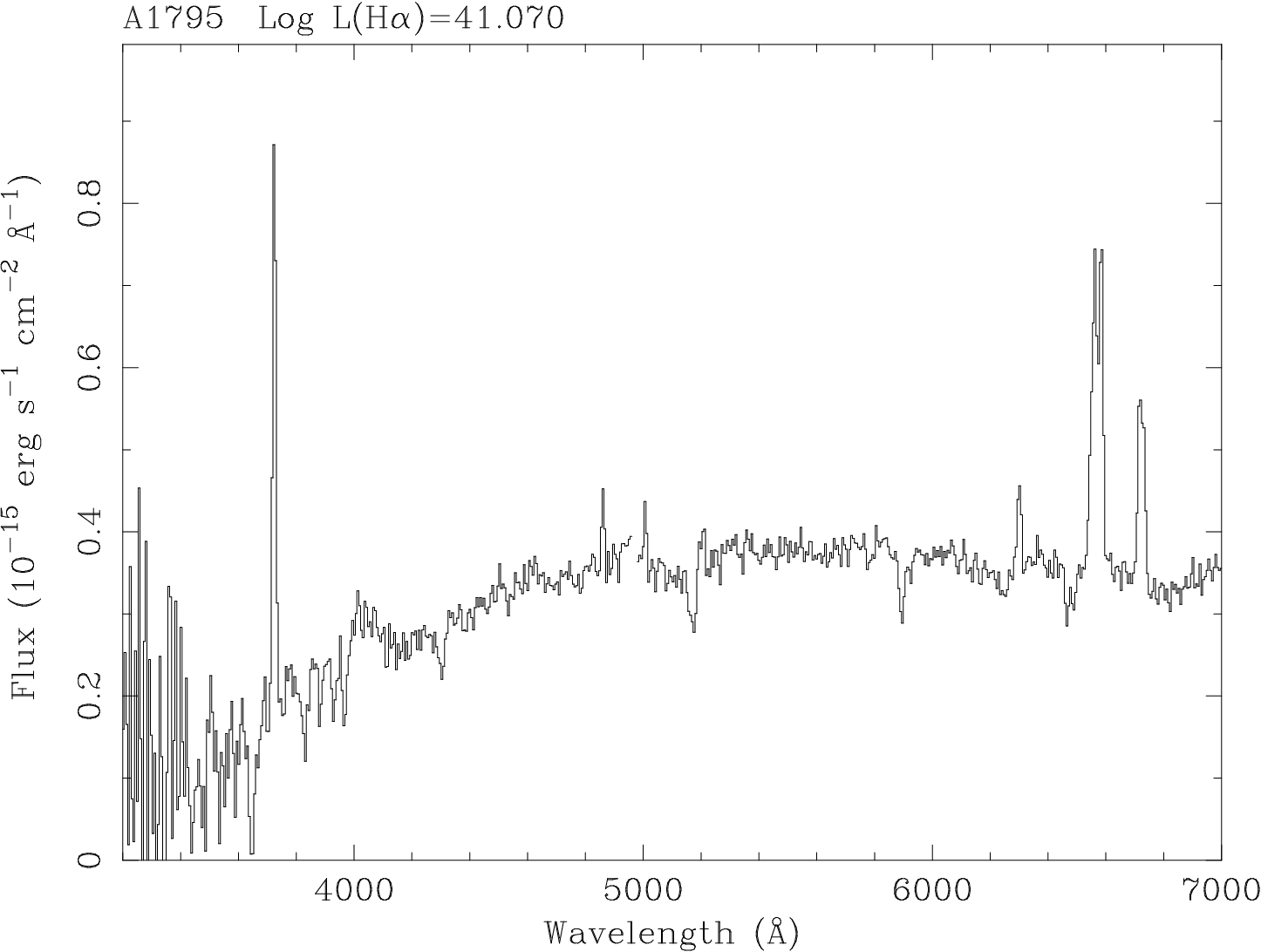,width=0.425\textwidth,angle=270}
\psfig{figure=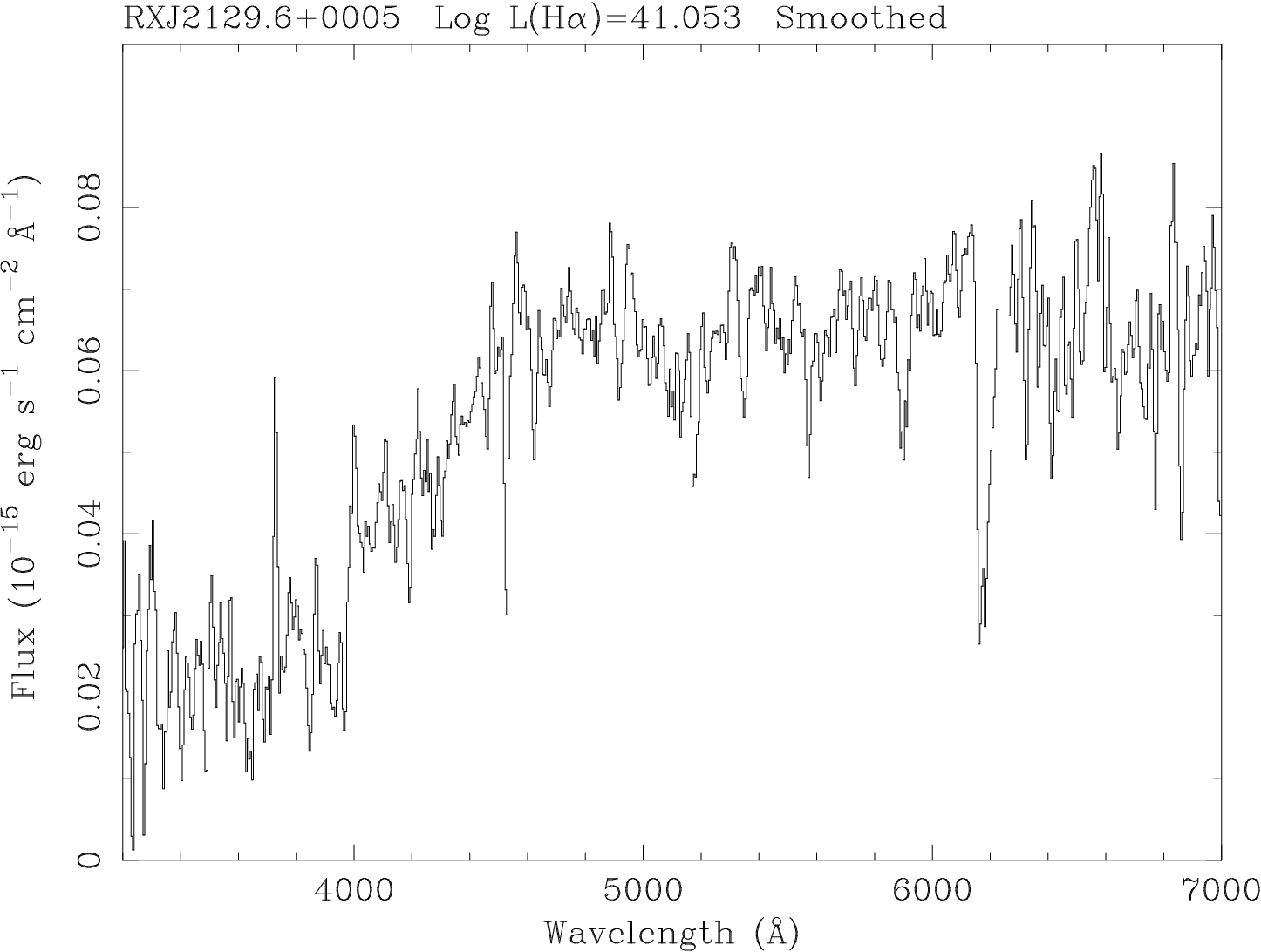,width=0.425\textwidth,angle=270}
}\hbox{
\psfig{figure=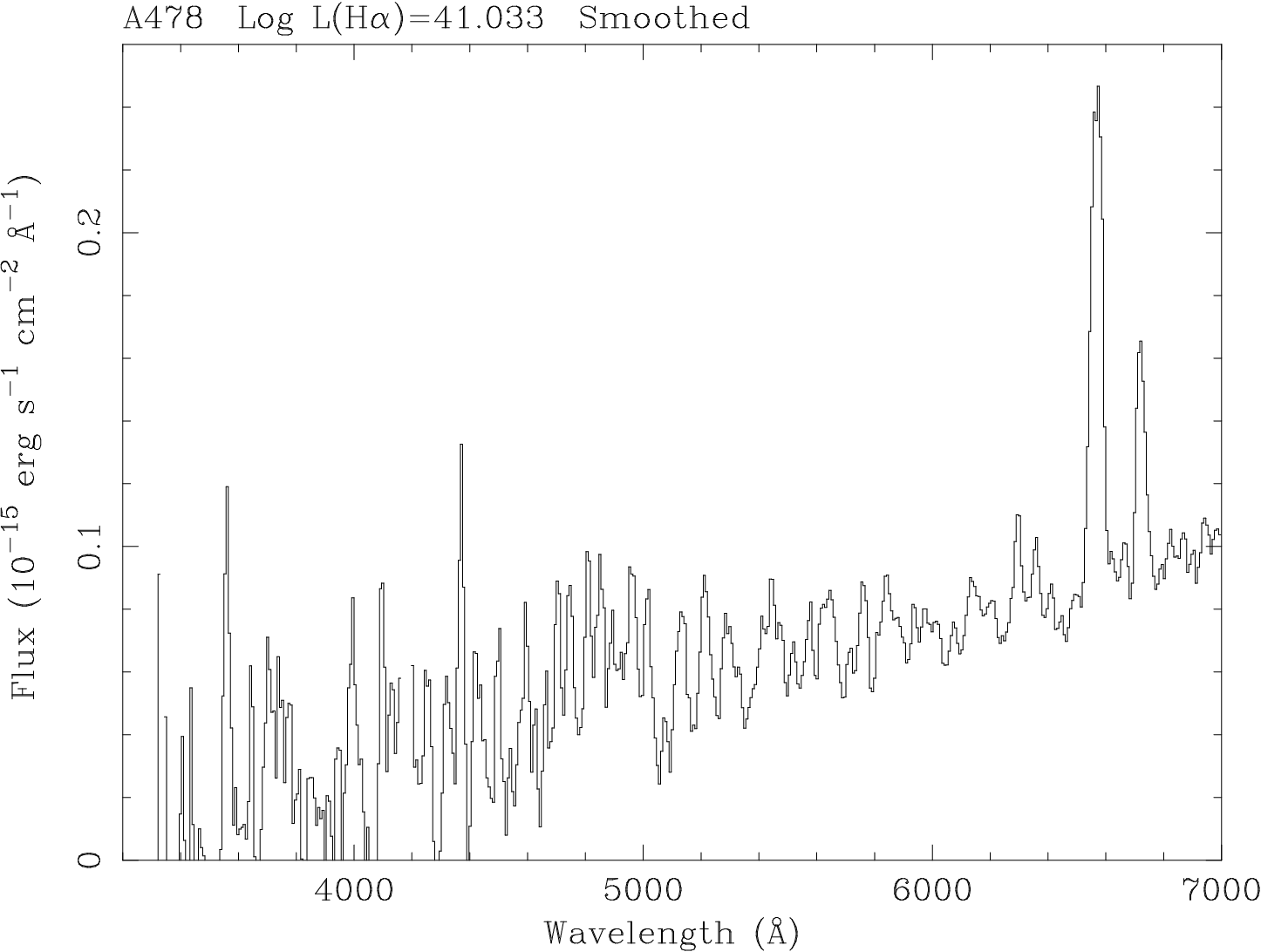,width=0.425\textwidth,angle=270}
\psfig{figure=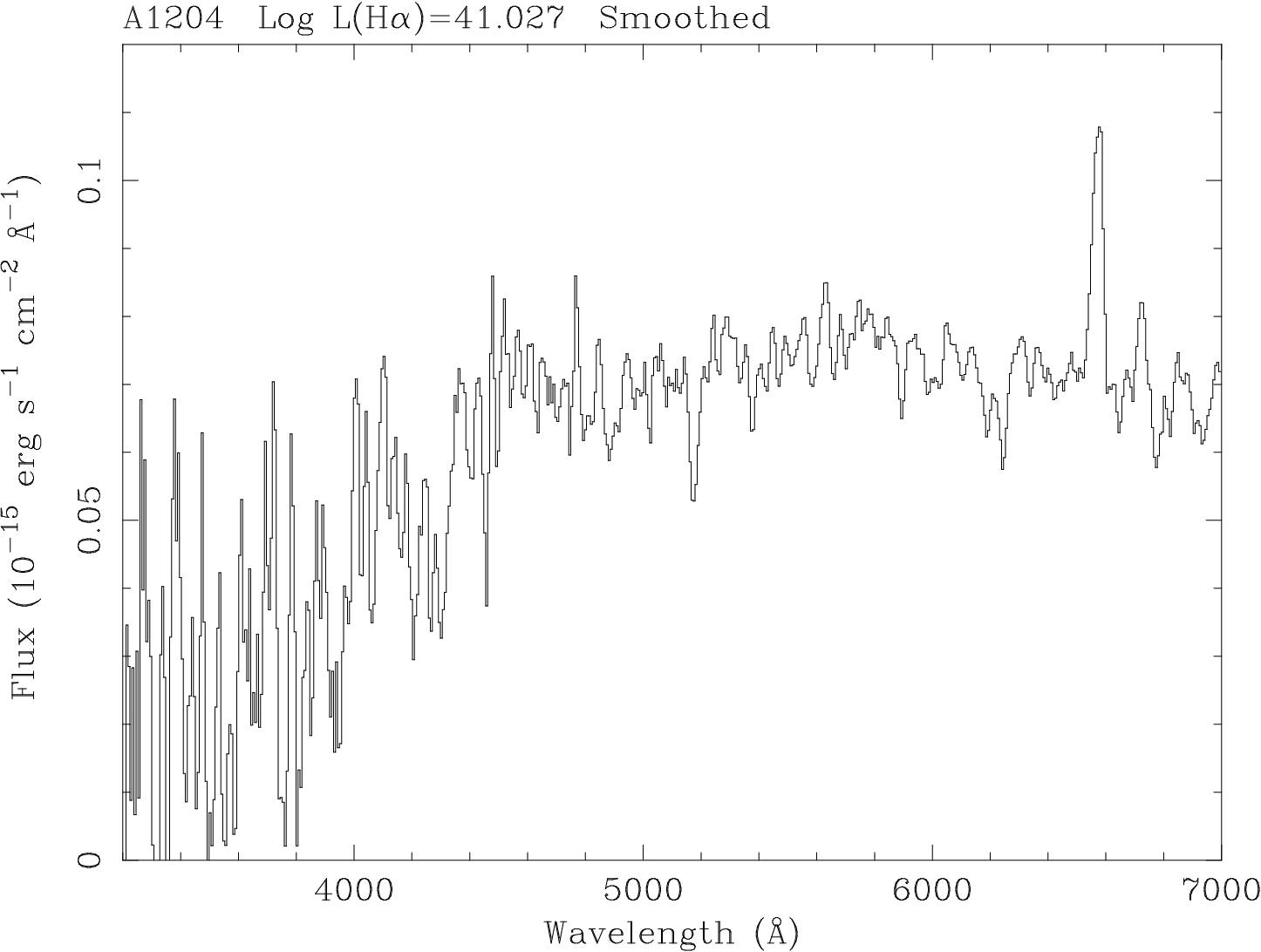,width=0.425\textwidth,angle=270}
}\hbox{
\psfig{figure=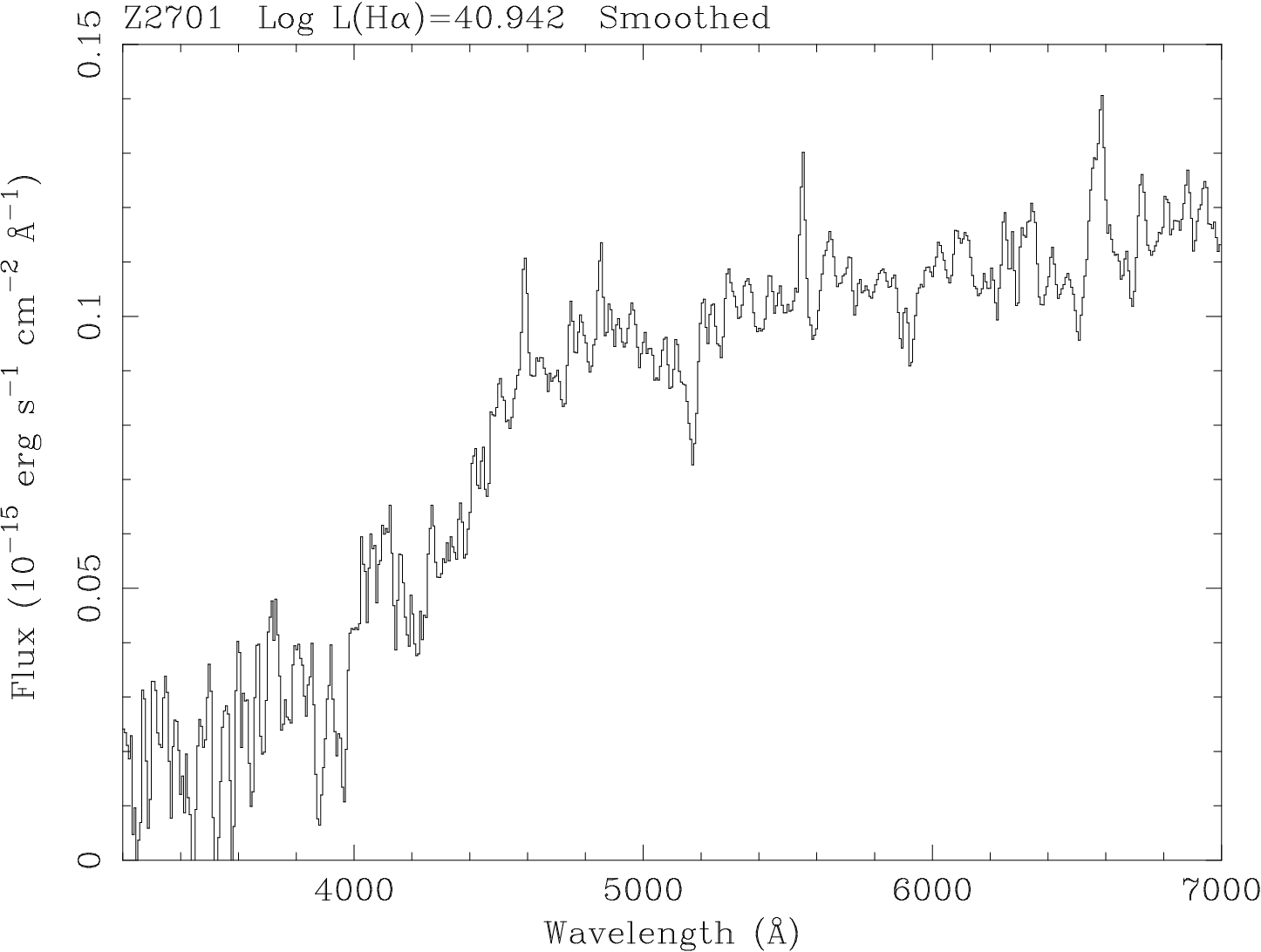,width=0.425\textwidth,angle=270}
\psfig{figure=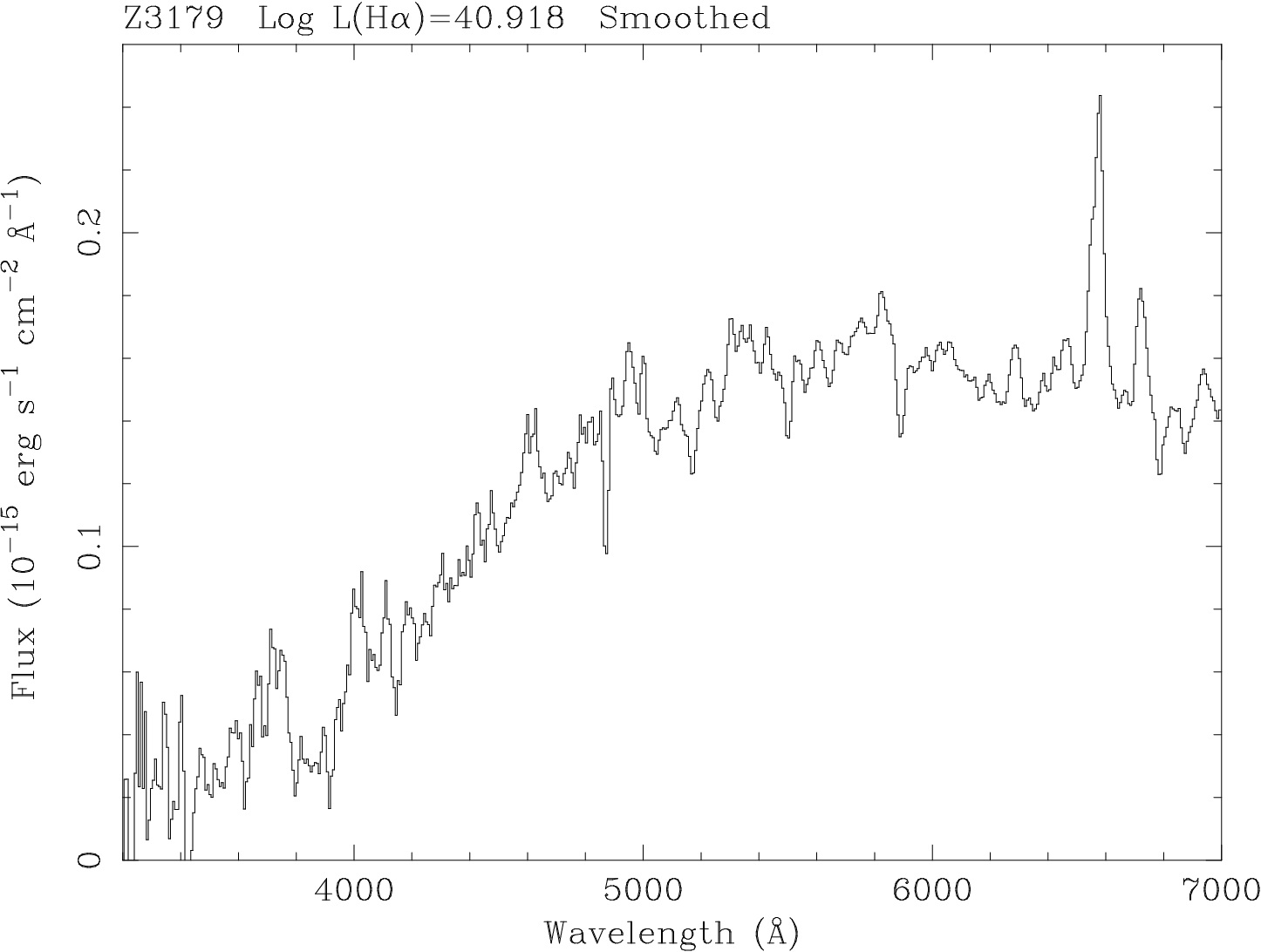,width=0.425\textwidth,angle=270}
}\caption{The spectra of line-emitting central cluster galaxies 
 continued; those with log L(\Han) in the range  41.13-40.91. }
}\end{figure}\vfill\eject

\addtocounter{figure}{-1}
\begin{figure}
\centering
\vbox{ 
\hbox{
\psfig{figure=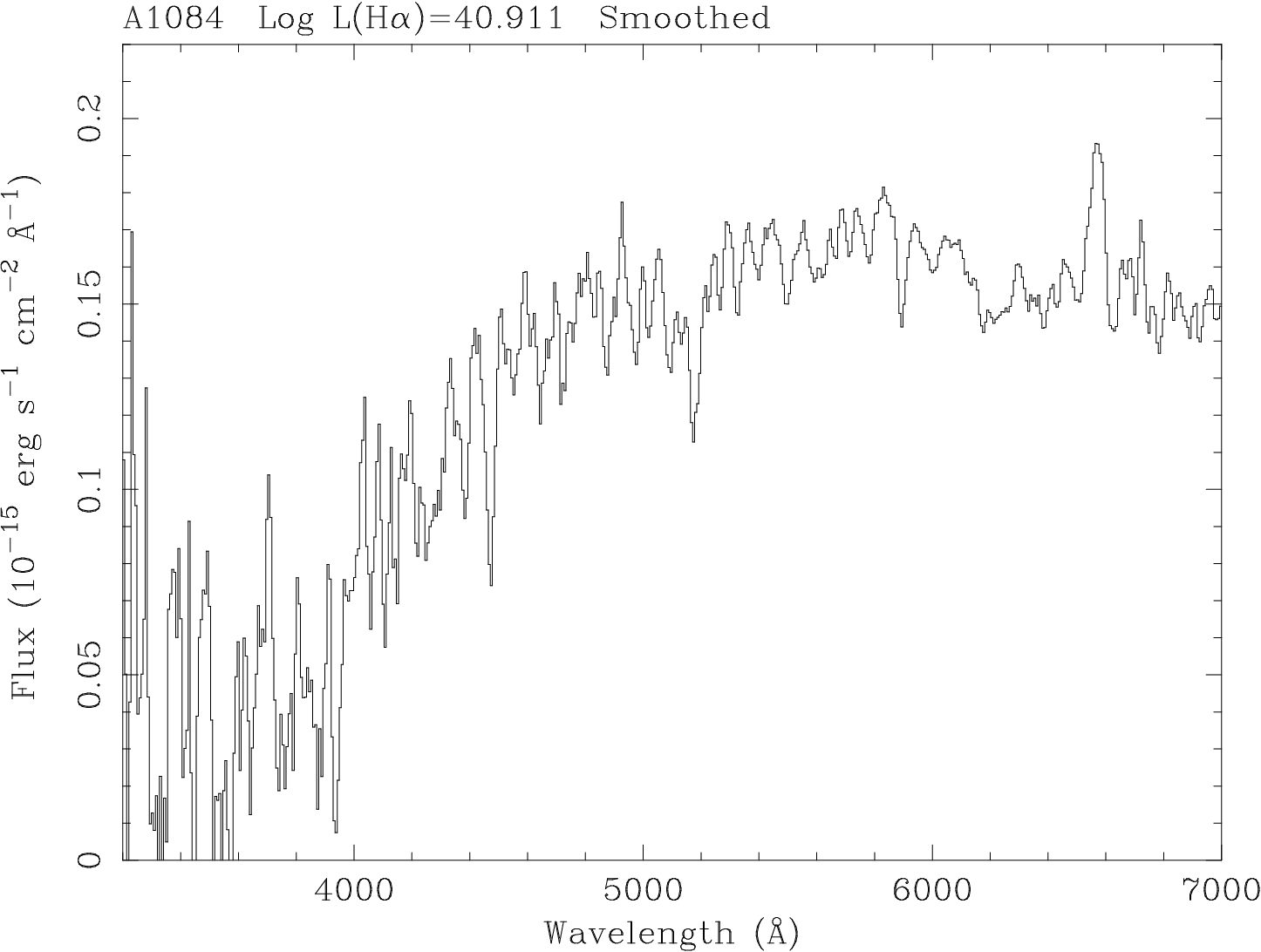,width=0.425\textwidth,angle=270}
\psfig{figure=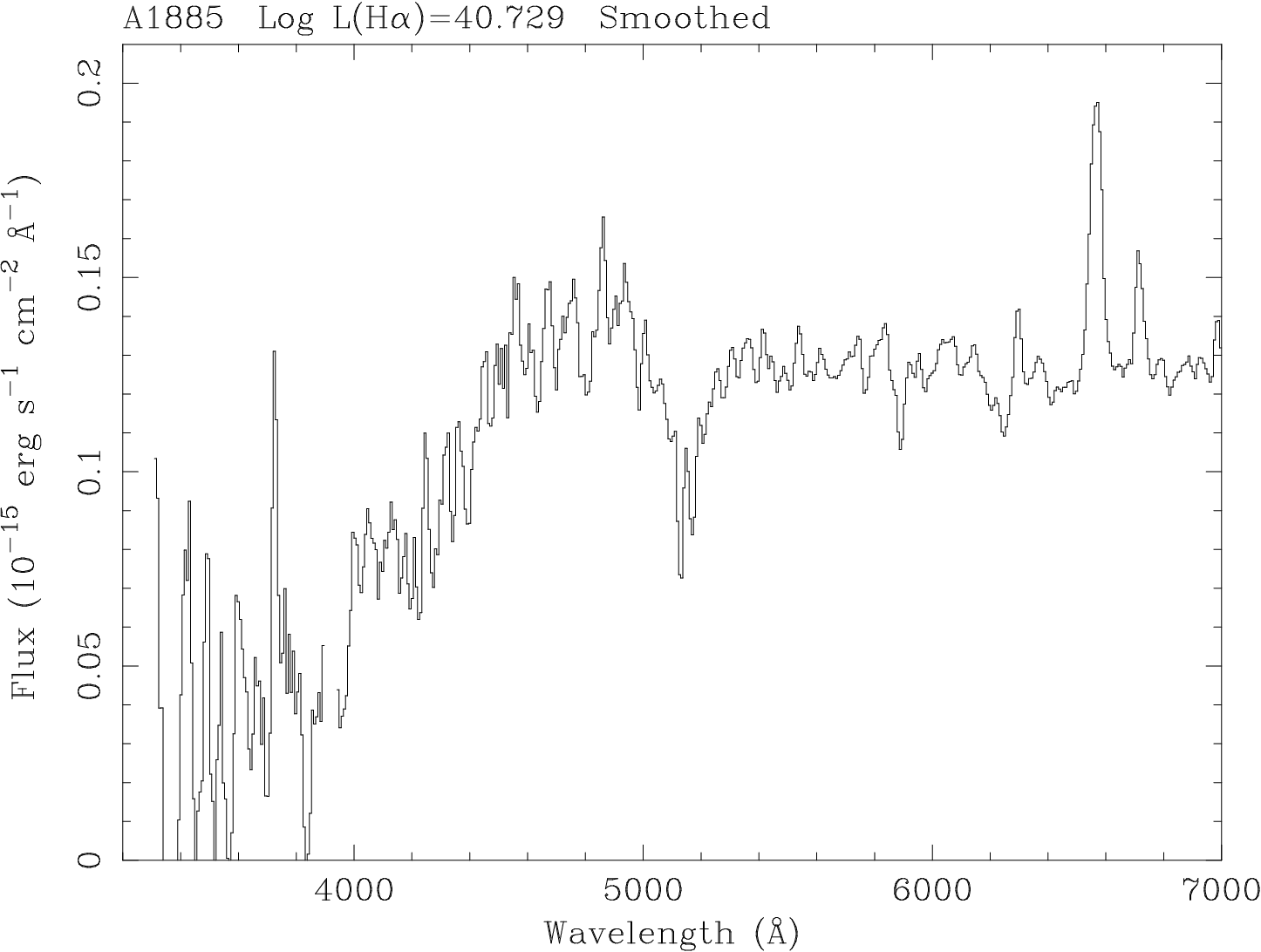,width=0.425\textwidth,angle=270}
}\hbox{
\psfig{figure=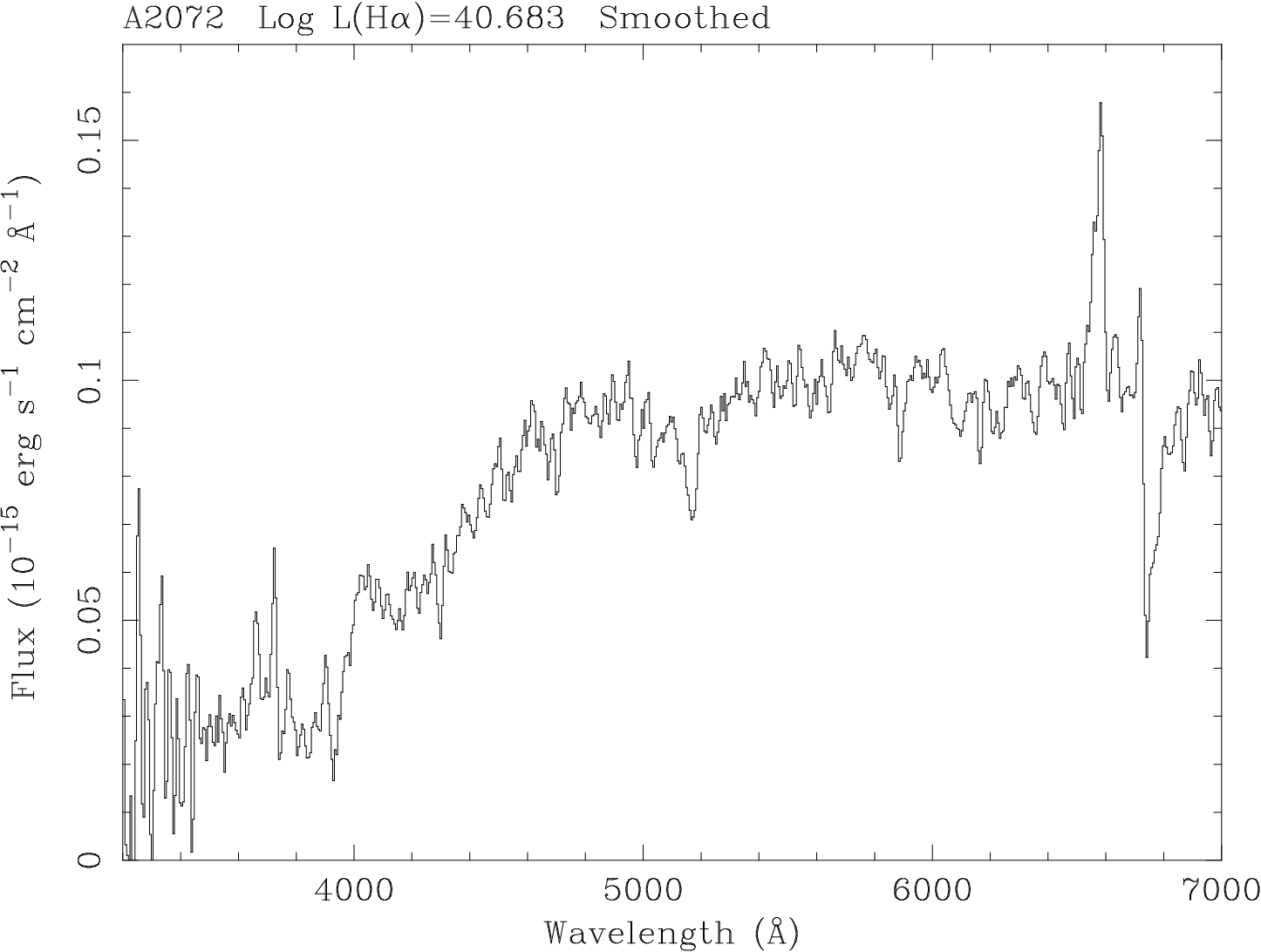,width=0.425\textwidth,angle=270}
\psfig{figure=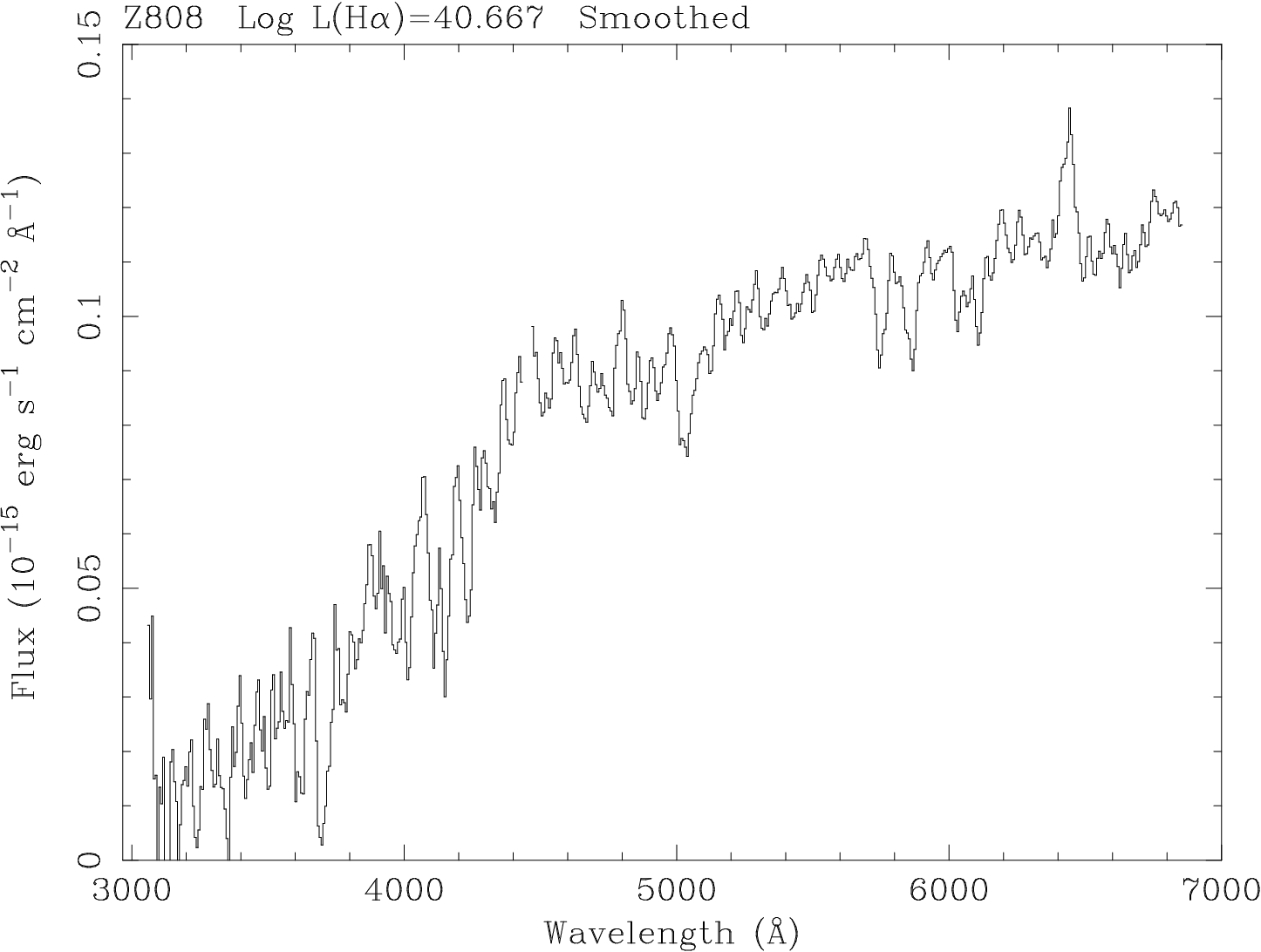,width=0.425\textwidth,angle=270}
}\hbox{
\psfig{figure=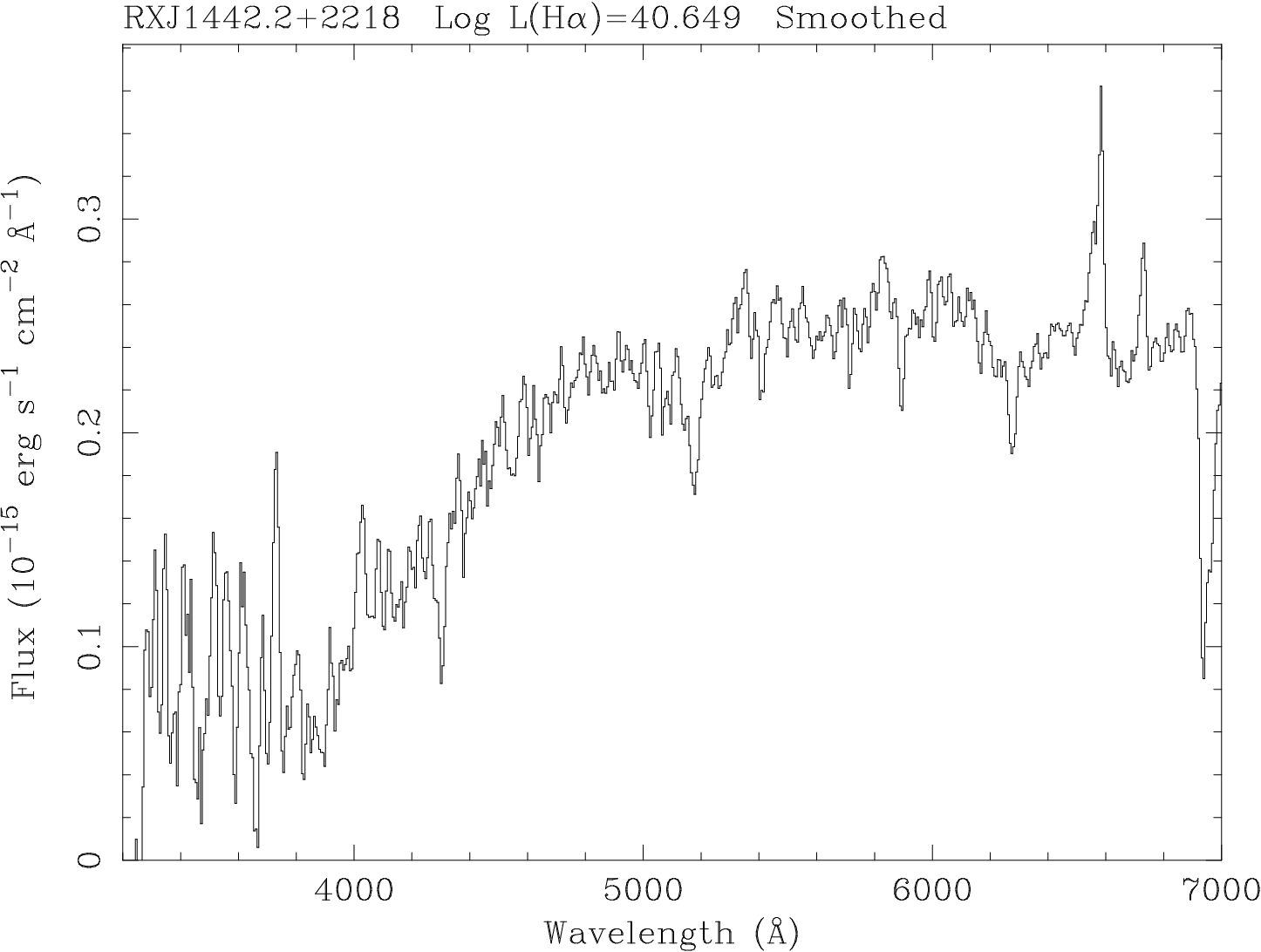,width=0.425\textwidth,angle=270}
\psfig{figure=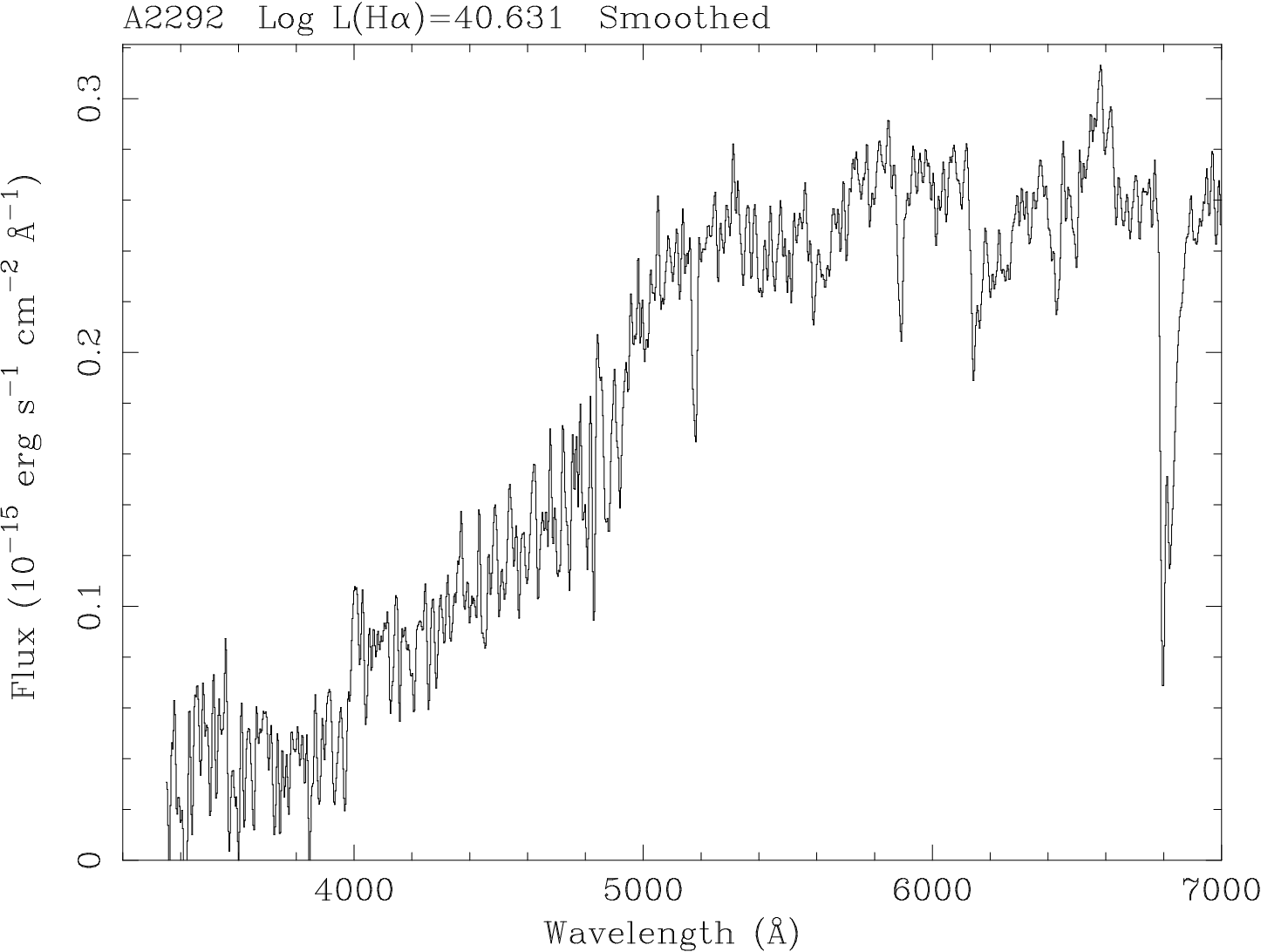,width=0.425\textwidth,angle=270}
}\hbox{
\psfig{figure=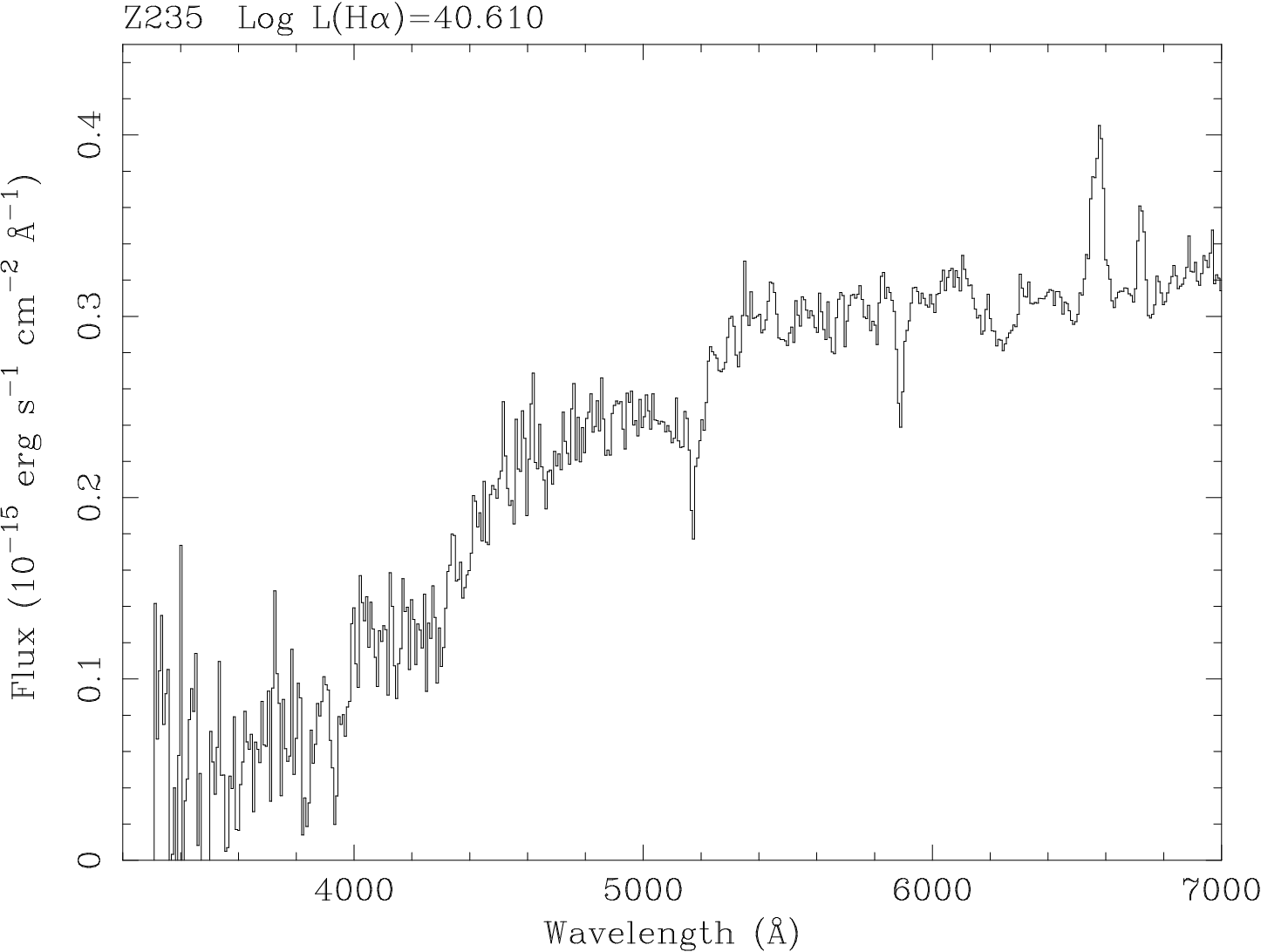,width=0.425\textwidth,angle=270}
\psfig{figure=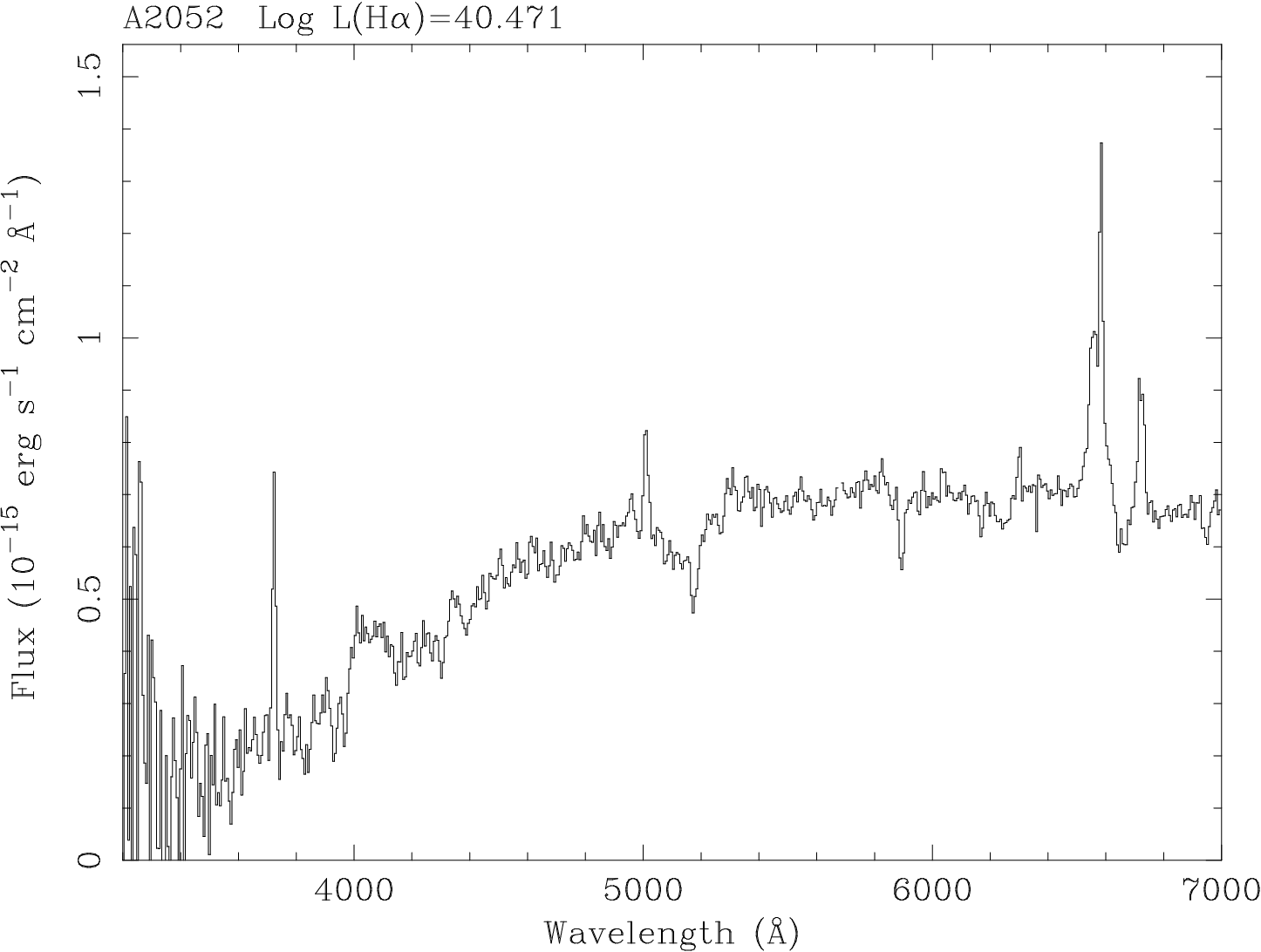,width=0.425\textwidth,angle=270}
}\caption{The spectra of line-emitting central cluster galaxies 
 continued; those with log L(\Han) in the range  40.91--40.47. }
}\end{figure}
\vfill\eject

\addtocounter{figure}{-1}
\begin{figure}
\centering
\vbox{
\hbox{
\psfig{figure=fig2p6a.ps,width=0.425\textwidth,angle=270}
\psfig{figure=fig2p6b.ps,width=0.425\textwidth,angle=270}
}\hbox{
\psfig{figure=fig2p6c.ps,width=0.425\textwidth,angle=270}
\psfig{figure=fig2p6d.ps,width=0.425\textwidth,angle=270}
}\hbox{
\psfig{figure=fig2p6e.ps,width=0.425\textwidth,angle=270}
\psfig{figure=fig2p6f.ps,width=0.425\textwidth,angle=270}
}\hbox{
\psfig{figure=fig2p6g.ps,width=0.425\textwidth,angle=270}
\psfig{figure=fig2p6h.ps,width=0.425\textwidth,angle=270}
}\caption{The spectra of line-emitting central cluster galaxies 
 continued; those with log L(\Han) in the range  40.45--40.08. }
}\end{figure}\vfill\eject

\addtocounter{figure}{-1}
\begin{figure}
\centering
\vbox{
\hbox{
\psfig{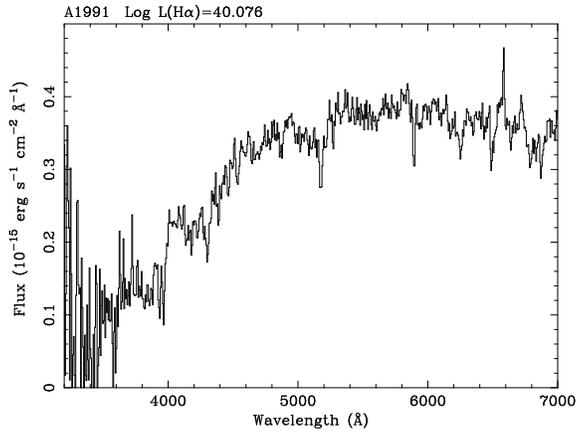}
} \caption{The spectra of line-emitting central cluster galaxies; 
A1991 with  log L(\Han) of 40.07. }}\end{figure}

\begin{figure}
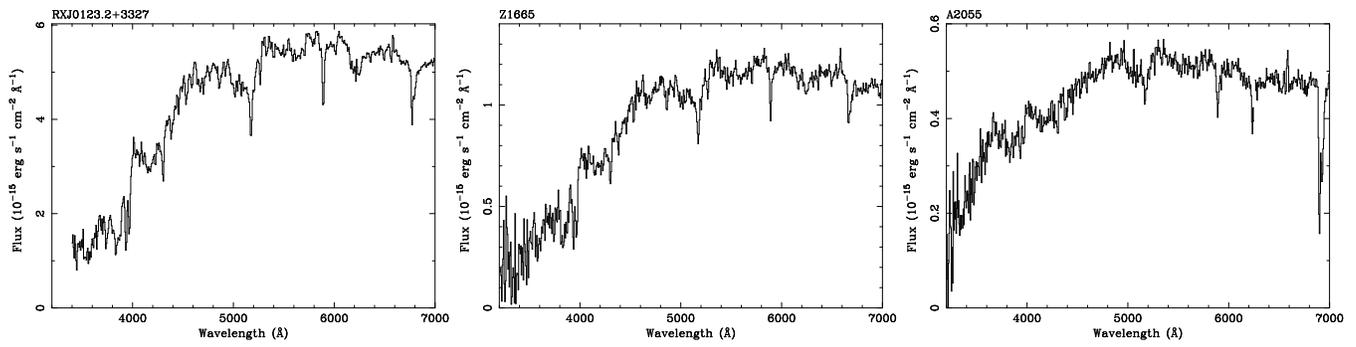

\centering
\vbox{\hbox{
\psfig{figure=fig3a.ps,width=0.33\textwidth,angle=270}
\psfig{figure=fig3b.ps,width=0.33\textwidth,angle=270}
\psfig{figure=fig3c.ps,width=0.33\textwidth,angle=270}
}\caption{
\label{fig:nemspectra}
Three examples (RXJ0123.2+3327, Z1665 and A2055) of central cluster galaxies with spectra showing only low level 
\niib\ line emission. }
}\end{figure}
\vfill\eject

\twocolumn

\begin{figure}
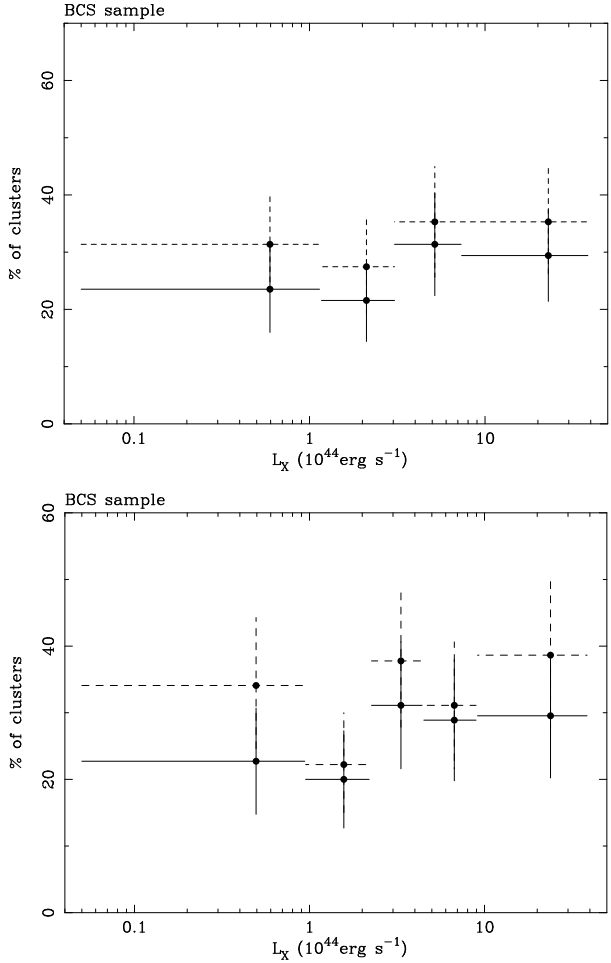

\vbox{
\psfig{figure=fig4top.ps,width=0.45\textwidth,angle=270}
\vspace{0.25cm}
\vbox{
\psfig{figure=fig4bot.ps,width=0.45\textwidth,angle=270}}
\caption{
\label{fig:lxfreq}
Frequency of line emitters as a function of 0.1-2.4\keV\ X-ray luminosity:
(top) for those  galaxies in the BCS (ie Tables~\ref{tab:bcslog}  and
~\ref{tab:bcsnotlog}; assuming the five clusters
whose spectral properties
are not known are non-line emitters) and (bottom) for all the
galaxies included in this paper for which we have
the X-ray luminosity (ie members of the BCS, \lq sub'-BCS and XBACS).
In both figures, markers with solid error bars show only the line
emitters
marked by a $\surd$ in Tables~\ref{tab:bcslog}--~\ref{tab:bcsnotlog}, whilst
the markers with dotted error bars  include
those with low-level \niib\ and no \Han. The errors assume $\surd n$
statistics.}}
\end{figure}

\begin{figure}
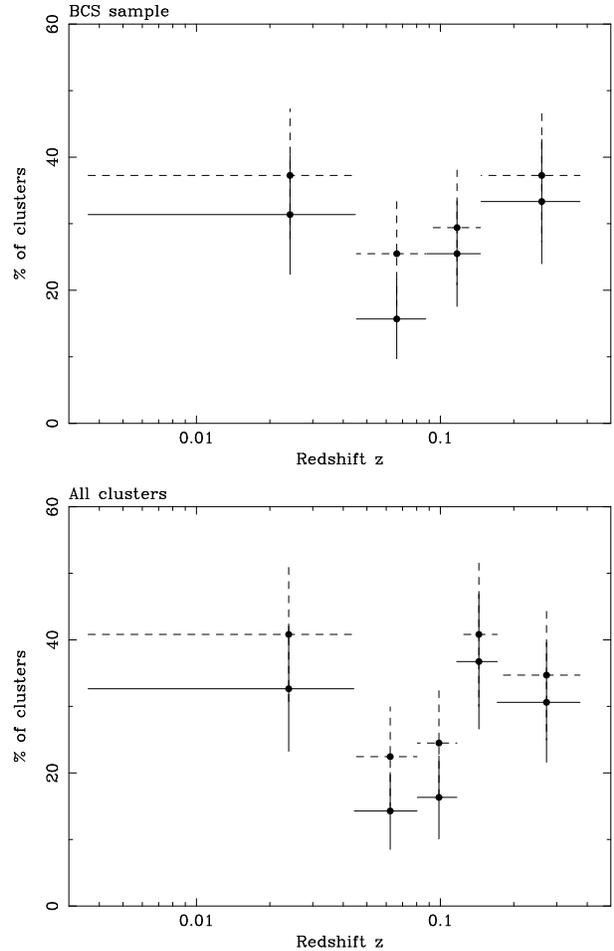

\vbox{
\psfig{figure=fig5top.ps,width=0.45\textwidth,angle=270}
\vspace{0.25cm}
\psfig{figure=fig5bot.ps,width=0.45\textwidth,angle=270}}
\caption{\label{fig:zfreq}
Frequency of line emitting central cluster galaxies  as a function of redshift for
(top) all the BCS sample (ie Tables ~\ref{tab:bcslog}  and
\ref{tab:bcsnotlog}; assuming the five  clusters whose spectral properties
are not known are non-line emitters) and (bottom) for all clusters included in the
paper. Markers and error bars as in Fig~\ref{fig:lxfreq}.
 }
\end{figure}

\subsection{Comparison of X-ray and optical centroids}
\label{sec:comparison}

We have calculated the projected separation between the X-ray centroid of each cluster
detection in the RASS (from Paper~I) and the optical position of
its BCG. Where identification of a single dominant galaxy is not obvious, we
have used the co-ordinates of the galaxy nearer to the X-ray centroid. We
have also excluded galaxies with AGN-type line intensity ratios (see
section~\ref{sec:lineint}), as there is a small possibility that an X-ray
centroid may be skewed towards a BCG if it is itself an X-ray emitter
(although ROSAT HRI images demonstrate that point source contamination is in
general 
not a problem). Although the coordinates of each optical galaxy are very
well determined from the digitized POSS
(Tables~\ref{tab:bcslog}--~\ref{tab:bcsnotlog}), the positions for the RASS
X-ray sources are far less well determined due to the broadness of the RASS
point-spread function. To make things worse, the attitude solution for the
RASS is less accurate than the one for pointed ROSAT observations, the
photon statistics in the RASS are generally poor and faint blends are
harder to recognize, all of which effects combine to give 
 X-ray positional errors of up to about 1 arcmin. A few sources may
have larger errors on the X-ray centroid if the cluster contains an AGN or
substantial substructure.

Even with these caveats on the accuracy of the X-ray position, we find that
the projected separation between the X-ray centroid of the cluster and the
optical position of the dominant galaxy  differs
according to the emission-line properties of the BCG. The line emitters
(including the [NII]-only emitters) show a much smaller average separation
between the BCG and cluster X-ray centroid (with a mean of 29$\pm$6\kpc), than do the
others (a mean of 89$\pm$8\kpc).  Assuming that the majority of the line emitting
galaxies are contained in cooling flows, this is consistent 
with the results of both Peres \etal (1998) and Allen (1998)
that the offset between the X-ray centroid and the brightest central galaxy,
or the gravitational lensing centre are larger for non-cooling flow clusters. 
A precise positional coincidence is not necessarily expected, especially if
the cluster contains a strong central radio source that may have displaced
the cooling gas at the very centre, eg 3C84 in the Perseus cluster
(B\"ohringer \etal 1993). We will address this issue further in the next
paper in the series, using better X-ray centroids obtained from pointed
ROSAT observations of this sample and also X-ray confirmation of whether
an individual cluster contains a cooling flow.

\subsection{Continuum spectral indices}
\label{sec:contmSI}
The ultraviolet/blue continuum sometimes seen in BCG is usually ascribed
to the light from an excess young stellar population (eg JFN;
McNamara \& O'Connell 1989, 1992, 1993; Allen 1995; McNamara \etal 1998). The level of massive
star formation can be assessed from its effect on absorption features in the
spectrum of the galaxy (JFN, Cardiel, Gorgas \& Aragon-Salamanca 1995,
1998). The excess blue light is concentrated towards the central 5-10 kpc of the
galaxy, and outside of this region the stellar spectral indices are very
similar to that of an ordinary giant elliptical galaxy (Cardiel \etal 1998).
Thus to make a meaningful comparison between galaxies in our sample, where
possible we measure the continuum spectral indices in a spectrum taken from
the same central region of a galaxy.

A spectrum for each galaxy was extracted from a projected aperture of
diameter 10\kpc\ along the slit. This extraction is straightforward
for the lower-redshift objects, whereas the accuracy of obtaining a 10\kpc\ aperture at the
more distant galaxies was dictated by whether the intensity peak fell on one
or between 2 rows of the detector, and the need to extract a spectrum using
an integer number of rows. In practise the spectra are extracted from
aperture sizes varying only between 8.6 and 11.2\kpc; we have listed the
aperture used in column 2 of Table~\ref{tab:stelind}  where the stellar indices derived from
these spectra are tabulated. Selecting a central aperture is only possible
for the {\it new} IDS observations presented in this paper. For the FOS
spectra taken from A92 and C95  we cannot specify the aperture (other
than an upper limit taken from the slit length of 6 arcsec) as the spectra
were extracted using an automatic optimal extraction technique that does not
yield spatially extended information. The aperture taken to produce the
spectra from the WHT observations in C95 is unknown.

The common presence of Balmer line emission in the blue continuum-excess
 galaxies precludes the use of standard Balmer absorption line indices to
 assess the stellar population. Instead we measure the strength of  two other 
 stellar features in our spectra: the Mg$_2$ index (which assesses the depth
 of the MgH and Mg{\it b} molecular bands at 5177\AA; as defined by Faber
 \etal 1985) and the \lq 4000\AA-break':
$$ D_{4000} = { {\int_{4050}^{4250}f_{\lambda}d\lambda}
\over  { {\int_{3750}^{3950}f_{\lambda}d\lambda} } }.$$
\noindent The 4000\AA-break was originally defined from spectra in 
$f_{\nu}$ (Bruzual 1983) but note that here we measure it from an
$f_{\lambda}$ spectrum. 
We also use a wider colour ratio, $\delta$BR, more suited to testing the
presence of an excess blue continuum, as defined in A92 and C95: 
$$ \delta BR =
{ {\int_{3500}^{3650}f_{\lambda}d\lambda}
\over {{\int_{5800}^{6200}f_{\lambda}d\lambda} } }$$
\noindent
Unlike both D$_{4000}$ and the Mg$_2$ index, $\delta$BR uses bandpasses free
from line emission, although it is more sensitive to the presence of internal
reddening (see section~\ref{sec:reddening}). D$_{4000}$ can be affected by
the emission lines [NeIII]$\lambda3869$, [SII]$\lambda\lambda$4069/4076 and
H$\delta$, and the Mg$_2$ index by [NI]$\lambda$5199 and
[OIII]$\lambda$4959. In practise, these lines of [NeIII] and [SII] are
rarely present in the BCG spectra, but contamination of Mg$_2$ by [OIII] and
[NI] is more commonplace. Where present, we remove  emission lines by
fitting them in a  \lq residual' spectrum created by subtracting a scaled non-emission-line
galaxy template from the BCG  (see section~\ref{sec:emilines}). This 
emission-line model  is then subtracted from the original galaxy spectrum,
and the
D$_{4000}$ and Mg$_2$ stellar indices are measured from this line-corrected
spectrum. 
 The stellar indices for each galaxy are shown in
Table~\ref{tab:stelind}, in RA order. Column~1 gives the name, and column~2
the projected spatial size of the aperture from which the given spectral
indices are measured. Columns 3, 4 and 5 give the Mg$_2$ index, D$_{4000}$
and $\delta$BR respectively. Column~6 presents the \Ha\ luminosity {\em from
the given aperture} -- note that as the emission lines are often spatially
extended beyond a 10\kpc\ diameter, this may represent an underestimate of the {\sl total} \Ha\
luminosity of the galaxy. Columns~7--9 give the estimate of internal
reddening from the Balmer emission lines, and the corrected values of
$\delta$BR and \Ha luminosity (see section~\ref{sec:reddening}). 
Bracketed values of D$_{4000}$ in column 4
represent uncertain measures where examination of the spectrum suggests the
abnormally high value is caused by a loss of blue light from the slit.

\onecolumn 

\begin{table}
\caption{ Stellar indices and \Ha luminosities from the $\sim10$~kpc 
galaxy apertures. \label{tab:stelind}}

\begin{tabular}{lcrcccrrrr}
           &   &          &         &         &            &        & &   &  \\
Cluster    &   & aperture & Mg$_2$  & 4000\AA & $\delta$BR & L(H$\alpha$) & E(B-V) & revised  & revised L(H$\alpha$) \\
           &   & size     & index   & break   & &(10$^{40}$erg/s) &   & $\delta$BR & (10$^{40}$erg/s)   \\
           &   & (kpc)         &         &         &            &        & &   &  \\
A7         &     & 10.5 & 0.31   & 1.92   & 0.11 & --- & --- & --- & ---  \\
A21 &a           & 10.1 & 0.30   & 2.12   & 0.05 & --- & --- & --- & ---  \\
    &b           & 10.1 & 0.31   & 2.09   & 0.08 & --- & --- & --- & ---  \\
A76  &           & 10.0 & 0.32   & 1.91   & 0.08 & --- & --- & --- & ---  \\
Z235           & & $<$12.6 & 0.29   & 1.90   & 0.06 & 4.1$_{-0.6}^{+0.6}$ & --- & --- & ---  \\
A115 &         &  8.8 & 0.33 & 1.60   & 0.14 & 12.7$_{-1.8}^{+1.9}$  & --- & --- & ---  \\ 
RXJ0107.4+3227 & & 10.4 & 0.31   & 1.78   & 0.09 & 0.5$_{-0.1}^{+0.1}$ &--- & --- & ---  \\
Z353 $^1$           & & $<$15.9 & 0.24   & (7.1)& 0.02 & ---  & ---   & --- & ---  \\
A168           & & 10.1 & 0.30   & 1.89   & 0.11 & --- & --- & --- & ---  \\
RXJ0123.2+3327 & &  9.8 & 0.30   & 1.86   & 0.09 & --- & --- & --- & ---  \\
RXJ0123.6+3315 & & 10.4 & 0.25   & 1.88   & 0.09 & --- & --- & --- & ---  \\
A193           & & 10.0 & 0.29   & 1.77   & 0.09 & --- & --- & --- & ---  \\
A267           & & $<$27.9 & 0.31   & 1.91   & 0.08 & ---  & ---  & --- & ---  \\
A262           & & 10.2 & 0.28 & 1.95   & 0.08 & 0.6$_{-0.1}^{+0.1}$ & --- & --- & --- \\
A272           & &  9.5 & 0.34   & 1.85   & 0.10 & --- & --- & --- & ---  \\
A291           & & $<$25.1 & 0.28 & 1.56   & 0.23 & 45.9$_{-1.9}^{+2.2}$ & ---  &---  & --- \\ 
RXJ0228.2+2811 & & 10.1 & 0.30   & 1.85   & 0.09 & --- & --- & --- & ---  \\
A376           & &  9.9 & 0.30   & 1.92   & 0.09 & --- & --- & --- & --- \\
A400 &a          & 10.2 & 0.31   & 1.90   & 0.07 & --- & --- & --- & --- \\
     &b          & 10.9 & 0.29   & 1.86   & 0.07 & --- & --- & --- & ---  \\
A399  &          & 10.1 & 0.29   & 1.68   & 0.14 & --- & --- & --- & ---  \\
A401  &          & 10.6 & 0.28   & 1.58   & 0.13 & --- & --- & --- & ---  \\
Z808            & & $<$22.4 & 0.23 & 2.02   & 0.08 & 4.6$_{-1.3}^{+1.2}$ & ---  & ---   & --- \\
A407  &          & 10.9 & 0.27   & 1.79   & 0.10 & --- & --- & --- & --- \\
A409 $^2$           & & $<$20.8 & 0.28  & 2.47   & 0.08 & ---   & ---   & --- & --- \\
RXJ0338.7+0958 & &  9.7 & 0.28   & 1.37   & 0.11 & 10.3$_{-0.4}^{+0.5}$ & 0.43$^{+0.13}_{-0.15}$ &$0.24^{+0.07}_{-0.06}$ & $29.0^{+10.6}_{-8.8}$   \\
RXJ0341.3+1524 & & $<$4.8 & 0.28  & 2.37 & 0.05 & ---  & ---   & --- & ---  \\%
RXJ0352.9+1941 $^1$  & & $<$15.9 & 0.27  & (2.81)  & 0.04 &
58.2$_{-1.6}^{+1.3}$ & 0.43$_{-0.07}^{+0.07}$ & $0.08^{+0.01}_{-0.01}$   &
$148.7^{+27.3}_{-23.0}$    \\
A478 $^2$           & & $<$13.0 & 0.24 & 2.08   & 0.11 & 10.8$_{-0.5}^{+0.5}$ &
0.29$^{+0.25}_{-0.29}$ & 0.20$^{+0.11}_{-0.09}$  & 20.1$^{+16.8}_{-9.3}$  \\
RXJ0419.6+0225 & & 10.6 & 0.31   & 1.80   & 0.10 & --- & --- & --- &---  \\
RXJ0439.0+0715 $^2$& & 10.2 & 0.24   & 1.76   & 0.17  & --- & --- & --- &--- \\
RXJ0439.0+0520 $^1$  & & $<$26.1 & 0.35 & 1.32   & 0.09
&110.7$_{-4.5}^{+3.0}$ & 0.40$_{-0.08}^{+0.07}$   & $0.19^{+0.03}_{-0.03}$
& $239.7^{+42.3}_{-44.0}$    \\
A520 $^2$          & &  8.9 & 0.22   & 2.02   &  0.13 & --- & --- & ---& --- \\
A523           & & 10.7 & 0.32   & 1.82   &  0.12 & --- & --- & ---& ---  \\
A531             & & $<$14.0 & 0.29 & 1.66   & 0.09 & ---  & ---  &--- & --- \\
RXJ0503.1+0608  & & $<$13.3 & 0.30 & 2.19 & 0.04 & ---  &---    &---&--- \\
RXJ0510.7--0801 & & $<$26.8  & 0.33 & 1.65 & 0.05 & --- & ---   & ---&---  \\
Z1121                 & & $<$12.6 & 0.18      &   1.13 & 0.11 & ---  &---& --- & --- \\
Z1133                 & & $<$22.9 & 0.30   & 1.91   & 0.07 & ---  &--- & --- &---  \\
A566           & &  9.9 & 0.29   & 1.62   &  0.15 & --- & --- & --- &--- \\
A576 & a         &  9.9 & 0.28   & 1.70   &  0.10 & --- & --- & ---& ---  \\
     & b         & 10.8 & 0.30   & 1.89   &  0.10 & --- & --- & ---& --- \\
A586 &           & 10.5 & 0.28   & 1.73   &  0.17 & --- & --- & --- &--- \\
RXJ0740.9+5526 & & 10.5 & 0.29   & 1.76   &  0.07 & --- & --- & --- &--- \\
RXJ0751.3+5012 &a & 9.8 & 0.29   & 1.96   & 0.08  & --- & --- & ---&--- \\
               &b & 10.2 & 0.28  & 1.49   & 0.12  & 0.5$_{-0.1}^{+0.1}$ &--- & --- & --- \\
               &c & 10.1 & 0.29  & 1.82   & 0.10  & --- & --- & --- &--- \\
A602 &a          &  9.9 & 0.21   & 1.74   & 0.11 & ---  & --- & --- &---\\
     &b          &  9.9 & 0.31   & 1.53   & 0.12 & ---  & --- & --- &--- \\
A611 $^2$          & & $<$32.2 & 0.16   & 1.57   & 0.12 & ---  & ---  &--- &---\\
A621           & & $<$27.3 & 0.23   & 2.03    & 0.04 & ---  & ---  &---&---\\
RXJ0819.6+6336 & & 10.9 & 0.34   & 1.62   & 0.14 & ---  & --- & --- &---\\
RXJ0821.0+0752 & & $<$16.0 & 0.27   & 1.73  & 0.10 & 30.3$_{-1.9}^{+2.2}$ &
1.16$^{+0.33}_{-0.58}$  &  & \\ 
A646           & & 10.5 & 0.25 & 1.39  & 0.17 & 15.6$_{-1.1}^{+1.2}$& 0.06$^{+0.10}_{-0.06}$  & 
$0.18^{+0.04}_{-0.01}$  & $17.6^{+5.4}_{-2.0}$  \\
Z1665          & & 10.8 & 0.28   & 1.65   & 0.11 & --- & --- & --- &---\\
A655           & & 10.6 & 0.26   & 1.74   & 0.13 & --- & --- & --- & ---\\
A667            & & $<$20.0 & 0.31  & 1.92   & 0.07 & ---  & ---  &---&--- \\
A671           & &  9.5 & 0.28   & 1.75   & 0.09 & --- & --- & --- &---\\
A665           & & 11.1 & 0.19  & 1.76   & 0.12 & --- & --- & --- & --- \\
           &   &           &         &         &            &        & &   &  \\

\end{tabular}
\end{table}

\addtocounter{table}{-1}
\begin{table}
\caption{ Stellar indices and H$\alpha$ luminosities from the $\sim10$~kpc 
galaxy apertures -- continued }
\begin{tabular}{lcrcccrrrr}
           &   &          &         &         &            &        & &   &  \\
Cluster    &   & aperture & Mg$_2$  & 4000\AA & $\delta$BR & L(H$\alpha$) & E(B-V) & revised  & revised L(H$\alpha$) \\
           &   & size     & index   & break   & &(10$^{40}$erg/s) &   & $\delta$BR & (10$^{40}$erg/s)   \\
           &   & (kpc)          &         &         &            &        & &   &  \\
A689 $^2$         & & $<$31.0 & 0.04   & 1.06    &  0.38  & ---& ---  & --- & ---  \\
A697         & & $<$31.8 &  0.26    & 1.85     &  0.11 & ---& ---  &---& --- \\
A750 $^2$           & & 10.9 & 0.15   & 1.91   & 0.16 & --- & --- & ---&---\\
A761        & & $<$13.6 & 0.31 & 1.40    &  0.16 & ---  & ---& ---  & --- \\
A763           & &  9.4 & 0.26   & 1.80   & 0.12 & --- & --- & ---& ---  \\
A757           & &  9.7 & 0.30  & 1.87    & 0.10 & --- & --- & ---& ---  \\
A773 & a & $<$26.8 & 0.27 & 2.32 &  0.11 & ---  & ---& ---  & ---  \\
     & b & $<$27.4 &  0.31 & 1.90   &  0.10 & ---  & ---& ---  & ---  \\
A795           & &  8.8 & 0.32 & 1.61   & 0.14 & 12.9$_{-1.1}^{+1.0}$ &  --- & ---  &--- \\
Z2701          & & $<$26.7 & 0.28 & 1.81   & 0.06 & 8.7$_{-2.8}^{+2.8}$ &  ---  & ---  &---\\
RXJ1000.5+4409 $^2$ & a & $<$21.0 & 0.12 & 1.93   & 0.08 & 2.2$_{-0.8}^{+0.8}$ & ---  & ---  &--- \\
                    & b & $<$20.8 &  0.33 & 2.23 & 0.09 & ---  & ---  &---&---  \\
Z2844          & & 10.3 & 0.36  & 1.79   & 0.09 & --- & --- & --- & ---\\
A961 $^2$           & &  9.7 & 0.24  & 2.44   & 0.07 & --- & ---& --- & ---  \\
A963           & &  9.1 & 0.31  & 1.94   & 0.10 & --- & ---& --- & ---  \\
A971 & & --- & 0.20 & 1.96 & 0.08        & ---& --- & --- & ---   \\ 
A980           & & 10.0 & 0.30  & 1.79   & 0.11 & --- & --- & ---& ---   \\
Z3146          & & $<$32.0 & 0.22 & 1.14  & 0.42 & 704.7$_{-11.1}^{+14.8}$
&0.20$_{-0.03}^{+0.04}$    &$0.58^{+0.05}_{-0.03}$ & $881.5^{+88.5}_{-61.5}$  \\
A990 $^2$ & & --- & 0.36 & 2.13 & 0.12 & ---& --- & --- & ---   \\ 
Z3179          & & $<$19.8 & 0.22 & 1.90 & 0.07 & 9.1$_{-3.1}^{+2.0}$ & --- & --- & --- \\
A1023          & & $<$16.8 & 0.27  & 1.76   & 0.13 & --- & --- & --- &--- \\
A1033          & & $<$17.9 & 0.33  & 1.50   & 0.14 & --- &  --- & ---&--- \\
A1035          & & 11.2 & 0.32  & 1.73   & 0.10 & ---& --- & --- & --- \\
A1045          & & $<$19.2 & 0.28  & 1.55   & 0.14 & ---& --- & --- & ---\\
A1068          & & $<$19.3 & 0.27 & 1.24 & 0.24 & 172.6$_{-19.3}^{+20.1}$ &
0.39$_{-0.08}^{+0.07}$  & $0.50^{+0.07}_{-0.07}$  & $386.8^{+71.3}_{-67.8}$  \\
A1084 $^1$     & & $<$18.6 & 0.30  & 2.16   & 0.04 & 8.0$_{-1.3}^{+0.9}$ & --- & --- &---\\
A1132          & & 11.1 & 0.32  & 1.76   & 0.12 & ---& --- & --- & --- \\
A1173          & & 11.0 & 0.26  & 1.69   & 0.08 & ---& --- & --- & --- \\
A1177          & & 11.2 & 0.28  & 1.80   & 0.08 & --- & ---& --- & --- \\
A1185 &a         & 10.0 & 0.26  & 1.64   & 0.09 & --- & --- & ---& --- \\
      &b         & 10.8 & 0.25  & 1.84   & 0.08 & --- & --- & --- & --- \\
      &c         & 8.9  & 0.26  & 1.83   & 0.06 & --- & --- & --- & --- \\
A1190 &          & 11.0 & 0.26  & 1.63   & 0.05 & --- & --- & --- & ---\\
A1201 &          & 10.4 & 0.29  & 1.80   & 0.14 & --- & --- & --- & ---\\
A1204 $^2$                      &       & $<$22.6 & 0.26  & 1.72   & 0.12
&10.6$_{-1.3}^{+1.8}$ & --- & --- & ---\\
Z3916 $^2$ &       & ---  & 0.24 & 1.44 & 0.04 & 30.2$_{-3.6}^{+1.9}$ & ---& ---&--- \\
A1246    &       & $<$24.5  & 0.28  & 1.83   & 0.13  & ---& --- & --- &--- \\
A1302    &       &  9.7 & 0.34  & 1.78   & 0.10 & --- & ---& --- & --- \\
A1314    &       &  9.7 & 0.33  & 1.76   & 0.08 & --- & ---& --- & --- \\
A1361    &       & $<$16.8 & 0.35 & 1.48   & 0.11 & 13.5$_{-1.9}^{+3.5}$ & --- & --- &---\\
A1366    &       &  9.7 & 0.25  & 1.24   & 0.23 & ---& --- & --- & --- \\
A1413    &       &  9.2 & 0.34  & 1.75   & 0.10 & --- & --- & --- & --- \\
Z4673    &       & $<$19.6 & 0.22  & 1.66   & 0.14 & --- &  --- & --- &--- \\
A1423 $^2$ &       & --- & 0.31 & 2.44 & 0.04 & --- & ---& --- & ---   \\
A1437 &a         & 10.9 & 0.29  & 1.75   & 0.11 & ---& --- & --- & --- \\
      &b         & 10.8 & 0.29  & 1.74   & 0.11 & --- & ---& --- & --- \\
Z4803 &          &  9.6 & 0.29  & 1.86   & 0.07 & --- & ---& --- & --- \\
RXJ1205.1+3920 & & $<$6.1 & 0.16 & 1.88    &  0.07 & ---  & --- & --- &--- \\
RXJ1206.5+2810&  &  9.9 & 0.28  & 1.82   & 0.11 & 0.7$_{-0.1}^{+0.1}$ & --- & --- &---\\
Z4905        &   &  9.6 & 0.24  & 1.56   & 0.10 & ---& --- & --- & --- \\
Z5029        &   &  9.6 & 0.34  & 1.67   & 0.09 & --- & ---& --- & --- \\
RXJ1223.0+1037 & & 10.1 & 0.27  & 1.69   & 0.09 & 0.5$_{-0.2}^{+0.2}$ & --- & --- &---\\
A1553 &a         & 10.3 & 0.27  & 1.98   & 0.10 & ---& --- & --- & --- \\
      &b         & 10.6 & 0.32  & 1.72   & 0.13 & ---& --- & --- & --- \\
RXJ1230.7+1220&  &  9.9 & 0.32 & 1.72   & 0.11 & 0.1$_{-0.1}^{+0.0}$ & --- & --- &---\\
Z5247           & & 10.3 & 0.34 & 1.90 & 0.08 & ---& --- & --- & --- \\
A1589         &  & 10.2 & 0.26  & 1.80   & 0.10 & ---& --- & --- & --- \\
Z5604 $^2$          & & $<$27.7 & 0.23 & 1.26 & 0.04 & ---& ---   &---   & --- \\
A1651         &  & $<$13.0 & 0.39 & 1.94 & 0.06 & --- & ---& --- & ---  \\
A1656         &  & 10.2 & 0.29  & 1.94   & 0.09 & --- & --- & ---& --- \\
A1664         &  & $<$18.0 & 0.29 &  1.24 & 0.37 & 113.8$_{-2.2}^{+2.7}$
&0.46$_{-0.07}^{+0.06}$   & $0.91^{+0.11}_{-0.11}$ &$304.9 ^{+42.6}_{-47.4}$  \\
           &   &           &         &         &            &        & &   &  \\

\end{tabular}
\end{table}

\addtocounter{table}{-1}
\begin{table}
\caption{ Stellar indices and H$\alpha$ luminosities from the $\sim10$~kpc 
galaxy apertures -- continued }
\begin{tabular}{lcrcccrrrr}
           &   &          &         &         &            &        & &   &  \\
Cluster    &   & aperture & Mg$_2$  & 4000\AA & $\delta$BR & L(H$\alpha$) & E(B-V) & revised  & revised L(H$\alpha$) \\
           &   & size     & index   & break   & &(10$^{40}$erg/s) &   & $\delta$BR & (10$^{40}$erg/s)   \\
           &   & (kpc)          &         &         &            &        & &   &  \\
A1668         &  & 10.5 & 0.31 & 1.66   &  0.08 & 2.3$_{-0.4}^{+0.4}$ &  --- & --- & --- \\
A1672 $^2$ &  & $<$24.3 & 0.30  & 1.60   & 0.15 & --- & --- & --- &--- \\
Z5694         &   & $<$24.0     &  0.24 & 2.05      & 0.09 & ---    & ---& --- & --- \\
A1682         & a & $<$27.0 & 0.31  & 1.85   & 0.19 & --- & --- & --- &---\\
              & b & $<$28.2 & 0.33  & 1.50   & 0.14 & --- & --- & --- &---\\
A1703         & a & $<$31.9 & 0.25  & 2.22  & 0.08 & --- & --- & --- & ---\\
\hspace{0.8cm}$^2$ & b & $<$28.2 & 0.13  & 0.85  & 0.15 & --- & --- & --- & ---\\
RXJ1320.1+3308& a &  9.6 & 0.33  & 1.84  & 0.08 & 0.4$_{-0.3} ^{+0.3}$ &--- & --- & ---\\
              & b & 10.8 & 0.21  & 1.80  & 0.10 & --- & --- & --- & ---\\
RXJ1326.3+0013&  & 10.2 & 0.21  & 1.61   & 0.12 & --- & --- & --- & ---\\
A1758    &   & $<$31.6      & 0.26   & 1.78 & 0.07 & --- & --- & --- &--- \\
A1763         &  &  9.7 & 0.34  & 2.06   & 0.09 & --- & --- & --- & ---\\
A1767         &  & 10.3 & 0.29  & 1.76   & 0.08 & --- & --- & ---& ---\\
A1775 &a         &  8.9 & 0.24  & 1.72   & 0.09 & --- & --- & --- & ---\\
      &b         & 10.9 & 0.32  & 1.76   & 0.09 & --- & --- & --- & ---\\
A1773 &          &  9.5 & 0.23  & 1.62   & 0.12 & --- & --- & --- & ---\\
A1795 &          & 10.3 & 0.24  & 1.27   & 0.13 & 11.3$_{-0.3}^{+0.4}$ &0.15$_{-0.15}^{+0.15}$ & $0.18^{+0.06}_{-0.05}$ &$16.1^{+7.0}_{-4.8}$ \\
A1800 &          &  9.4 & 0.26  & 1.68   & 0.10 & --- & --- & --- & --- \\
A1809 &          &  9.9 & 0.30  & 1.81   & 0.10 & --- & --- & --- & --- \\
A1831 &          & 12.2 & 0.28  & 1.75   & 0.11 & --- & --- & --- & --- \\
A1835 &          & 10.4 & 0.10 & 1.16   & 0.33 & 163.9$_{-3.6}^{+2.1}$ &0.40$_{-0.06}^{+0.06}$ &$0.71^{+0.09}_{-0.08}$  &$430.3^{+67.0}_{-58.4}$ \\
      &          & 10.4 & 0.14 & 1.15   & 0.29 & 144.0$_{-2.9}^{+2.9}$ & 0.55$_{-0.12}^{+0.15}$ & $0.93^{+0.31}_{-0.19}$   &$494.9^{+218.7}_{-124.5}$ \\
      &          & $<$29.7 & 0.27 & 1.11  & 0.27 & 318.9$_{-6.0}^{+6.2}$
&0.38$_{-0.04}^{+0.04}$  & $0.54^{+0.09}_{-0.09}$ &$784.4^{+161.2}_{-147.4}$ \\
A1885 $^2$ &          & $<$13.5 & 0.47 & 1.81    & 0.09 &5.4$_{-0.4}^{+0.4}$ & ---  & ---  &---\\ 
Z6718 $^2$&          & $<$11.0 & 0.34  & 1.90   & 0.05 & ---  & ---  & ---&---\\
A1902 $^2$&          & $<$21.5 & 0.15  & 1.46   & 0.23 & ---  & ---  & ---&---\\
A1914 &          & 10.5 & 0.32  & 1.77   & 0.11 & --- & --- & --- &---\\
A1918 &          & 10.1 & 0.28  & 1.78   & 0.09 & --- & --- & --- &---\\
A1927 &          & 10.1 & 0.32  & 1.84   & 0.07 & --- & --- & --- &---\\
A1930 &          & 10.8 & 0.30  & 1.67   & 0.12 & 2.4$_{-0.6}^{+0.6}$ &--- & --- &---\\
RXJ1440.6+0327 &a & 9.9 & 0.28  & 1.80   & 0.08 & --- & --- & --- &---\\
               &b & 9.9 & 0.28  & 1.79   & 0.07 & --- & --- & --- &---\\
RXJ1442.2+2218 & & 10.1 & 0.32  & 1.92   & 0.11 &3.7$_{-0.4}^{+0.4}$ & --- & --- &---\\
RXJ1449.5+2746 & & 10.2 & 0.33  & 1.80   & 0.10 & --- & --- &--- &---\\
A1978          & &  9.4 & 0.32  & 1.78   & 0.09 & --- & --- & --- &---\\
A1983 &a         & 10.0 & 0.26  & 1.50   & 0.12 & --- & --- & --- &---\\
      &b         & 10.4 & 0.24  & 2.02   & 0.06 &  --- & --- & --- &---\\
A1991 &          & 9.9  & 0.27 & 1.62   & 0.10 &1.1$_{-0.3}^{+0.3}$ &---& --- &---\\
Z7160 &          & $<$30.1 & 0.31 & 1.58& 0.10 &50.3$_{-4.7}^{+4.6}$ &
0.24$_{-0.11}^{+0.13}$   & $0.49^{+0.23}_{-0.34}$  & $153.1^{+113.7}_{-85.8}$ \\
A2009 &          & 9.7  & 0.19  & 1.60   & 0.13 & 6.1$_{-1.1}^{+1.2}$ &--- & --- &--- \\
A2034       & a  & $<$16.6 & 0.28  & 2.10 & 0.05 & ---  & ---  &---  &---\\
            & b  & $<$16.1 & 0.26  & 1.69 & 0.08 & ---  & ---  & --- &---\\
A2029 &          & 9.8  & 0.28  & 1.72   & 0.10 & --- & --- & --- &--- \\
A2033 &          & 9.7 & 0.32  & 1.70   & 0.12 & ---  & --- & --- & --- \\
A2050  &         & 9.7 & 0.34  & 2.19   & 0.10 &  --- & --- & --- &--- \\
A2052    &       & 9.4  & 0.26 & 1.64   & 0.10 &2.6$_{-0.4}^{+0.4}$&$0.22^{+0.16}_{-0.21}$ &  $0.15^{+0.07}_{-0.05}$ &$4.8^{+2.1}_{-1.8}$  \\
A2055 &a         & 8.8  & 0.16  & 1.14   & 0.25 &  --- & --- & ---&--- \\
      &b         & 8.8  & 0.25  & 1.82   & 0.09 &  --- & --- & --- &--- \\
      &c         & 10.8 & 0.25  & 1.67   & 0.11 & --- & --- & --- &--- \\
A2064 &          &  9.3 & 0.31 & 1.77   & 0.10 & --- & --- & --- &---  \\
A2061 & a        & 9.5  & 0.26  & 1.69   & 0.11 &  --- & --- & --- &--- \\
      & b        & 9.6  & 0.30  & 1.75   & 0.10 & --- & --- & --- &--- \\
RXJ1522.0+0741&  & 10.2 & 0.28  & 1.75   & 0.08 & --- & --- & --- &--- \\
A2065         & a &  8.8  & 0.28  & 1.83   & 0.07 & --- & --- & --- &--- \\
  & b & 10.2 & 0.24  & 1.79   & 0.10 & --- & --- & --- &--- \\
 & c &  9.4  & 0.33 & 1.77   & 0.08 & --- & --- & --- &--- \\
A2063         &  & 9.9  & 0.25  & 1.75   & 0.08 & --- & --- & --- &--- \\
A2069 &a         & 9.6  & 0.32  & 1.86   & 0.09 & --- & --- & --- &--- \\
      &b         & 9.3  & 0.27  & 1.76   & 0.09 & --- & --- & --- &--- \\
      &c         & 9.6  & 0.27  & 1.88   & 0.07  & --- & --- & --- &--- \\
           &   &           &         &         &            &        & &   &  \\

\end{tabular}
\end{table}

\addtocounter{table}{-1}
\begin{table}
\caption{ Stellar indices and H$\alpha$ luminosities from the $\sim10$~kpc 
galaxy apertures -- continued }
\begin{tabular}{lcrcccrrrr}
           &   &          &         &         &            &        & &   &  \\
Cluster    &   & aperture & Mg$_2$  & 4000\AA & $\delta$BR & L(H$\alpha$) & E(B-V) & revised  & revised L(H$\alpha$) \\
           &   & size     & index   & break   & &(10$^{40}$erg/s) &   & $\delta$BR & (10$^{40}$erg/s)   \\
           &   & (kpc)          &         &         &            &        & &   &  \\
A2072 &          & 10.5 & 0.31 & 1.93   & 0.11 &4.8$_{-1.0}^{+0.9}$ & --- & --- & --- \\
RXJ1532.9+3021 & &  8.6 & 0.14 & 1.18   & 0.32 & 415.5$_{-19.5}^{+20.8}$
&0.21$_{-0.03}^{+0.03}$ & $0.48^{+0.03}_{-0.03}$ & $689.4^{+51.6}_{-50.3}$ \\
A2107          & & 10.4 & 0.29  & 1.79   & 0.08 &  --- & --- & --- &--- \\
A2111 $^2$ &a         &  9.8 & 0.26  & 2.27   & 0.07 & --- & --- & --- &---\\
\hspace{0.9cm}$^2$ &b         &  9.8 & 0.33  & 1.86   & 0.12 & --- & --- & --- &---\\
A2110 &          & 10.1 & 0.28  & 1.74   & 0.10 & --- & --- & --- &---\\
A2104 $^2$&          & $<$21.1 & 0.31  & 1.53   & 0.18 &  --- & --- & --- &---\\
A2124 &          &  9.7 & 0.28  & 1.70   & 0.10 & --- & --- & --- &--- \\
A2146          & & $<$28.3 & 0.30   & 1.38   & 0.14 & 141.3$_{-4.8}^{+4.6}$
& 0.20$_{-0.08}^{+0.07}$  & $0.21^{+0.03}_{-0.03}$  & $185.8^{+34.4}_{-32.7}$   \\
A2142 &a         & 10.9 & 0.32 & 1.80   & 0.13 & --- &--- & --- & --- \\
      &b         & 9.9  & 0.35  & 1.68   & 0.11 & --- & --- & --- & ---\\
A2147 &          & 10.3 & 0.30  & 1.76   & 0.09 & --- & --- & --- & ---\\
A2151a &          & 8.7  & 0.32  & 1.88   & 0.09 & --- & --- & --- & ---\\
RXJ1604.9+2356& & 9.4  & 0.31  & 1.77  & 0.08 & --- & --- & --- & ---\\
Z7833          & & $<$26.5 & 0.25  & 1.79  & 0.11 & --- & --- & --- & ---\\
A2175          & & 10.0 & 0.33  & 1.75   & 0.09 & --- & --- & --- & ---\\
A2187 $^2$          & & $<$23.9 & 0.23 &1.91  & 0.10 & --- & --- & --- & ---\\
A2201 $^2$   & & --- & 0.17 & 1.95 &0.05 & --- & --- & --- & ---\\%
A2199          & & 10.2 & 0.29  & 1.73  & 0.08 & 2.7$_{-0.3}^{+0.2}$
&0.10$_{-0.10}^{+0.09}$ & $0.10^{+0.01}_{-0.01}$   & $3.5^{+0.8}_{-0.8}$ \\
A2208          & & $<$18.6 & 0.28  & 1.73  & 0.11 &  --- & --- & --- &---\\
A2204          & &  9.6  & 0.23  &  1.08 & 0.31 &159.4$_{-3.4}^{+4.1}$  &--- & --- & ---\\
A2219 $^2$       & a & $<$27.5 & 0.43  & 2.22 & 0.13 & --- & --- & --- &---\\
\hspace{0.8cm} $^2$       & b & $<$27.5 & 0.34  & 1.44 & 0.04 & --- & --- & --- &---\\
\hspace{0.8cm}      $^2$       & c & $<$28.3  & 0.22  & 1.56 & 0.14 &--- & --- & --- &---\\
A2228 $^2$         & & $<$14.9 & 0.46  & 1.89 & 0.08 & --- & --- & --- &---\\
 RXJ1657.8+2751 &    & 9.0 & 0.28 & 2.07 & 0.14 & --- & --- & --- &---\\
A2244          & & 10.3 & 0.27  & 1.68   & 0.10 & --- & --- & --- &---\\
A2256 &a         & 9.1  & 0.28  & 1.67   & 0.09 & --- & --- & --- &---\\
      &b         & 9.4  & 0.30  & 1.94   & 0.06 & --- & --- & --- &---\\
      &c         & 8.8  & 0.27  & 1.84   & 0.03 & --- & --- & --- &---\\
      &d         & 9.9  & 0.27  & 1.79 & 0.08 & --- & --- & --- &---\\
A2249 &          & 9.2  & 0.28  & 1.69   & 0.10 & --- & --- & --- &---\\
A2255 &a         & 10.5 & 0.33  & 1.71   & 0.10 & --- & --- & --- &---\\
      &b         & 9.3  & 0.25  & 1.74   & 0.08 & --- & --- & --- &---\\
RXJ1715.3+5725 & & 9.3  & 0.27  & 1.65   & 0.11 & 1.2$_{-0.2}^{+0.2}$ & ---&--- &---\\
Z8193          & & $<$23.1 &0.18  & 1.49   & 0.19 & 153.1$_{-5.8}^{+4.5}$ & 0.57$^{+0.13}_{-0.15}$& $0.56^{+0.16}_{-0.14}$ &$605.8^{+220.0}_{-42.3}$\\
A2254 $^2$                  & & ---  & 0.36 & 2.05 & 0.04& --- & ---& --- & ---\\
Z8197         &  & 9.6  & 0.26   & 1.42   & 0.13 &16.0$_{-0.8}^{+0.8}$ &0.33$_{-0.10}^{+0.12}$& $0.25^{+0.06}_{-0.04}$ & $35.1^{+11.6}_{-7.4}$ \\
A2259  $^2$                & & ---  & 0.24  & 2.15  & 0.07 & --- & --- & --- &---\\
RXJ1720.1+2638&  & 8.8  & 0.31 & 1.25   & 0.14 & [12.7$_{-3.2}^{+2.7}$] & --- & --- &---\\
A2261 $^{2}$                  & & ---  & 0.35 & 2.48 & 0.04 & --- & --- &--- &---\\
A2294 $^{2}$                  & & ---  & --0.02 & 1.56 & --0.03 &30.2$_{-2.3}^{+2.3}$ & --- & --- &---\\
RXJ1733.0+4345 & & 9.6  & 0.29 & 1.78   & 0.09 &0.2$_{-0.1}^{+0.1}$ & --- & --- &---\\
RXJ1740.5+3539 &a & 10.3 & 0.26  & 1.62   & 0.07 &  --- & --- & --- &---\\
               &b & 10.2 & 0.30  & 1.84   & 0.08 & --- & --- & --- &---\\
Z8276          & & 9.4  & 0.28  & 1.72   & 0.10 &12.5$_{-0.7}^{+0.8}$&0.24$_{-0.10}^{+0.09}$ & $0.16^{+0.02}_{-0.03}$ & $22.0^{+5.2}_{-4.7}$ \\
Z8338          & & 9.8  & 0.32 & 2.12 & 0.12 &  --- & --- & --- &---\\
RXJ1750.2+3505 & & 10.6 & 0.29  & 1.84   & 0.11 & 15.0$_{-3.8}^{+3.7}$$\ast$ & --- & --- &---\\
A2292 $^2$   & & --- &0.09        & 2.10   & 0.06 & 3.0$_{-0.9}^{+0.9}$ & --- & --- &---\\
Z8451   &      & $<$13.3 & 0.41  & 1.61 & 0.14 & --- & --- & --- &---\\%
RXJ2114.1+0234 & & 10.2 & 0.30   & 1.90   & 0.10 & --- & --- & --- &---\\
Z8503          & & $<$19.7 & 0.25   & 2.55   & 0.13 & --- & --- & --- &---\\
RXJ2129.6+0005 $^2$ & & 9.9  & 0.24   & 1.66   &  0.13 & 5.8$_{-1.9}^{+2.0}$ &---& --- & ---\\
A2390    &     & 9.9  & 0.31 & 1.35  & 0.19 & 61.6$_{-3.3}^{+3.0}$&0.22$^{+0.10}_{-0.11}$ & $0.29^{+0.06}_{-0.06}$ & $104.7^{+28.7}_{-23.78}$\\
A2409 $^2$   &     &  $<$20.2 & 0.25 & 1.60  & 0.14 & --- & --- & --- &---\\
A2426    &     &  $<$14.6 & 0.29 &  1.21 & 0.12 & --- & --- & --- &---\\
A2428    &     &  $<$12.8 & 0.35 &  1.22 & 0.16 & --- & --- & --- &---\\
A2443    &     & 11.2 & 0.30 & 1.97  & 0.09 & --- & --- & --- &---\\
A2457    &     &  9.8 & 0.31 & 2.04  & 0.06 & --- & --- & --- &---\\

           &   &           &         &         &            &        & &   &  \\

\end{tabular}
\end{table}

\addtocounter{table}{-1}
\begin{table}
\caption{ Stellar indices and H$\alpha$ luminosities from the $\sim10$~kpc 
galaxy apertures -- continued }
\begin{tabular}{lcrcccrrrr}
           &   &          &         &         &            &        & &   &  \\
Cluster    &   & aperture & Mg$_2$  & 4000\AA & $\delta$BR & L(H$\alpha$) & E(B-V) & revised  & revised L(H$\alpha$) \\
           &   & size     & index   & break   & &(10$^{40}$erg/s) &   & $\delta$BR & (10$^{40}$erg/s)   \\
           &   & (kpc)          &         &         &            &        & &   &  \\
A2495    &     & 10.1 & 0.17 & 1.65 & 0.12 & 2.0$_{-0.5}^{+0.5}$ & --- & --- & ---\\
Z8852    &     &  9.9 & 0.30 & 1.92   & 0.09  & --- & --- & --- &---\\
A2572a   & a    & 9.1  & 0.33  & 2.04   & 0.06& --- & --- & --- &---\\
         & b    & 9.9  & 0.33  & 2.00   & 0.07 & --- & --- & --- &---\\
A2572b   &     & 9.2  & 0.28  & 1.86   & 0.09 & --- & --- & --- &---\\
A2589    &     & 9.1  & 0.31  & 1.89   & 0.09 & --- & --- & --- &---\\
A2593    &     & 9.6  & 0.31  & 1.91   & 0.09 & --- & --- & --- &---\\
A2626 & a      & 10.3 & 0.27  & 1.87   & 0.08 & --- & --- & --- &---\\
      & b      & 10.2 & 0.29  & 1.95   & 0.08 & 1.1$_{-0.4}^{+0.4}$ &--- & --- & ---\\
A2627  & a  & $<$18.0 & 0.14 & 1.22   & 0.27 & ---  &---   & --- &---\\
           & b  & $<$17.4 & 0.29 & 2.28   & 0.07 & ---  &---   & --- &---\\
A2631      &    & $<$31.5 & 0.24 & 2.12   & 0.10 & ---  & ---  & --- &---\\
A2634  &       &  9.8 & 0.32  & 1.85   & 0.09 & 1.3$_{-0.6}^{+0.4}$ & --- &--- & ---\\
A2657  &       &  9.9 & 0.27  & 1.83   & 0.10 & --- & --- & --- &---\\
A2665  &       & 10.5 & 0.29  & 1.82   & 0.09 & [0.6$_{-0.3}^{+0.3}$] &--- & --- & ---\\
A2675  &       & 10.7 & 0.25  & 1.91   & 0.09 & --- & --- & --- &---\\

           &   &           &         &         &            &        & &   &  \\

\end{tabular}

\end{table}
\begin{table}

Notes:\\
1. FOS spectrum with very little flux below 4000\AA, suggesting
that the slit was not at the parallactic angle (see also spectra in Figs 3
and 4). \\
2. The spectrum is very noisy. \\
$\ast$ The \Ha\ affected by a cosmic ray hit.\\

The FOS spectra in A92 and C95  were reduced using an optimal
extraction technique, so the precise aperture from which the spectrum is taken
is not known for each galaxy; we give the upper limit from the slit length
of 6 arcsec in column 2. The aperture used for extracting the
spectra in C95 taken with the ISIS on the WHT 
is not known. \\ 
The 4000\AA\ break in column 4 is measured from an $f_{\lambda}$ spectrum
rather than $f_{\nu}$. Bracketed values mark where an abnormal value is 
due to loss of blue light from the slit. \\ 
The \Ha line luminosity in column 6  is 
measured for the spectrum taken from the given aperture and does {\sl 
not} represent the total
line luminosity from the galaxy. A bracketed line luminosity shows 
where the line flux may have been reduced by atmospheric absorption. \\
The E(B-V) in column 7 has been estimated from the
H$\alpha$/H$\beta$ ratio in the spectrum from the given aperture, and {\it
not} from the total line fluxes given in Tables~\ref{tab:redlines} and
 \ref{tab:bluelines}. \\ 
The \lq revised'
$\delta$BR in column 8, and the \lq revised' \Ha luminosity in column 9 
are after the spectrum has been corrected for internal
reddening by the amount shown in column 7. \\

{\bf Notes on individual galaxies:} \\

A1437 There is overlap between the 10kpc apertures taken for each
     component of the dumbbell galaxy \\
RXJ1206.5+2810 The centre of the aperture was taken as the midway between
the two galaxy components. \\
RXJ1750.2+3505 The \Ha emission is badly affected by a cosmic ray hit. \\
A2626 The 10kpc aperture for each galaxy overlaps.\\

\end{table}

\twocolumn

\subsection{Emission-line fitting}
\label{sec:emilines}

We estimate the spatial diameter of the line emission along the slit by
extracting a profile of the galaxy centred on the observed [NII] and \Ha\
emission, and subtracting the profile of the galaxy in nearby line-free
continuum (\eg 6400--6500\AA). The nebulae often extend well beyond the
$\sim10$\kpc\ aperture used to find the stellar indices in
section~\ref{sec:contmSI}, so in order to measure the {\em total} line
luminosities from the galaxy in the slit, we have extracted a spectrum from
the total spatial region spanned by the emission lines. We cannot rule out
the possibility that the nebulae could also be extended in a direction perpendicular to the
slit position, so these line fluxes may sometimes give underestimates of the
{\sl total} line luminosity of the galaxy.

We fit the emission lines in close sets, such as the
\Ha+\niib+[SII]$\lambda\lambda$6717,6731+[OI]$\lambda\lambda$6300,6363 complex.
Within such a set, the individual emission lines are modelled as gaussians
fixed to be at the same redshift and velocity width as each other, and the
underlying continuum is modelled by a linear function. The \Ha+\niib\ blend
is sufficiently resolved to give good fits for individual line components
(eg Z3146, RXJ0439.0+0520, Z8197 in Fig~\ref{fig:emspectra}; although note
that many other spectra in this figure are shown after smoothing and belie
the ease of resolving the lines).
The resulting fluxes for the red emission lines of galaxies are given in
Table~\ref{tab:redlines}; lines that are partially affected, or even
completely absorbed by the Earth's atmosphere are marked by square brackets.
The errors given on the fluxes are estimated from the $\Delta\chi^2=1$
confidence
limits to the gaussian intensity in each line,
 assuming the fit to the (unweighted) spectrum has a reduced  $\chi^2=1$.
Thus these errors do not
take into account any systematics introduced by the data reduction. $2\sigma$ upper limits
are estimated for lines not detected, assuming them to have the same
redshift and velocity width as the rest of the complex. We also assume that
any \Ha in absorption due to the underlying stellar continuum is negligible
in comparison to the \Ha visible in emission (this point is assessed further
in section~\ref{sec:reddening}).

We are unable to make such an assumption about stellar \Hb absorption in
the underlying continuum, however, as it is expected to be stronger relative
to the \Hb emission than is the case for \Han. In order to take the effects
of stellar \Hb absorption into account, we created a template galaxy from
the averaged rest-frame spectra of 24 galaxies in our sample. These galaxies
were chosen to have good signal-to-noise, no line emission, and \lq average'
stellar indices -- ie Mg$_2$ in the range 0.25--0.32 and D$_{4000}$ between
1.65--1.95; the final template had an Mg$_2$ index of 0.29 and D$_{4000}$ of
1.9. The template was then normalized to each line-emitting central cluster
galaxy at the region 5050--5100\AA\ (in the rest frame; this comprises the
flat left-hand shoulder of the Mg absorption feature) and subtracted to give
a residual spectrum. The [OIII]$\lambda\lambda$4959,5007, \Hb and
[NI]$\lambda$5199\ emission lines were then fit together in the residual
spectrum, again with each line modelled by a gaussian of the same redshift
and velocity width as each other. The fluxes of these blue emission lines
are tabulated in Table~\ref{tab:bluelines}, along with the flux of the \otw\
doublet which was also fit by a gaussian (a satisfactory fit at this
spectral resolution).  In Table~\ref{tab:bluelines} we list first the fluxes of \Hbn,
[OIII] and [NI] from the fits to the {\em non}-template-subtracted spectrum,
and second in each column we list the fits from the residual spectrum
in square brackets. If no template-subtracted result is presented, it is no
improvement on the previous fit. The correction for the underlying continuum
is not very significant for all the strong line emitters (eg Z3146, A1835).
Although one might expect the depths of the Mg$_2$ and \Hb absorption
features to be well correlated for an old stellar spectrum, this might not
be the case for galaxies containing a blue light component. Galaxies with
the strong line emission are in fact those that also contain the excess blue
continuum (A92, C95,  Allen 1995), for which the template subtraction
given here may not be the most appropriate. However, it is clear that the
line emission in these galaxies is so strong that errors due to
inaccurate stellar \Hb absorption are minimal. Where the template
subtraction is in danger of being most in error is if the galaxy contains a
fading starburst with Balmer absorption lines at a maximum. 
 Whilst this may not be a principal component of the most
line-luminous systems, it may be relevant for lower-luminosity systems.

Most of the galaxies newly observed for this paper are either non-emitters
or show lower-luminosity line emission than those in A92 and C95; only
three of the new systems show a slit \Ha luminosity above
$3\times10^{41}$\ergps (RXJ1532.9+3021, A2204 and A2390).  As well as being
the second highest redshift cluster in the sample, RXJ1532.9+3021 contains an
exceptionally line-luminous BCG, second only to that in Z3146.

\begin{figure}
\psfig{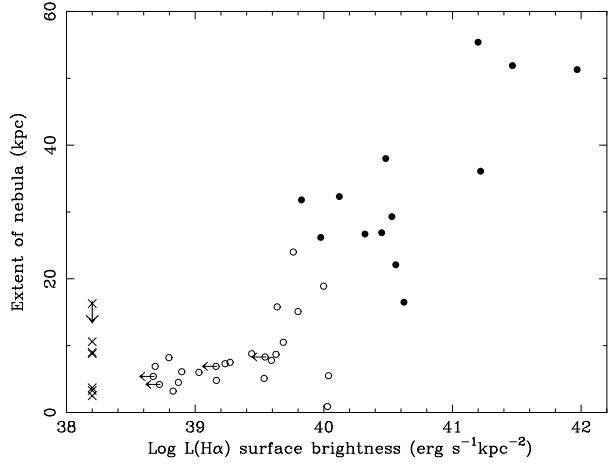}
\caption{
\label{fig:lhavsd}
\Ha surface brightness  plotted against the diameter of the nebula
for all galaxies with spatially resolved spectra.
High \Ha luminosity systems (L(\Han)$>10^{41}$\ergps) are shown by solid
circles, and the lower
luminosity line emitters  by open circles. The crosses shown at an arbitrary 
value of \Ha surface brightness of $10^{38.2}$\ergpspkpcsq show the diameters of the [NII]-{\em only}
emitters. }
\end{figure}

The luminosity of the \Ha line emission correlates with the size of the
nebula (for those galaxies with spatially resolved spectra only), in the
sense that the more luminous systems are larger. Fig~\ref{fig:lhavsd} shows
a plot of the observed \Ha surface brightness against the diameter of the
emission-line nebula. The systems at higher luminosity are clearly larger
(mean diameter of 34\kpc, compared to 8\kpc\ for the lower-luminosity
nebulae). These numbers are not precise, as the diameters are measured from 
slit positions placed at random over what may be an asymmetric nebula; this
may be responsible for some of the scatter in the plot.  Few of
the lower-luminosity systems are any larger than those showing only [NII]
line emission.

We find no
correlation between the velocity width (taken from the fit to the \Ha+[NII]
emission line complex and corrected for the instrumental resolution) and the
\Ha luminosity  for all the emission-line galaxies.

\onecolumn 

\begin{table}
\caption{Red emission line fluxes (in units of
$10^{-15}\ergpspcmsq$).\label{tab:redlines}}
\begin{tabular}{lrrrrrrrr}

& & & & & & & & \\
Cluster        & log L(H$\alpha$)  & Size (kpc) & H$\alpha$              & [NII]$\lambda$6584      &
[SII]$\lambda$6717     & [SII]$\lambda6731$      & [OI]$\lambda$6300 & [OI]$\lambda$6363 \\
& & & & & & & &\\

Z235 & 40.61 & --- & 1.3$^{+0.2}_{-0.2}$ & 2.3$^{+0.2}_{-0.2}$ & 0.9$^{+0.3}_{-0.3}$ & 0.6$^{+0.3}_{-0.3}$ & $<$0.1 & $<$0.1 \\

A115 & 41.33 & 38.0 & 1.2$_{-0.2}^{+0.2}$ & 1.0$_{-0.2}^{+0.2}$ & 0.2$_{-0.2}^{+0.2}$ & 0.6$_{-0.2}^{+0.2}$ & $<$0.1 & [---] \\

RXJ0107.4+3227 & 39.72 & 4.5 & 3.9$_{-0.9}^{+0.9}$ & 7.9$_{-1.0}^{+0.4}$ & 1.2$_{-0.9}^{+1.1}$ & [1.4$_{-1.2}^{+0.6}$] & $<$0.1 & $<$0.2 \\

RXJ0123.2+3327 & --- & 3.3 & $<$0 & 7.0$_{-1.6}^{+2.0}$ & --- & --- & --- & --- \\

A262           & 39.74  & 6.9 &  4.6$_{-0.4}^{+0.4}$ & 13.3$_{-0.5}^{+0.5}$${\ast}$ & 4.2$_{-0.5}^{+0.5}$ & [2.9$_{-0.4}^{+0.5}$] &
0.3$_{-0.3}^{+0.4}$ & $<$0.1 \\

A291 & 41.66 & --- & 2.5$_{-0.1}^{+0.1}$ & 2.0$_{-0.1}^{+0.1}$ &
1.5$_{-0.1}^{+0.1}$ & 
0.7$_{-0.1}^{+0.1}$ & 0.6$_{-0.1}^{+0.1}$  & 0.2$_{-0.1}^{+0.1}$   \\

A400 a  & --- & 3.7 & $<$0  &  7.4$_{-1.6}^{+1.8}$ & [---]             & [---]                   & ---                        & --- \\

A400 b         & --- & 2.5 & $<$0                    &  3.5$_{-0.9}^{+1.3}$  & 
 [---]             & [---]                   & ---                     & --- \\

A401 & --- & 10.6 & $<$0 & 0.9$_{-0.3}^{+0.3}$ & --- & --- & --- & --- \\

Z808 &  40.67 & --- & 0.4$^{+0.1}_{-0.4}$ & 0.8$^{+0.2}_{-0.1}$ &
0.2$^{+0.1}_{-0.1}$  & $<$0.1 & $<$0.1 & $<$0.1 \\
 
RXJ0338.7+0958 & 41.25 & 
31.8 &  35.3$_{-1.2}^{+1.3}$ & 42.7$_{-1.3}^{+1.3}$ & 
14.8$_{-1.3}^{+1.3}$ & 11.2$_{-1.4}^{+1.3}$ & 9.2$_{-1.2}^{+1.2}$ & 4.6$_{-1.2}^{+1.2}$ \\

RXJ0352.9+1941 &  41.77 &--- & 10.7$^{+0.2}_{-0.3}$ &
9.8$^{+0.3}_{-0.2}$ & 5.8$^{+0.3}_{-0.3}$ & 3.9$^{+0.3}_{-0.2}$ &
2.5$^{+0.2}_{-0.2}$ & 0.9$^{+0.2}_{-0.2}$ \\

A478 & 41.03 & --- & 3.3$^{+0.2}_{-0.2}$ & 3.7$^{+0.2}_{-0.2}$ &
1.5$^{+0.2}_{-0.2}$ & 1.1$^{+0.2}_{-0.2}$ & 0.9$^{+0.2}_{-0.2}$ & 0.5$^{+0.2}_{-0.2}$ \\

RXJ0439.0+0520 & 42.04 &---& 5.4$^{+0.1}_{-0.2}$ & 8.1$^{+0.2}_{-0.1}$
& 3.1$^{+0.2}_{-0.2}$ & 1.9$^{+0.2}_{-0.1}$ & 0.9$^{+0.1}_{-0.1}$ &0.4$^{+0.1}_{-0.1}$ \\ 

Z1121 &  40.16 & --- &   0.5$_{-0.1}^{+0.1}$ & 0.7$_{-0.1}^{+0.1}$ & $<$0.1 & $<$0.1 & $<$0.1 & $<$0.1 \\

RXJ0751.3+5012 b & 39.81 & 8.2 & 2.7$_{-1.6}^{+0.9}$ &
8.9$_{-1.1}^{+1.2}$ & [---] & [---] & $<$0.4 & $<$0.2 \\

A611 & --- & --- & $<$0 &  0.4$_{-0.1}^{+ 0.2}$ & $<$0.1 & $<$0.1 & $<$0.1 & $<$0.1
\\ 

RXJ0821.0$+$0752 & 41.48 &---&   5.5$_{-0.2}^{+0.2}$  & 2.7$_{-0.1}^{+0.2}$ &
1.1$_{-0.2}^{+0.2}$ &   0.5$_{-0.2}^{+0.2}$ & $<$0.2 & $<$0.1 \\ 

A646           & 41.41 & 29.3 & 3.5$_{-0.2}^{+0.2}$ & 3.8$_{-0.2}^{+0.2}$ &
1.5$_{-0.2}^{+0.2}$ & [0.5$_{-0.2}^{+0.2}$] & 0.9$_{-0.2}^{+0.2}$ &
0.4$_{-0.2}^{+0.2}$ \\

Z1665          & $<$39.36 & 5.4 & $<$0.5 & 3.1$_{-0.5}^{+0.4}$  & [---] & --- & --- & --- \\

A671           & --- & --- & [$<$0]                   & [3.0$_{-0.8}^{+0.9}$]    & --- & --- & --- & --- \\

A750          & ---  &  $<$16.3 & $<$0                      & 0.4$_{-0.1}^{+0.1}$ & --- & --- & --- & --- \\

A795           & 41.29 & 22.1 & 2.3$_{-0.3}^{+0.2}$ & 3.8$_{-0.3}^{+0.3}$    &
[---] & [---] &  1.1$_{-0.2}^{+0.2}$ & 0.3$_{-0.2}^{+0.2}$ \\

Z2701 &  40.94 & --- & 0.4$^{+0.1}_{-0.1}$ & 0.8$^{+0.2}_{-0.1}$ &
0.3$^{+0.2}_{-0.1}$ & 0.2$^{+0.2}_{-0.2}$ & $<$0.2 & $<$0.2 \\

RXJ1000.5+4409 a & 40.35 & --- &   0.2$_{-0.1}^{+0.1}$  &
0.5$_{-0.1}^{+0.1}$  & 0.2$_{-0.1}^{+0.1}$ & 
0.1$_{-0.1}^{+0.1}$ & 0.1$_{-0.1}^{+0.1}$    & $<$0.1 \\

RXJ1000.5+4409 b &  40.34 & --- & 0.2$_{-0.1}^{+0.1}$ &  0.3$_{-0.1}^{+0.1}$  & $<$0.1 & $<$0.1 &  $<$0.1 & $<$0.1 \\

Z3146 &  42.85 & --- & 17.1$^{+0.4}_{-0.3}$ & 13.5$^{+0.3}_{-0.3}$ & 6.5$^{+0.4}_{-0.3}$ & 5.3$^{+0.4}_{-0.4}$ & 3.5$^{+0.3}_{-0.3}$ & 0.9$^{+0.3}_{-0.3}$ \\

Z3179  & 40.92 &--- &  0.9$^{+0.2}_{-0.4}$ & 3.2$^{+0.4}_{-0.3}$ &
0.6$^{+0.3}_{-0.3}$ & 0.8$^{+0.3}_{-0.2}$ & 0.5$^{+0.2}_{-0.2}$ &
$<$0.1 \\

A1023 & --- & --- & $<$0 &  1.0$_{ -0.3}^{+ 0.3}$ & --- & --- & --- & --- \\

A1068 & 42.24 & --- & 19.5$^{+0.6}_{-0.4}$ & 24.6$^{+0.3}_{-0.6}$ &
5.2$^{+0.6}_{-0.4}$ & 4.4$^{+0.4}_{-0.6}$ & 2.8$^{+0.4}_{-0.4}$ &
1.2$^{+0.4}_{-0.4}$ \\

A1084 &  40.91 & --- &  1.0$_{-0.2}^{+0.2}$        &
1.2$_{-0.2}^{+0.2}$ &  0.6$_{-0.3}^{+0.3}$ &  0.1$_{-0.1}^{+0.3}$ &  0.4$_{-0.1}^{+
0.1}$    & 0.1$_{-0.1}^{+   0.1}$    \\

A1204 & 41.03 & --- &0.8$^{+0.1}_{-0.1}$  & 1.1$^{+0.1}_{-0.1}$ &
0.3$^{+0.1}_{-0.1}$ & 0.3$^{+0.1}_{-0.1}$ & 0.3$^{+0.1}_{-0.1}$ &
0.2$^{+0.1}_{-0.1}$ \\%

Z3916 &  41.48 & --- & 1.5$_{-0.1}^{+0.1 }$  & 1.7$_{-0.1}^{+0.1}$ & 1.1$_{-0.1 }^{+ 0.1}$  &    0.5$_{-0.1}^{+0.1}$    & [---] & [---] \\ 

A1361 & 41.13  & --- & 2.2$^{+0.2}_{-0.1}$ & 3.6$^{+0.2}_{-0.1}$$\ast$ &
1.4$^{+0.1}_{-0.2}$ & 0.6$^{+0.2}_{-0.1}$ & 0.4$^{+0.1}_{-0.1}$ &
0.4$^{+0.1}_{-0.1}$ \\



RXJ1206.5+2810 & 39.80 
&6.0  & 1.8$_{-0.3}^{+0.3}$ & 3.4$_{-0.3}^{+0.3}$
& [---] & [---] & $<$0.1 & $<$0.1 \\

Z4905          & $<$39.60 & 6.9 & $<$0.2                  &
0.9$_{-0.2}^{+0.2}$  & --- & [---] & --- & --- \\

RXJ1223.0+1037 & 39.71 
&6.1  & 1.8$_{-0.3}^{+0.3}$ &
4.4$_{-0.3}^{+0.3}$  & [---] & [---] & $0.1_{-0.1}^{+0.3}$ & $<$0.1 \\

RXJ1230.7+1220 & 39.11 
& 0.9 & 23.2$_{-9.2}^{+7.8}$ & 142.2$_{-8.6}^{+10.1}$ & 20.3$_{-8.3}^{+7.8} $  
& 38.3$_{-7.9}^{+8.9}$ & [$<$1.6] & $<$1.9 \\

A1664 & 42.06 & --- &  15.2$_{-0.3}^{+0.4}$   & 10.5$_{-0.3}^{+
0.3}$   & 5.7$_{-0.3}^{+  0.3 }$   & 
4.7$_{-0.3}^{+0.3 }$  & 2.4$_{-0.2}^{+0.2}$   & 0.8$_{-0.2}^{+
0.2 }$ \\

A1668          & 40.37 &10.5 & 
 1.3$_{-0.2}^{+0.2}$ & 3.9$_{-0.2}^{+0.2}$ &
1.2$_{-0.2}^{+0.2}$ & 0.5$_{-0.2}^{+0.2}$ & 0.6$_{-0.2}^{+0.2}$ &
0.1$_{-0.2}^{+0.2}$ \\


RXJ1320.1+3308 a & 39.74  & 4.8 & 1.0$_{-0.4}^{+0.4}$ & 3.0$_{-0.4}^{+0.4}$
& [$<$0.3] & 0.9$_{-0.4}^{+0.4}$ & $<$0.2 & $<$0.1 \\

A1767          & --- & 9.0 & $<$0                                     & 1.8$_{-0.4}^{+0.5}$ & [---] & [---] & --- & ---  \\

A1795          & 41.07 & 26.2 
& 6.9$_{-0.2}^{+0.2}$ & 7.9$_{-0.2}^{+0.2}$ & 4.0$_{-0.2}^{+0.3}$ &
2.9$_{-0.3}^{+0.3}$ & 1.9$_{-0.2}^{+0.2}$ &  0.7$_{-0.2}^{+0.2}$ \\

A1835          & 42.43 &48.5  
& 8.8$_{-0.2}^{+0.2}$ & 6.1$_{-0.2}^{+0.2}$ & 2.7$_{-0.2}^{+0.2}$ &
2.4$_{-0.2}^{+0.2}$ & 2.1$_{-0.2}^{+0.2}$& 0.6$_{-0.2}^{+0.2}$  \\

& 42.37 & 51.9 
& 7.7$_{-0.2}^{+0.2}$ & 5.1$_{-0.2}^{+0.2}$ & 2.7$_{-0.2}^{+0.2}$ &
2.2$_{-0.2}^{+0.2}$ & 1.8$_{-0.2}^{+0.2}$ & 0.3$_{-0.2}^{+0.2}$ \\

& 42.50 & --- & 10.4$_{-0.2}^{+0.2 }$& 7.7$_{-0.2 }^{+0.2}$  & 3.4$_{-0.2}^{+0.2}$   & 2.2$_{-0.2}^{+0.2}$ & 1.9$_{-0.2}^{+0.2}$   & 0.7$_{-0.18}^{+0.18 }$ \\

A1885 &  40.73 & --- &   1.5$_{-0.1}^{+0.1}$  &  1.7$_{-0.1}^{+0.1}$ & 0.8$_{-0.1}^{+0.1}$   &  0.3$_{-0.1}^{+0.1 }$ &   0.5$_{-0.1}^{+0.1}$ &  0.1$_{-0.1}^{+0.1}$  \\   

A1930          & 40.38 & 15.1 
& 0.3$_{-0.1}^{+0.1}$ & 0.8$_{-0.1}^{+0.1}$ & [---]
& [---] & 0.1$_{-0.1}^{+0.1}$ & $<$0.1 \\

RXJ1442.2+2218 & 40.65 & 24.0 & 1.1$_{-0.2}^{+0.2}$ & 2.7$_{-0.2}^{+0.2}$ &
0.2$_{-0.2}^{+0.2}$ & 1.1$_{-0.2}^{+0.2}$ & [---] &
0.1$_{-0.2}^{+0.2}$ \\

RXJ1449.5+2746 & $<$39.30 & 4.2 & $<$0.5                &
5.2$_{-1.1}^{+1.3}$ & [---] & --- & --- & --- \\

A1991          & 40.08 & 8.8 
& 0.8$_{-0.3}^{+0.2}$ &  2.1$_{-0.3}^{+0.2}$ &
1.1$_{-0.3}^{+0.3}$ & 0.7$_{-0.3}^{+0.3}$ &
[0.4$_{-0.2}^{+0.3}$] & 0.4$_{-0.2}^{+0.3}$  \\

Z7160        &  41.70 & --- & 1.6$_{-0.1}^{+0.2}$ & 1.4$_{-0.1}^{+0.1}$  & 0.8$_{-0.1}^{+0.2 }$   &   1.0$_{-0.2}^{+0.1}$ & 0.6$_{-0.1}^{+0.1}$   & 0.3$_{-0.1}^{+0.1}$  \\

A2009          & 41.09 & 26.7
&1.1$_{-0.2}^{+0.2}$ &
[1.6$_{-0.2}^{+0.2}$] & 1.0$_{-0.2}^{+0.3}$ &
0.3$_{-0.3}^{+0.2}$ & 0.2$_{-0.3}^{+0.3}$ &  $<$0.1 \\

A2033          & $<$40.05 & 8.3 & $<$0.4 &  1.5$_{-0.2}^{+0.2}$ & --- & --- & --- & --- \\
\end{tabular}
\end{table}
\vfill\eject

\addtocounter{table}{-1}
\begin{table}
\caption{Red emission line fluxes -- continued }
\begin{tabular}{lrrrrrrrr}
& & & & & & & & \\
Cluster        & log L(H$\alpha$)  & Size (kpc) & H$\alpha$              & [NII]$\lambda$6584      &
[SII]$\lambda$6717     & [SII]$\lambda6731$      & [OI]$\lambda$6300 & [OI]$\lambda$6363 \\
& & & & & & & &\\

A2052          & 40.47 & 8.7 
& 5.5$_{-0.5}^{+0.4}$ & 12.3$_{-0.5}^{+0.5}$ &
3.9$_{-0.5}^{+0.4}$ & 2.6$_{-0.4}^{+0.5}$ & 1.6$_{-0.4}^{+0.4}$ & 0.1$_{-0.4}^{+0.4}$  \\

A2055   a       & --- & 8.8 & $<$0 & 1.1$_{-0.2}^{+0.2}$ & --- & --- & --- & ---  \\

A2072          & 40.68 & 18.9 
& 0.7$_{-0.1}^{+0.1}$ &  1.3$_{-0.1}^{+0.1}$ &
0.4$_{-0.1}^{+0.1}$ & [---] & 0.2$_{-0.1}^{+0.1}$ & $<$0.1 \\

RXJ1532.9+3021 & 42.78 & 51.3 
&  9.2$_{-0.3}^{+0.5}$ &  6.5$_{-0.3}^{+0.4}$ & 3.7$_{-0.4}^{+0.4}$ &
2.9$_{-0.4}^{+0.4}$ & 1.9$_{-0.4}^{+0.4}$ & 0.6$_{-0.4}^{+0.4}$  \\


 A2146 &  42.15 & --- &  5.4$_{-0.2}^{+0.2 }$  & 9.2$_{-0.2}^{+  0.2}$ & 
2.2$_{-0.2}^{+ 0.2 }$  &  2.8$_{-0.2}^{+0.2 }$ & 0.9$_{-0.1}^{+0.1}$ & 0.4$_{-0.1}^{+0.1}$   \\

A2199          & 40.44 & 7.8 & 
  6.5$_{-0.6}^{+0.4}$ & 14.9$_{-0.5}^{+0.7}$ &
[3.1$_{-0.6}^{+0.5}$] & 2.2$_{-0.5}^{+0.6}$ & 0.9$_{-0.4}^{+0.4}$ & 0.2$_{-0.4}^{+0.4}$ \\

A2204          & 42.29 & 55.4 
& 18.5$_{-1.3}^{+1.5}$ & [23.3$_{-1.4}^{+1.4}$] & 8.7$_{-1.4}^{+1.6}$  &
9.6$_{-1.6}^{+1.5}$ & 5.0$_{-1.3}^{+1.4}$ &  2.3$_{-1.3}^{+1.4}$  \\

RXJ1715.3+5725 &  40.09 & 5.5
& 3.5$_{-0.6}^{+0.6}$ & 12.7$_{-0.7}^{+0.7}$ &
[---] & 1.6$_{-0.7}^{+0.5}$ & [$<$0.3] & $<$0.2 \\

Z8193 & 42.19 & --- &  10.7$_{-0.3}^{+ 0.2 }$   & 6.2$_{-0.2}^{+
0.3}$ & 3.4$_{-0.3}^{+0.2}$ & 
2.1$_{-0.2}^{+0.3}$ & 1.9$_{-0.2}^{+0.2}$ & 0.7$_{-0.2}^{+
0.2}$   \\

Z8197          &  41.33 & 26.9 
& 3.6$_{-0.2}^{+0.3}$ &  4.5$_{-0.2}^{+0.1}$ & 1.6$_{-0.2}^{+0.2}$ &
1.4$_{-0.2}^{+0.2}$ & 0.6$_{-0.2}^{+0.2}$ & 0.3$_{-0.2}^{+0.2}$  \\

RXJ1720.1+2638 &  [41.18] & 15.1 
& [1.3$_{-0.6}^{+0.6}$] &  [0.5$_{-0.1}^{+0.2}$] & 1.1$_{-0.1}^{+0.2}$$^*$ &
0.6$_{-0.2}^{+0.1}$ & 0.4$_{-0.1}^{+0.2}$ & 0.4$_{-0.1}^{+0.2}$ \\

A2294 &    41.48 & --- &  2.0$_{-0.1}^{+0.1}$ &     1.0$_{-0.1
}^{+0.1}$  &    0.3$_{-0.1}^{+0.1}$ &
  0.3$_{-0.1}^{+0.1 }$     & 0.1$_{-0.1}^{+0.1}$ & $<$0.1 \\

RXJ1733.0+4345 & 39.26 & 3.2 
&  0.4$_{-0.3}^{+0.3}$ &  2.2$_{-0.3}^{+0.3}$
& [0.5$_{-0.3}^{+0.3}$] & 0.4$_{-0.3}^{+0.3}$ & $<$0.1 & $<$0.1 \\

A2292 &  40.63 & --- &  0.7$_{-0.2}^{+0.2 }$  &  1.4$_{-0.3}^{+0.1 }$  & $<$0.1 & $<$0.2 & $<$0.3  & $<$0.4 \\

Z8276          &  41.23 & 32.3 & 
 6.8$_{-0.3}^{+0.3}$ & 11.4$_{-0.3}^{+0.3}$ &
4.6$_{-0.4}^{+0.3}$ & 2.7$_{-0.3}^{+0.4}$ & 1.8$_{-0.3}^{+0.3}$
& 0.3$_{-0.3}^{+0.3}$  \\

RXJ1750.2+3505 &  41.23 & 15.8 
& 1.2$_{-0.1}^{+0.2}$${\ast}$ &  2.9$_{-0.1}^{+0.1}$ &
0.7$_{-0.1}^{+0.2}$ & 
0.2$_{-0.2}^{+0.2}$  & 0.3$_{-0.2}^{+1.3}$ & 0.2$^{+1.3}_{-0.2}$ \\

RXJ2129.6+0005 &  41.05 & 16.5 
& 0.4$_{-0.1}^{+0.2}$ &
0.6$_{-0.1}^{+0.1}$ & $<$0.1 & $<$0.1 & 0.2$_{-0.1}^{+0.1}$ &
$<$0.1 \\

A2390          &  41.99 & 36.1 
& 3.8$_{-0.3}^{+0.3}$ &  4.3$_{-0.3}^{+0.3}$ & 1.9$_{-0.3}^{+0.3}$ & 
0.5$_{-0.3}^{+0.3}$ & 1.0$_{-0.3}^{+0.3}$ & 0.1$_{-0.3}^{+0.3}$   \\

A2495          &  40.41 & 15.8 
& 0.9$_{-0.2}^{+0.2}$ &
1.4$_{-0.3}^{+0.2}$ & $<$0.2 & $<$0.2 & 0.2$_{-0.2}^{+0.2}$ & [---]\\



A2626   b       &  [39.96] & 5.1 
& [0.7$_{-0.1}^{+0.2}$] &
1.8$_{-0.2}^{+0.2}$ & 0.6$_{-0.2}^{+0.1}$ & 0.7$_{-0.1}^{+0.2}$
& $<$0.1 & $<$0.1 \\

A2627 a &---  & --- &  $<$0 &  0.5$_{-0.1}^{+ 0.1}$ & ---$^*$ & ---$^*$ &  $<$0.1 & $<$0.1 \\

A2627 b & --- & --- &  0.1$_{-0.1}^{+0.1 }$&  0.7$_{-0.2}^{+0.1 }$   & $<$0.3 & $<$0.3 & $<$0.1 & $<$0.1 \\ 

A2634          &  40.12 & 7.5 
& 3.4$_{-1.3}^{+1.1}$ &
18.1$_{-1.1}^{+1.0}$ &  [---] & 0.5$_{-0.5}^{+0.3}$ & $<$0.2 &
$<$0.3 \\

A2665          &  39.81 & 7.3 
& 0.4$_{-0.2}^{+0.2}$ &  1.7$_{-0.2}^{+0.2}$ & 0.3$_{-0.2}^{+0.2}$ &
0.5$_{-0.2}^{+0.2}$ &  
$<$0.2 & $<$0.1 \\

\end{tabular}
\onecolumn
Notes:\\
\noindent H$\alpha$ luminosities are expressed in units of \ergps, and are
the {\em total} slit luminosities,  ie {\em not} those from the \lq 10kpc
aperture'.  \\
\noindent All fluxes are expressed in units of $10^{-15}$\ergpspcmsq, and
are the total flux from the slit.\\
\noindent Values in  square brackets are affected by atmospheric absorption. \\
\noindent H$\alpha$ in absorption is marked as $<$0. \\
\noindent Values marked by an asterisk are affected by a cosmic ray hit \\
\end{table}
\vfill\eject

\begin{table}
\caption{Blue emission line fluxes (in units of
$10^{-15}$\ergpspcmsq).\label{tab:bluelines} }
\begin{tabular}{lrrrrrrrrr}

& & & & & & & \\
Cluster        & [OII]$\lambda3727$ & H$\beta$ && [OIII]$\lambda$5007 && [NI]$\lambda5199$ & \\
& & & & & & & \\

Z235 & 1.5$_{-0.5}^{+0.5}$ & $<$0.4 & [0.5$_{-0.2}^{+0.2}$] & $<$0.6 & [0.4$_{-0.2}^{+0.2}$] & $<$0.1 & --- \\ 

A115   &  1.4$_{-0.1}^{+0.1}$ & 0.2$_{-0.1}^{+0.1}$ & [0.5$_{-0.2}^{+ 0.2}$]
& $<$0.1  &[0.2$_{-0.2}^{+0.2}$] &  --- $\ast$ & --- $\ast$\\ 

RXJ0107.4+3227 &  $<$0.7 & $<$0.2 & [1.7$_{-0.5}^{+0.6}$] & $<$0.7 & [2.1$_{-0.5}^{+0.5}$] & $<$0.3  &---\\ 

A262    &  6.1$_{-0.8}^{+1.0}$ &  $<$0.5  &[2.0$_{-0.4}^{+0.5}$]  & 1.0$_{-0.7}^{+0.7}$  &[2.2$_{-0.4}^{+0.4}$] & $<$0.6 & ---\\

A291  &  2.1$_{-0.4}^{+0.4 }$ & 0.9$_{-0.2}^{+0.2 }$ & [1.0$_{-0.1}^{+0.1 }$] &  0.3$_{-0.1}^{+0.2 }$  &[0.3$_{-0.1 }^{+0.1 }$] & $<$0.1   &[0.2$_{-0.1  }^{+0.1 }$] \\

Z808 &$<$0.7  & $<$0.1 & --- & $<$0.2  & --- & $<$0.1 & ---\\

RXJ0338.7+0958 &  57.1$_{-8.9}^{+9.3}$ &  4.7$_{-1.6}^{+1.8}$ & [7.4$_{-1.7}^{+1.9}$] & 3.2$_{-1.4}^{+1.6 }$ &[4.4$_{-1.5}^{+1.6}$]  & 4.4$_{-1.5}^{+1.5}$  & [3.1$_{-1.6}^{+1.9}$] \\

RXJ0352.9+1941 & 5.0$_{-0.3}^{+0.3}$   & 2.0$_{-0.2}^{+0.2}$ & [2.2$_{-0.2 }^{+0.2}$] & 1.4$_{-0.2}^{+0.2}$  &[1.4$_{-0.2}^{+0.2 }$] &   0.7$_{-0.2}^{+0.2}$  &[1.0$_{-0.1}^{+0.1}$]\\ 

A478 & $<$0.5 & 0.9$_{-0.4}^{+0.4}$  &[0.8$_{-0.3}^{+0.3}$] &0.5$_{-0.4}^{+0.4}$     &[0.4$_{-0.3}^{+0.3}$] & $<$0.5& 
[0.2$_{-0.2}^{+0.2}$]  \\

RXJ0439.0+0520 &   2.2$_{-0.3}^{+0.3}$ & 0.9$_{-0.1}^{+0.1}$ & [1.1$_{-0.1
}^{+0.1 }$] &  1.2$_{-0.1}^{+0.1}$  &[1.3$_{-0.1 }^{+0.1}$] &
0.3$_{-0.2}^{+0.2}$ & [0.3$_{-0.1}^{+0.1}$]\\

Z1121 & 0.2$_{-0.2}^{+ 0.4 }$ & $<$0.5 & --- & $<$0.5 & --- & $<$0.1 &---\\

RXJ0751.3+5012 b & 7.0$_{-2.9}^{+3.6}$  & $<$0.4  &[$<$1.2] & 1.4$_{-0.9}^{+0.9}$  &[1.6$_{-0.6}^{+0.6}$]  & $<$0.2 &---\\


RXJ0821.0$+$0752 &   0.4$_{-0.4}^{+0.4}$ & 0.2$_{-0.2 }^{+  0.2}$
 &[0.4$_{-0.2 }^{+0.2 }$]
& $<$0.2  &[1.0$_{-0.2 }^{+0.2 }$]& $<$0.19 &---\\ 

A646           &  4.6$_{-0.5}^{+0.5}$ & 0.9$_{-0.2}^{+0.3 }$  &[1.1$_{-0.3}^{+0.2}$] & 0.3$_{-0.2}^{+0.2 }$  &[0.3$_{-0.2}^{+0.2}$] &  0.7$_{-0.3 }^{+0.3}$ & [0.4$_{-0.2}^{+0.2}$]    \\ 

A795           & 3.3$_{-0.3}^{+0.3}$ &   0.2$_{-0.2}^{+0.2 }$ & [0.4$_{-0.2}^{+0.2}$]  &  0.5$_{-0.2}^{+0.2 }$  &[0.7$_{-0.2}^{+0.2}$]   & $<$0.1 & [$<$0.1]     \\

Z2701 & 0.9$_{-0.5}^{+0.5}$     &  0.5$_{-0.1}^{+0.1}$  &[0.6$_{-0.1}^{+0.1}$] & 0.1$_{-0.1}^{+0.1}$   &[0.1$_{-0.1}^{+0.1}$]
& 0.2$_{-0.1}^{+0.2}$ &[0.4$_{-0.1}^{+0.1}$] \\

RXJ1000.5+4409 a  &   0.5$_{-0.3 }^{+0.3 }$      & 0.1$_{-0.1 }^{+0.1}$  & [0.1$_{-0.1}^{+0.1}$] & 0.2$_{-0.1}^{+0.1}$  &[0.2$_{-0.1 }^{+0.1}$]& $<$0.2 &---\\

RXJ1000.5+4409 b  &  0.5$_{-0.3}^{+0.4}$ & $<$0.1 & --- & $<$0.1 & --- & $<$0.1  &---\\

Z3146 &  15.0$_{-0.4 }^{+0.4}$   &  4.6$_{-0.2}^{+0.2}$  &[4.6$_{-0.2}^{+0.2}$]   & 2.6$_{-0.2 }^{+0.2}$  &[2.5$_{-0.2 }^{+0.2}$]    & 1.4$_{-0.1 }^{+0.1}$   &[2.0$_{-0.2}^{+     0.2}$]\\

Z3179 & 0.8$_{-0.3}^{+0.3}$  & 0.1$_{-0.1}^{+0.2}$  &[0.2$_{-0.2}^{+0.2}$]
& 0.8$_{-0.3}^{+0.3}$ &[0.8$_{-0.2}^{+0.3}$] & $<$0.1 & [$<$0.2]\\

A1068 & 9.7$_{-0.4 }^{+0.4 }$& 3.6$_{-0.4}^{+0.4}$ & [4.1$_{-0.4}^{+0.4}$] & 12.2$_{-0.5}^{+0.5}$  &[12.6$_{-0.4}^{+0.5}$] & 1.6$_{-0.2 }^{+0.2}$  &[1.8$_{-0.3}^{+0.4}$] \\ 

A1084 & $<$0.7 & $<$0.2 &---& $<$0.2 &---& $<$0.1&--- \\

A1204 &  1.1$_{-0.4}^{+0.5}$ &$<$0.1  &[0.2$_{-0.1}^{+0.1}$] & 0.1$_{-0.1}^{+0.1}$ &[0.1$_{-0.1}^{+0.2}$] & $<$0.1 &[$<$0.1] \\

Z3916 & 0.4$_{-0.3}^{+0.4 }$ & 0.3$_{-0.1}^{+0.2}$  &[0.4$_{-0.2}^{+
0.2 }$] & 0.1$_{-0.1}^{+0.2}$  &[0.2$_{-0.1}^{+0.1}$] & $<$0.1 &---\\

A1361 &  3.4$_{-0.5 }^{+0.5 }$ &  0.8$_{-0.2}^{+0.2}$  &[1.0$_{-0.2 }^{+0.2
}$]  & 0.9$_{-0.2 }^{+0.2}$  &[0.9$_{-0.2 }^{+0.2}$] & $<$0.3  &---\\


RXJ1206.5+2810 & $<$0.8  & $<$0.5 & --- & $<$0.5 & --- &  $<$0.2 &---\\ 
RXJ1223.0+1037 & $<$2.6 & $<$1.0 & --- & $<$0.5 & --- & $<$0.6 &---\\

RXJ1230.7+1220 &  84.0$_{-19.4}^{+22.0}$ & $<$3.6  &[20.8$_{-3.1}^{+3.3}$]& 7.6$_{-7.3}^{+ 7.0}$  &[21.0$_{-3.4}^{+3.6}$]    & $<$7.5  &---\\

A1664 & 10.9$_{-1.1 }^{+ 1.0 }$   & 2.7$_{-0.2 }^{+0.2 }$   & [2.9
$_{-0.2 }^{+0.2}$]    &  1.4$_{-0.2}^{+ 0.2 }$  & [1.5$_{-0.2}^{+0.2}$] & 0.6$_{-0.2}^{+0.2 }$  &[1.1$_{-0.3}^{+0.3}$] \\

A1668          & 3.2$_{-0.6}^{+0.6}$  & 0.5$_{-0.3}^{+0.3}$
 &[1.0$_{-0.2}^{+0.2}$] &  1.6$_{-0.3}^{+0.3}$  & [1.7$_{-0.2}^{+0.2}$]    & 0.5$_{-0.3}^{+0.4}$  &[0.5$_{-0.1}^{+0.1}$]    \\

RXJ1320.1+3308 a & $<$1.5 & $<$0.5  &[0.6$_{-0.4}^{+0.5}$] &   1.4$_{-0.6}^{+ 0.6}$ & [1.2$_{-0.3}^{+0.3}$]  & $<$0.5 &---\\

A1795          & 11.4$_{-0.9}^{+0.9}$  &  1.4$_{-0.2}^{+0.3}$ & [2.0$_{-0.5}^{+0.6}$]    & 1.2$_{-0.2}^{+0.2 }$  & [2.0$_{-0.6}^{+0.7}$]      & 1.2$_{-0.4}^{+0.4 }$   &[1.6$_{-0.2}^{+0.2}$]    \\

A1835          &  6.7$_{-0.2}^{+0.2}$  &  1.8$_{-0.1}^{+0.1}$  & [2.1$_{-0.1}^{+0.1}$]    &
  1.3$_{-0.1 }^{+0.1 }$ & [1.5$_{-0.1}^{+0.1}$]    &  0.5$_{-0.2}^{+0.1}$  & [0.7$_{-0.2}^{+0.1}$]    \\

               &  5.6$_{-0.3}^{+0.3}$  & 1.1$_{-0.1}^{+0.1}$  & [1.2$_{-0.1}^{+0.1}$]    &  0.9$_{-0.1}^{+0.1 }$  &[1.0$_{-0.1}^{+0.2}$]    & 0.4$_{-0.2}^{+0.2}$ &  [0.4$_{-0.2}^{+0.2}$] \\

&  6.3$_{-0.4 }^{+0.4}$     & 1.8$_{-0.1}^{+0.1}$  &[2.2$_{-0.1}^{+0.1}$]  & 1.6$_{-0.1}^{+0.1}$ & [1.7$_{-0.1 }^{+0.1 }$] & 0.6$_{-0.2 }^{+0.2}$  & [1.6$_{-0.2}^{+0.2}$]\\

A1885 &   2.2$_{-0.5 }^{+0.5}$ & 1.0$_{-0.3}^{+0.3}$ &  [1.2$_{-0.2 }^{+0.2 }$] &   0.5$_{-0.2}^{+0.3}$  &[0.4$_{-0.2}^{+0.2}$]&  0.2$_{-0.2 }^{+0.2 }$  &[0.3$_{-0.1}^{+0.1}$]    \\

A1930  & $<$0.4 & $<$0.3  &--- & $<$0.3  &---& $<$0.1  &---\\

RXJ1442.2+2218 & 3.1$_{-0.7}^{+0.7}$ & $<$0.2  &[0.4$_{-0.3 }^{+0.4}$] & 0.4$_{-0.3 }^{+0.3 }$  &[0.7$_{-0.3}^{+0.4}$]  & $<$0.2  &---\\

RXJ1449.5+2746 & 2.7$_{-0.9}^{+0.9}$ & $<$0.2 & [$<$0.9] &
0.3$_{-0.3}^{+0.2}$  &[1.6$_{-0.4}^{+0.4}$]    & $<$0.5  &---\\

A1991          & 1.1$_{-0.3}^{+0.3}$  & $<$0.1 & [$<$0.1] &
 0.3$_{-0.3 }^{+0.2}$  &[0.4$_{-0.3}^{+0.4}$]    & $<$0.2 &---\\

Z7160 &  1.3$_{-0.1}^{+0.1}$ &  0.1$_{-0.1}^{+0.1}$  &[0.3$_{-0.1 }^{+0.1}$]& $<$0.1  &[$<$0.1] & $<$0.1  &---\\

A2009          & 1.2$_{-0.3}^{+0.3}$ &  0.2$_{-0.2}^{+0.2}$  &  [0.4$_{-0.3}^{+0.1}$] &
  0.1$_{-0.2}^{+0.2}$  &[0.3$_{-0.3}^{+0.2}$]  & 0.2$_{-0.2}^{+0.2}$  &[0.2$_{-0.1}^{+0.1}$] \\

A2052          & 7.5$_{-0.8}^{+0.8}$  & 0.6$_{-0.5 }^{+0.5 }$  &[1.4$_{-0.3}^{+0.3}$]    &
 5.7$_{-0.6 }^{+  0.6}$  &[5.9$_{-0.3}^{+0.3}$]     & $<$0.3  &[0.7$_{-0.2}^{+0.2}$] \\

A2072          & 0.7$_{-0.2}^{+0.2}$ & 0.1$_{-0.1 }^{+0.1}$  &[0.3$_{-0.1}^{+0.1}$]
&  0.1$_{-0.1 }^{+0.1}$  &[0.2$_{-0.1}^{+0.1}$]    & $<$0.1  &[0.2$_{-0.1}^{+0.1}$]    \\

RXJ1532.9+3021 &   9.2$_{-0.2}^{+0.2}$  &   2.5$_{-0.1 }^{+0.1}$ & [2.5$_{-0.1}^{+0.1}$]     &
  1.4$_{-0.1 }^{+0.1}$  &[1.4$_{-0.1}^{+0.1}$]     &   0.7$_{-0.1 }^{+0.1 }$  &[0.9$_{-0.1}^{+0.1}$]    \\


A2146 & 4.6$_{-0.6}^{+0.7 }$ & 
1.4$_{-0.2}^{+0.2}$  &[1.5$_{-0.1}^{+0.1 }$] & 10.7$_{-0.2
}^{+0.3}$  & [10.6$_{-0.2 }^{+0.2}$]   & 
0.4$_{-0.1}^{+0.1}$  &[0.7$_{-0.1}^{+0.1 }$]  \\

A2199          &   8.9$_{-1.2}^{+1.3}$  & 0.7$_{-0.3}^{+0.3}$  &[2.0$_{-0.2}^{+0.2}$]    &
 1.6$_{-0.3 }^{+0.3}$  &[2.4$_{-0.3}^{+0.3}$]    &  $<$0.1  &[0.8$_{-0.1}^{+0.1}$]    \\

A2204          & 25.5$_{-0.6}^{+0.6}$  & 4.9$_{-0.4}^{+0.4 }$ &[5.6$_{-0.2}^{+0.2}$]
&   3.0$_{-0.4 }^{+0.4}$  &[3.8$_{-0.2}^{+0.2}$]    &   1.5$_{-0.5}^{+0.5 }$   &[2.7$_{-0.3}^{+0.3}$]    \\

RXJ1715.3+5726 &   3.8$_{-1.0}^{+1.1}$ & $<$0.3  &[1.8$_{-0.4}^{+0.4}$]    &
 0.7$_{-0.8 }^{+0.8}$  &[1.4$_{-0.3}^{+0.3}$]    & $<$0.2  &[1.0$_{-0.3}^{+0.3}$] \\

Z8193 & 8.5$_{-0.6}^{+0.7}$  & 2.2$_{-0.4 }^{+0.4 }$   &[1.8$_{-0.3}^{+ 0.3}$] & 1.4$_{-0.3}^{+ 0.3}$  & [1.4$_{-0.3}^{+0.3}$] &  1.2$_{-0.4}^{+0.5 }$ &[$<$0.2] \\

Z8197          &   4.4$_{-0.3}^{+0.3}$  &  0.6$_{-0.2 }^{+0.1}$ & [0.8$_{-0.1}^{+0.1}$]    &
0.5$_{-0.2}^{+0.1 }$ &[0.5$_{-0.1}^{+0.1}$]    & $<$0.2  &[0.3$_{-0.1}^{+0.1}$]    \\

RXJ1720.1+2638 &   2.1$_{-0.2}^{+0.2}$  & 0.5$_{-0.1}^{+0.2}$
 & [0.7$_{-0.1}^{+0.1}$]    & $<$0.2  &[0.2$_{-0.1}^{+0.1}$]    &  0.2$_{-0.1 }^{+0.1}$   &[0.3$_{-0.1}^{+0.1}$]    \\ 

\end{tabular}
\end{table}
\vfill\eject

\addtocounter{table}{-1}

\begin{table}
\caption{Blue emission line fluxes  -- continued }
\begin{tabular}{lrrrrrrrrr}

& & & & & & & \\
Cluster        & [OII]$\lambda3727$ & H$\beta$ && [OIII]$\lambda$5007 && [NI]$\lambda5199$ & \\
& & & & & & & \\

A2294 & 0.4$_{-0.1}^{+0.1}$  & 0.6$_{-0.2}^{+0.2}$  &[0.9$_{-0.2
 }^{+0.2}$]  & 0.9$_{-0.2 }^{+ 0.1}$  &[1.2$_{-0.3}^{+0.3 }$]   &
 0.2$_{-0.1}^{+0.1}$  & [0.3$_{-0.2}^{+  0.2 }$] \\

RXJ1733.0+4345 & $<$0.7 & $<$0.1  &[0.4$_{-0.1}^{+0.1}$]  &
 0.3$_{-0.3 }^{+   0.3}$  & [0.6$_{-0.2}^{+0.2}$] & $<$0.1 & [0.1$_{-0.1}^{+0.1}$]  \\

A2292 & $<$0.1 & 0.4$_{-0.3 }^{+    0.3}$  &[0.6$_{-0.4}^{+0.4}$] &  0.1$_{-0.1}^{+0.1}$ 
 &[0.2$_{-0.2}^{+0.2}$] &   $<$0.1  &---\\

Z8276          &  9.0$_{-0.7}^{+0.7}$  & 1.7$_{-0.3}^{+ 0.3}$  &[2.0$_{-0.2}^{+0.2}$]     &
  1.6$_{-0.3}^{+0.3}$    &[1.6$_{-0.2}^{+0.2}$]     &   0.3$_{-0.3 }^{+0.3}$
 &[0.4$_{-0.2}^{+0.2}$]    \\

RXJ1750.2+3505 &  1.5$_{-0.2}^{+0.1}$ & $<$0.3  &[0.5$_{-0.2}^{+0.2}$]    &  0.7$_{ -0.2 }^{+0.2}$ & [0.7$_{ -0.2}^{+0.2}$] & $<$0.10  &[0.4$_{-0.1}^{+0.1}$] \\

RXJ2129.6+0005 &  0.7$_{-0.2}^{+0.1}$ & $<$0.2  &[0.2$_{-0.1}^{+0.1}$] & $<$
 0.1  &[$<$0.2] & $<$0.1  &[0.2$_{-0.1}^{+0.1}$] \\

A2390          &  4.3$_{-0.3}^{+0.3}$ &  0.2$_{-0.2}^{+0.3 }$ &[1.1$_{-0.2}^{+0.2}$]    &
  0.6$_{-0.3}^{+0.3}$ & [1.0$_{-0.2}^{+0.2}$]    & $<$0.2 & [0.4$_{-0.4}^{+0.5}$]    \\

A2495 & $<$0.7 & $<$0.3 & [$<$0.2] & $<$0.2 & [$<$0.1] & $<$0.1 & ---  \\


A2634 & $<$3.2 &
 $<$0.6  &[0.9$_{-0.9}^{+0.9}$]$^*$  
& 0.9$_{-0.9}^{+0.9}$  &[1.9$_{-0.9}^{+0.9}$]  & $<$0.5 & --- \\

A2665 & 1.4$_{-0.5}^{+0.5}$   & $<$0.2  &[0.5$_{-0.1}^{+0.1 }$]   & $<$0.2  &[
 0.1$_{-0.1}^{+0.1 }$] & $<$0.2  &---\\

\end{tabular}
\onecolumn
Notes:\\
We have tabulated values and upper limits for the blue emission line
fluxes in those clusters only where 
H$\alpha$ or \otw\ are definitely detected. \\ 
The values of H$\beta$, \oth\ and [NI] tabulated in square brackets are 
measured in the  `residual' spectrum (ie after a scaled template 
galaxy has been subtracted). Where there is no value shown for a line flux  for the residual
spectrum, the values/limits are not improved  from the original fit. \\
All fluxes are expressed in units of $10^{-15}$\ergpspcmsq and are {\it
total} fluxes from the slit. \\
Values marked by an asterisk are affected by a cosmic ray hit. \\
\end{table}
\vfill\eject 
\twocolumn

\subsection{Line intensity ratios}
\label{sec:lineint}

We have calculated line intensity ratios, using the template-subtracted
values of \Hb and [OIII] line flux where relevant. The errors on the line
ratios are formally propagated through from the errors on the line intensity from the
fitting, and do not represent extrema of the division. We do not include the red emission
line ratios for RXJ1720.1+2638, as line fluxes of [NII] and \Ha were badly
affected by atmospheric absorption, and [SII] by a cosmic ray hit.
RXJ1750.2+3505 is also excluded because the \Ha line is affected by a
cosmic ray hit.

\begin{figure}
\psfig{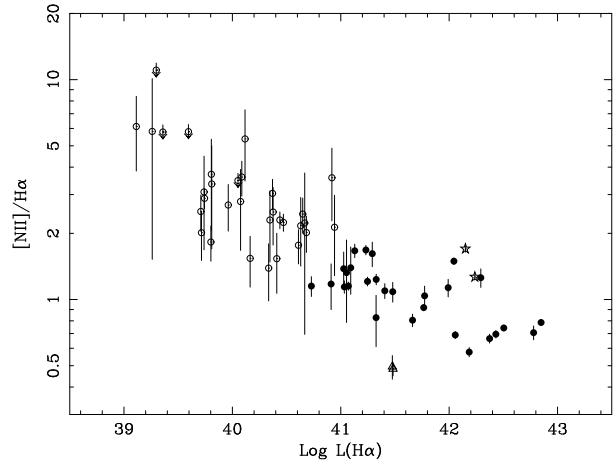}
\caption{\label{fig:lhavsn2oha}
Plot of the H$\alpha$ slit luminosity against the 
[NII]$\lambda$6584/H$\alpha$ line intensity ratio, for all 
emission-line objects in this paper. 
High \Ha luminosity systems are marked by solid
circles, and lower-luminosity ones by open circles. 
Outliers from  the general trend 
 are marked by stars  (at higher ionization) or by open triangles (at
lower ionization; two marker which overlap).  
}\end{figure}

\begin{figure}
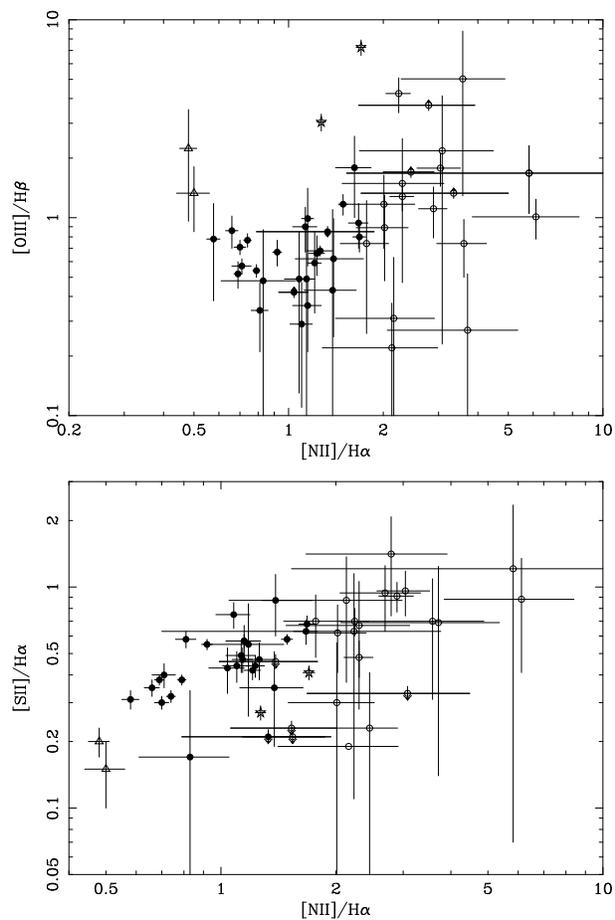

\vbox{
\psfig{figure=fig8top.ps,width=0.45\textwidth,angle=270}
\vspace{0.25cm}
\psfig{figure=fig8bot.ps,width=0.45\textwidth,angle=270}}
\caption{
\label{fig:ratios}
Plot of the line intensity ratios
[OIII]$\lambda$5007/\Hb (top)  and [SII]$\lambda$6717/\Ha (bottom) against
[NII]$\lambda$6584/\Han. Symbols as in Fig~\ref{fig:lhavsn2oha}. 
}\end{figure}

\begin{figure}
\psfig{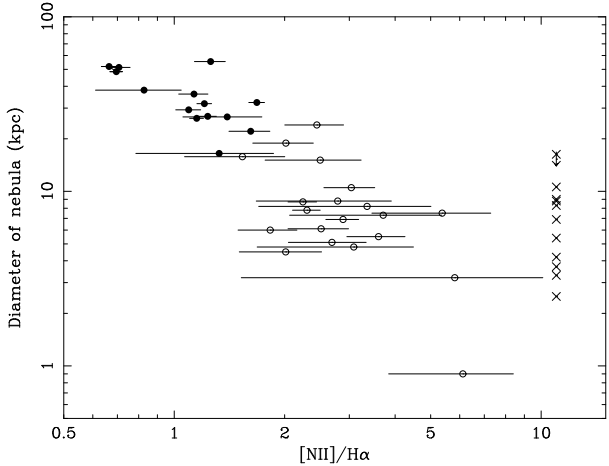}
\caption{\label{fig:n2ohavsd}
Ratio of [NII]$\lambda$6584/\Ha plotted against the diameter of the nebula
for all galaxies where this could be determined. Symbols are the same 
as in Fig~\ref{fig:lhavsn2oha}, with the [NII]-only emitting galaxies shown
at an arbitrary value of [NII]/\Ha of 11. }
\end{figure}

We plot the ratio of [NII]$\lambda6584$/\Ha against the luminosity of \Ha in
Fig~\ref{fig:lhavsn2oha}, and confirm the trend for the more luminous
systems to show a larger ratio of Balmer line emission to forbidden line
emission (as in Heckman \etal 1989, A92, C95).  We note that the
galaxies form a continuous distribution of [NII]/\Ha through four decades in
\Ha luminosity, with only a few obvious outliers. Whilst the majority of the
clusters selected in an X-ray flux-limited sample will contain cooling
flows, the emission-line BCG in this paper may not be a heterogeneous set.
Follow-up pointed X-ray data will detail exactly which clusters contain what
strength of cooling flow (Allen \etal 1999; Edge \etal 1999d; Crawford \etal
1999), but it is possible to mark out exceptions from the optical
emission-line spectrum alone. We broadly split the galaxies within the
[NII]/\Ha versus L(\Han) correlation shown in Fig~\ref{fig:lhavsn2oha} into
high \Ha luminosity systems (L(\Ha)$>10^{41}$\ergps) marked by solid circles
(eg A1835, Z3146, Z8193) and those of lower \Ha luminosity (eg A262, A1991,
A2199) marked by open circles. Note that whilst this division resembles the classification of
cooling flow nebulae into \lq class~I'/\lq class~II' systems of Heckman
\etal (1989), the clusters show a {\em continuous} distribution between the
two. In particular, this continuous behaviour makes the division between the two types fairly
arbitrary, and there is some overlap at L(\Ha)$\sim10^{41}$\ergps. Here the
classification of each individual galaxy  was also based on
other line ratios (Fig\ref{fig:ratios}). The separation of the sample into
different symbols in the figures is less an attempt to classify the observed
systems, and more to separate out the behaviour of each end of the
distribution in the diagrams in this paper.  A few emission line systems lie
outside the general trend; these are marked either by stars for two systems
with higher [NII]/\Ha than expected for their \Ha\ luminosity (A1068 and
A2146) or by open triangles for the two systems with systematically lower
[NII]/\Ha at a range of \Ha\ luminosity (A2294 and RXJ0821.0+0752; note that
the position of these last two galaxies in Fig~\ref{fig:lhavsn2oha}
overlap). Although the central galaxies in A2204 and RXJ0439.0+0520 appear
to have properties similar to the anomalously high [NII]/\Ha systems (stars)
in Fig~\ref{fig:lhavsn2oha}, their behaviour in all other plots fits better
with the other high-luminosity systems (Fig~\ref{fig:ratios}). The line
intensity ratio diagrams of [SII]$\lambda$6717/\Ha and
[OIII]$\lambda$5007/\Hb against [NII]$\lambda$6584/\Ha are shown in
Fig~\ref{fig:ratios}. We choose these diagrams as they are the least subject
to internal reddening (see section~\ref{sec:reddening}). The ionization
state of the nebula is also correlated with its size
(Fig~\ref{fig:n2ohavsd}).

\subsection{Spatially-extended emission line ratios}
\label{sec:spatext}

The line ratio plots in Figs~\ref{fig:lhavsn2oha}--\ref{fig:n2ohavsd} are
constructed from the flux-weighted {\em average} values, as given in
Tables~\ref{tab:redlines} and
\ref{tab:bluelines}. It is possible, however,
for the line ratios to vary within an individual galaxy. Such behaviour has
been noted before for the central galaxy in S1101 (Crawford \& Fabian 1992),
where the relative intensities of [NII]$\lambda$6584 and
\Ha swap between the   nucleus  and an extended filament.
Several of our emission-line galaxies are sufficiently extended that we are
able to sample the line ratios at several points in the nebula. We find that
the line ratios commonly change away from the centre of the galaxy, but not
in a heterogeneous manner.  For example, the line ratio of [NII]/\Ha can
either increase (eg A1835, Fig~\ref{fig:extratios}) or decrease (eg A1795,
Fig~\ref{fig:extratios}) between
the galaxy and the extended parts of the nebula. The line ratio of
[SII]$\lambda6717$/\Ha behaves in a similar manner to the [NII]/\Han. The
changes in ionization are not simply accounted for by an increased stellar
\Ha absorption in the galaxy continuum decreasing the amount of measured \Ha
line emission (and thus increasing the value of [NII]/\Ha derived). Firstly,
the change in the [NII]/\Ha  line ratio with radius is in some galaxies positive
(Figs~\ref{fig:extratios} and \ref{fig:extratiosall}).
Secondly, we have estimated the effect of stellar absorption on the
[NII]/\Ha ratio for all these galaxies with spatially extended emission, by
subtracting off the non-emitting template galaxy, scaled to match the depth
of the (nearby) NaD absorption line. We find that the value of [NII]/\Ha
measured for the galaxy centre is decreased by at most 7 per cent (in
RXJ0338.7+0958 and A1795), and more commonly by only 1-2 per cent. Thus the
change in line ratio appears to be a real change in ionization.

\begin{figure}
\vbox{

\psfig{figure=fig10top.ps,width=0.45\textwidth,angle=270}
\vspace{0.25cm}
\psfig{figure=fig10bot.ps,width=0.45\textwidth,angle=270}}
\caption{ \label{fig:extratios}
The spatial profile of the \Ha (solid line), [NII]$\lambda6584$ (dot-dash line)
and red continuum (dashed line) along the slit for the central galaxy in
A1835 (top) and A1795 (bottom).
 The y axis is in units of $10^{-15}$\ergpspcmsq for line intensities,
and of $10^{-16}$\ergpspcmsq (A1835) and 
 $5\times10^{-15}$\ergpspcmsq (A1795) for the continuum. 
The crosses give the values of 
the intensity ratio [NII]/H$\alpha$ at each row along the slit. 
}\end{figure}

\begin{figure}
\psfig{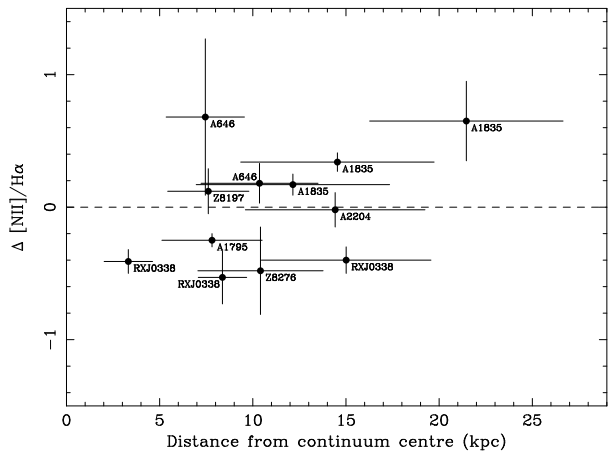}
\caption{\label{fig:extratiosall}
 The difference in the intensity ratio of
 [NII]/H$\alpha$ from the central value ($\Delta$[NII]/H$\alpha$), as a function of
 distance from the continuum centre. The dashed line marks
 $\Delta$[NII]/H$\alpha$ of zero, ie no change in value between the centre
of the galaxy and the spatially extended line emission.
}\end{figure}

\subsection{Intrinsic reddening}
\label{sec:reddening}

For several of the galaxies, the \Han/\Hb intensity ratio is significantly
greater than the expected canonical value of 2.86, suggesting that the
observed spectra are reddened along the line of sight. We have already
corrected the spectra for the extinction expected from our Galaxy, and so
assume any further reddening is intrinsic to the central cluster galaxy. In
column 7 of Table~\ref{tab:stelind}, we list the E(B-V) derived from the
observed
\Han/\Hb intensity ratio in the $\sim$10\kpc\ aperture (and from the total
aperture for spectra from A92 and C95). The intensity of \Hb is measured
by again constructing a difference spectrum (see Section~\ref{sec:emilines}; but for
the 10kpc apertures now) to approximate  \Hb in absorption from
the underlying  galaxy continuum. We only estimate E(B-V) where 
\Hb was detected, or should have been significantly detected
above the noise if it were  35 per cent of the \Ha intensity. We also omit
galaxies with only marginal evidence for internal reddening as deduced from
a noisy line ratio (eg A1204,
Z3916).  The limits
shown on E(B-V) are propagated from the errors to the fits of the \Ha and
\Hb emission line fluxes, and do not take into account any systematics (for
example, the appropriateness of the template subtraction). The spectra of
the affected galaxies are then dereddened by the amount shown, assuming
R=3.2 and that the dust forms a uniform screen at the redshift of the
galaxy.  We then measure \lq revised' values of the $\delta$BR index and \Ha
luminosity, shown in columns 8 and 9 of Table~\ref{tab:stelind}
respectively. The errors on the revised values of $\delta$BR and \Ha
luminosity are from the fits, and not due to the range of values given by
the errors on E(B-V). The Mg$_2$ index is not affected by reddening, and the
D$_{4000}$ break is less affected than $\delta$BR.

We have six spectra in common with Allen (1995), and find reduced values of
E(B-V) for some of these objects (eg A1664, A1068). Although we use the same spectra,
this difference is due to our template subtraction increasing the measured
\Hb flux and thus increasing the \Han/\Hb ratio.
Our errors on E(B-V) are also smaller than those of Allen, 
as we  propagate
the errors to the intensity from the fits to the emission lines
(which we also quote to 1$\sigma$  rather than to 90 per cent) 
 to give
errors on the \Han/\Hb ratio, rather than using the extrema of the fits.
There are several galaxies with low-level line emission for which the
\Han/\Hb ratio is significantly less than 2.86 (eg Z2701, RXJ1230.7+1220).
This suggests that for these galaxies it is necessary to also take into
account \Ha in absorption to give an accurate Balmer decrement. Several
galaxies have the lower limit to E(B-V) consistent with no internal
reddening (eg A478, A646, A1795, A2052 and A2199).

Placement of the spectrograph slit at the telescope depends on a (live)
direct image, and is thus optimized to the strong red continuum of an
elliptical galaxy. Thus if a galaxy is not observed at the parallactic
angle, we can expect the blue light to be preferentially lost from the slit
spectrum. This might affect the observations (in that the \Han/\Hb intensity
ratio is artifically increased due to preferential loss of \Hb from the
slit) of A291, RXJ0338.7+0958 and A1795, and does appear from the spectrum
(see Fig~\ref{fig:emspectra}) to be important for the observations of RXJ0352.9+1941 and
RXJ0439.0+0520.  However, we note that a galaxy marked as an 
`low-luminosity outlier' in Figs~\ref{fig:lhavsn2oha}--\ref{fig:n2ohavsd}
(RXJ0821.0+0752; the other A2294 is too noisy to tell) requires a exceptionally high level of
internal reddening (E(B-V)$>1.1$)  to interpret the lack of a strong \Hb detection. 
We note
though that this galaxy was not observed at the parallactic angle (and so
\Hb may have been preferentially been lost from the slit relative to \Han).

\subsection{Relation of the stellar indices to line luminosity}
\label{sec:relation}

\begin{figure}
\psfig{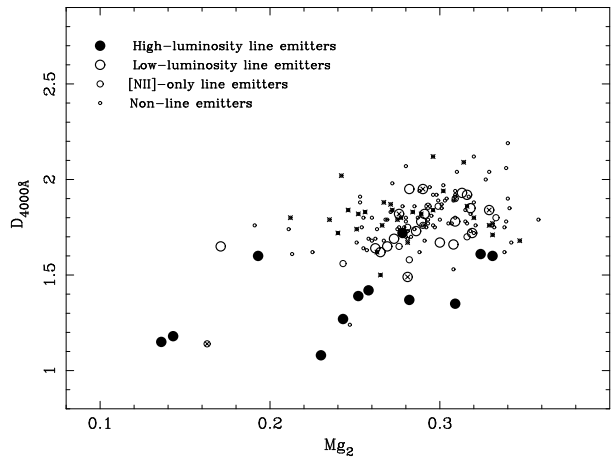}
\caption{\label{fig:mg2vsdcent}
Plot of the 4000\AA\ break (\D) against the Mg$_2$ index, as
measured from $\sim10$\kpc\ central apertures only. 
Symbols are similar to previous figures:  solid circles show  high-luminosity emitters; large open
circles show lower-luminosity systems, [NII]-only line emitters
and non-line emitters, dependent on the symbol size (see key). 
Galaxies that were not observed at the
parallactic angle are marked also by a cross. 
}\end{figure}

\begin{figure}
\psfig{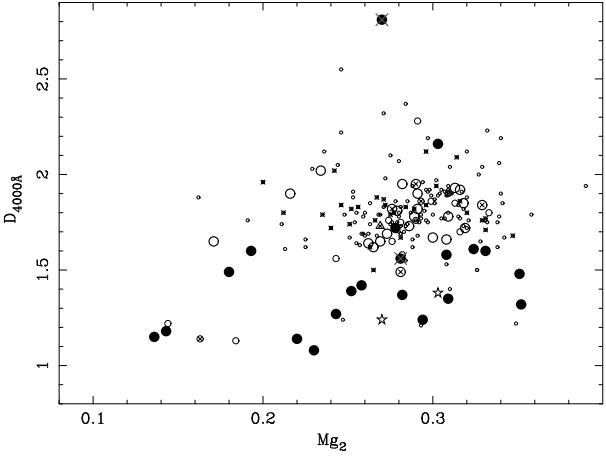}
\caption{\label{fig:mg2vsdall}
Plot of the 4000\AA\ break (\Dn) against the Mg$_2$ index, but now
including the data from A92 and C95, which are not all  from a  
10kpc aperture. Symbols as in previous figure. 
}\end{figure}

We plot the values of Mg$_2$ and D$_{4000}$ for the newly-observed 
galaxies (IDS spectra) in Fig~\ref{fig:mg2vsdcent}, excluding spectra marked as noisy in
Table~\ref{tab:stelind}. The
indices are all measured from a central $\sim$10\kpc\ aperture, and the
galaxies are marked by symbols dependent on both the presence and strength
of \Ha line emission in the spectrum. The  high luminosity systems
are marked by  solid circles, and the lower-luminosity systems by open
circles of the same size. [NII]-only and non-line emitters are also marked
by open circles, but of a smaller size as shown in the key to the figure.
Observations not taken at the parallactic angle (which may affect the
measured value of \Dn) are marked also by a cross.
Fig~\ref{fig:mg2vsdall} shows the same data, but now including stellar indices
measured from the spectra presented in A92 and C95, and again excluding 
 spectra marked as noisy in Table~\ref{tab:stelind}. The larger scatter within
the plot is mainly due to the variety of projected apertures from which the FOS and
WHT spectra are extracted. Nonetheless, the same trend as in
Fig~\ref{fig:mg2vsdcent} is
apparent; the stronger line emitters show a significantly bluer spectrum,
ie: a lower D$_{4000}$ for a given Mg$_2$ index. The lower-luminosity line
emitters show stellar indices little different from the general population of BCG. 

\begin{figure}
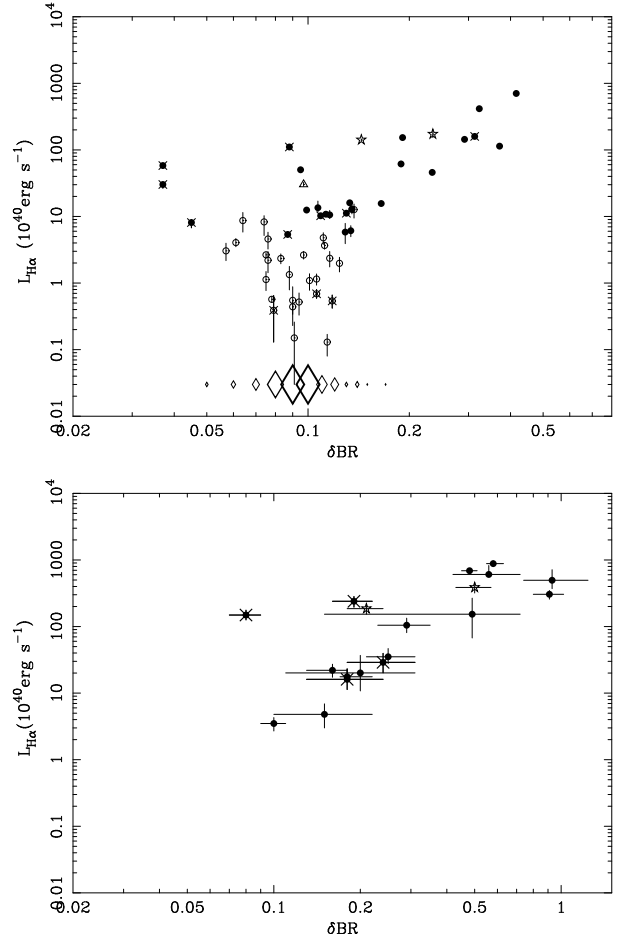

\vbox{
\psfig{figure=fig14top.ps,width=0.45\textwidth,angle=270}
\vspace{0.25cm}
\psfig{figure=fig14bot.ps,width=0.45\textwidth,angle=270}
}
\caption{ 
\label{fig:delbrvslha}
(Top) Plot of the $\delta$BR stellar index (see section~\ref{sec:contmSI}) against the \Ha
slit luminosity for all the \Ha line emitters in this paper, for the
apertures listed in Table~\ref{tab:stelind}.
Symbols are as in previous figures, with
crosses marking those galaxies not observed at the parallactic
angle (these might preferentially lose blue light from the slit and thus
have a decreased $\delta$BR than expected from the \Ha luminosity). The
distribution of $\delta$BR observed for the 10\kpc\ aperture spectra of the non-
line-emitting galaxies is shown by the diamonds at the arbitrary value of L(\Han)=$3\times10^{38}$\ergps, with the
size of the symbol proportional to the number of galaxies with that value of
$\delta$BR. 
(Bottom) The same plot, but now only for those galaxies showing internal
reddening (see section~\ref{sec:reddening}), using `corrected' values of both $\delta$BR and
L(\Han). }
\end{figure}

\begin{figure}
\psfig{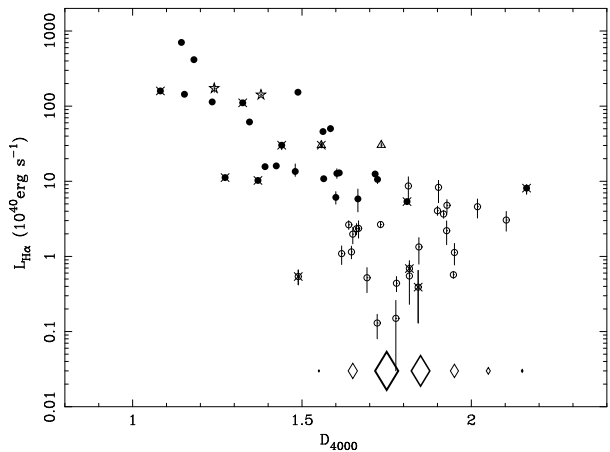}
\caption{\label{fig:dvslha}
Plot of the 4000\AA\ break (\Dn) against the \Ha
slit luminosity for all the \Ha line emitters in this paper (with the
exception of RXJ0352.9$+$1941 which, with a  D$_{4000}$ of 2.8 is off the
plot) for the apertures listed in Table~\ref{tab:stelind}.
Symbols are as in previous figures, with
crosses marking those galaxies not observed at the parallactic
angle. 
The distribution of  D$_{4000}$ observed for the 10\kpc\ aperture spectra of the non-
line-emitting galaxies is shown by the diamonds at an arbitrary value of L(\Han)=$3\times10^{38}$\ergps, with the
size of the symbol proportional to the number of galaxies with that value of
 D$_{4000}$. }
\end{figure}

The association of high-luminosity line emitters with the excess blue
continuum is also apparent from a plot of \Ha line luminosity against
$\delta$BR for all apertures (Fig~\ref{fig:delbrvslha}), whether these
quantities are corrected for internal reddening or not. The distribution of
$\delta$BR observed in the {\em non} line-emitting BCG in the sample is 
also plotted by diamond markers at the arbitrary value of
L(\Han)=$3\times10^{38}$\ergps, using only spectra from the $\sim$10\kpc\
apertures. The size of diamond marker plotted is directly proportional to
the number of non-emitting galaxies with that value of $\delta$BR, where the
majority lie in the range $0.08<\delta$BR$<0.11$. Of the three non-emitting
systems with $\delta$BR=0.05, two are not at the parallactic angle.  There
are a total of six non line-emitters with 0.14$<\delta$BR$<$0.20, and only
one above $\delta$BR of 0.20 (A1366 at $\delta$BR of 0.23). Similar values
are found for the $\delta$BR values for the data from A92 and C95,
although with a larger spread  probably due to the less accurate
sky-subtraction obtained with the shorter FOS slit. There are also two blue
non line-emitters from these data (A1682a, A1902 with $\delta$BR of 0.19 and
0.23 respectively and discarding A689 as a probable AGN).
Fig~\ref{fig:dvslha} shows that there is a similar trend of reduced \D with
increasing \Ha luminosity, again with the distribution of D$_{4000}$
for the 10-kpc aperture {\em non} line-emitting galaxies plotted at the
arbitrary value of L(\Han)=$3\times10^{38}$\ergps. Most of the non
line-emitting galaxies have stellar indices of $1.6<$\Dn$<2.0$, with only
A1366 an exception with a \D of 1.24. Six further non line-emitting galaxies have values of
\Dn$<1.5$ from the FOS data (A761, Z5604, A1703, A1902, A2426 and A2428),
but we note that these data are more suspect to inaccurate sky subtraction
due to the short slit, and thus the D$_{4000}$ is much less certain.

\subsection{Stellar populations}
\label{sec:stellarpop}

\begin{figure}
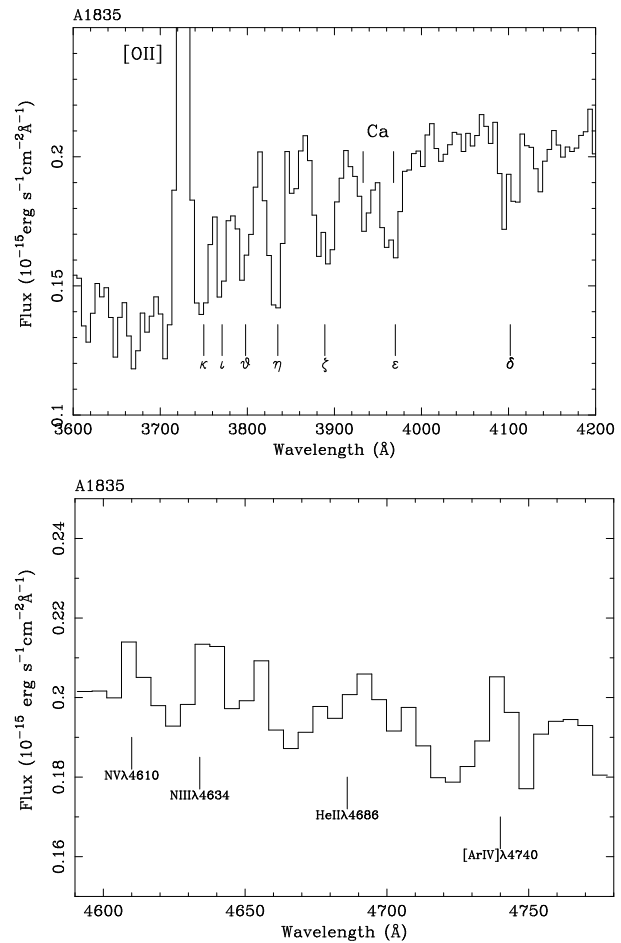

\vbox{
\psfig{figure=fig16top.ps,width=0.45\textwidth,angle=270}
\vspace{0.25cm}
\psfig{figure=fig16bot.ps,width=0.45\textwidth,angle=270}}
\caption{\label{fig:a1835stars}
(Top) 
Co-added IDS (rest-frame) spectra of A1835 showing the clear signature
of Balmer lines in absorption. [OII] in emission and the Ca absorption doublet 
are also marked. Note the false zero of the spectrum. 
(Bottom)  Co-added spectrum of A1835 (in the rest-frame) showing the $\sim$4650\AA\
Wolf-Rayet feature.  Note the false zero of the spectrum. 
}\end{figure}

The spectra of many of the line-luminous BCG show evidence of a young
stellar population; the best (and most extreme) example is the central
galaxy in A1835, which shows clear Balmer absorption lines
(Fig~\ref{fig:a1835stars}). 
We construct a co-added spectrum from the two {\em new} IDS observations of
this galaxy, and note that at the spectral resolution afforded by these data
we can independently confirm the  appearance of a broad
HeII$\lambda4686$ feature (Fig~\ref{fig:a1835stars}; cf. Allen
1995, where he also shows the central galaxy in A1068 to have Wolf-Rayet
features). 
We have searched for such features in other line-emitting
galaxies, but the signal-to-noise of most spectra precludes any other
definite detection. 

Once corrected for intrinsic reddening, many of the
line-emitting galaxies show very prominent UV/blue continua with a clear
Balmer break at 3646\AA, again both indicative of a population of massive
stars. Assuming that the excess blue continuum of the central cluster
galaxies is due to massive young stars rather than a power-law continuum, we
characterize the stellar population by a simple spectral analysis (cf.
Crawford \& Fabian 1993; Allen 1995). We examine all the BCG spectra
(extracted from the central apertures, shown in Tables~\ref{tab:stelind} and
\ref{tab:starcpts} and corrected for intrinsic reddening) with \Ha in
emission. The galaxy spectra are fit over the rest-frame wavelength range of
3300--5400\AA\ by the spectra of the template elliptical galaxy and model
stellar atmospheres (Kurucz 1979). For simplicity we use only templates of
O5, B5, A5 and F5 stellar types, and let the normalisation of each and the
elliptical galaxy vary as free parameters in the fitting. The emission lines
of [OII] and
\Hbn, [OIII] and [NI] were masked out of the BCG spectrum before fitting,
and the data are smoothed by 3 pixels. We do not attempt to fit those 
galaxies where the spectrum is either too noisy (A478) or shows a
marked loss of blue light from the slit (ie A2294, RXJ0338.7+0958 and
RXJ0352.9+1941, RXJ0439.0+0520 and Z3916). We include the high-ionization
systems (A2146 and A1068) but not the low-ionization system (RXJ0821.0+0752)
 with the very high value of intrinsic reddening. 

The composition of the best-fit stellar models are shown in
Table~\ref{tab:starcpts}, with each component given as a fraction of the
total continuum at (rest-frame) 4500\AA.  The errors on the components are
the $\Delta\chi^2=1$ confidence limits for the components in the model, and
do not encompass the range in the estimate of E(B-V) from
Table~\ref{tab:stelind}. Examples of the best-fit stellar models are shown
in Fig~\ref{fig:starfits}. The differing results between the three
observations of A1835 is most likely due to the long slits sampling a
slightly different position of the galaxy each time (the emission-line
nebulae and blue light distribution are not necessarily symmetric, eg
McNamara \& O'Connell 1993). Although a few of the spectra are the same as
those modelled in a similar manner by Allen (1995), we find a subtly
different balance of massive stars required by the fits, mainly due to both
the lesser values of E(B-V) inferred from our Balmer line ratios and the use
of a different elliptical galaxy template. The only galaxy for which we
could not achieve any satisfactory fit using stellar models was that in
Z8193. In all galaxies where the stellar components were fitted to a
spectrum extracted from a
$\sim10\kpc$ aperture, the monochromatic rest-wavelength 4500\AA\ luminosity
of underlying galaxy component is within the range of the twenty-four non
emission-line galaxies used to construct the elliptical template
($0.08-3.11\times10^{40}$\ergpspA from the 10\kpc\ aperture; with an average
of  $0.96\times10^{40}$\ergpspA).

Most of the best-fit models to the high \Han-luminosity systems 
require a fraction of O5 stars (from Z3146 at 58
per cent to Z7160 at 2 per cent), whereas few of the lower
\Han-luminosity systems require any O5 stars.
The fluxes of each
stellar component are converted to a number of stars, assuming monochromatic
4500\AA\ luminosities of $\sim2.7\times10^{34}$, $\sim2.5\times10^{32}$,
$\sim1.3\times10^{31}$, and $\sim1.6\times10^{30}$\ergpspA for O5, B5, A5 and
F5 stellar types respectively. The number of O5 stars inferred from the best
fits, as well as the number ratios of the different stellar
types are listed for  each best-fit
model in Table~{\ref{tab:starnos}.
The  rate of visible star formation inferred by our model fits is also shown
in Table~{\ref{tab:starnos}. We assume a constant rate of formation, and 
stellar lifetimes of 3.66$\times10^6$, 2.65$\times10^7$, 4.78$\times10^8$
and 2.11$\times10^9$ years for O5, B5, A5 and F5
stars respectively (from Shapiro \& Teukolsky 1983). The visible star
formation rates are quoted for the {\em total} area covered by the slit aperture
 (given in Table~{\ref{tab:starcpts}) -- conversion to a star formation rate
per kpc is non-trivial, and depends on the assumed radial profile of the
excess blue light. 
We confirm the finding of Allen (1995) that in all cases, the number
of lower-mass stars in relation to the O stars exceeds predictions from
either  continuous formation (at a constant rate) or a single burst of star formation with a \lq
normal' Scalo IMF. The exceptions to this are the  BCG whose spectra only
need an O star population in addition to the elliptical galaxy template,
which is compatible with a very recent single starburst. Over the wavelength
range being modelled, however, the spectrum of O stars are fairly
featureless, and the models can be equally well-fit by an elliptical
galaxy diluted by a power-law continuum, albeit with a steep 
$f_{\lambda}\propto\lambda^{-3.5}$ spectral slope. 
Note that the limits on optical polarization observed in the blue light
regions of  central galaxies in the A1795 and A2597 clusters appears to rule
out the possibility of a scattered nucleus as the origin of the blue light
 (MacNamara 1996a,c, 1998). 

As originally mooted by JFN, 
young O stars will provide a major source of photoionization for the 
 the emission-line nebula (the less massive stars do not contribute
significantly to the photoionization). 
We calculate the \Ha luminosity
resulting from photoionization by the O star component required by each best
fit stellar model given in Table~\ref{tab:starcpts}, assuming that each star
produces an \Ha luminosity of $5.5\times10^{36}$\ergps (following Allen
1995), and that the covering fraction of the stars by the gas is unity. 
The comparison between the predicted and observed (corrected for
intrinsic reddening) \Ha luminosities is shown in Fig~\ref{fig:oha}. Those
galaxies where the best-fit stellar population did not require any O stars
 are plotted at an arbitrary predicted \Ha
luminosity of $2\times10^{39}$\ergps. 
The errors on the predicted \Ha in this plot are derived only using the
errors on the stellar components in the spectral fitting,  whereas
there are also systematic uncertainties due to the amount of E(B-V) inferred
from the Balmer decrement. We repeat the stellar fits for each galaxy, now
sampling the full range of E(B-V) given in Table~\ref{tab:stelind} where
appropriate, and plot
the enlarged errors for the these points  by diamonds 
in Fig~\ref{fig:oha}. (A point with no error diamond around it is one
not requiring any dereddening from the Balmer decrement, eg A2204.)
Increasing the E(B-V) to the maximum shown in Table~\ref{tab:stelind} did
not permit any O stars as part of the stellar composition in the 
central galaxies which did not require them in the first place. 
The data are
broadly consistent with a large fraction of the observed \Ha luminosity
being provided by photoionization from a young stellar population (note that
there will be a range of O star sub-types present all contributing to the
photoionization).  Larger departures from this relationship may reflect the
range of covering fractions present in the galaxies. 
In all this analysis, the distribution of the dust is unknown, and a large
uncertainty remains about how well the reddening
inferred from the line emission can be applied  to the galaxy stellar
continuum. 

Surprisingly, a few of the lower \Han-luminosity systems are also close to this
relation, although six of them (and two of the high-luminosity objects) do
not have any predicted O5 stars to generate the observed line luminosity;
this agrees with higher line intensity ratios seen in such systems implying 
a different ionization source dominates such systems. 

We also carry out a simple stellar analysis for the only non \Han-emitting
galaxy out of the new IDS observations to show appreciable evidence for a
blue continuum, A1366 (see section~\ref{sec:contmSI}). It requires an excess
population of 12 per cent O5 stars and 3 per cent B5 at (rest-frame
4500\AA). It is also well-fit by a power-law component with slope of 
$f_{\lambda}\propto\lambda^{-3.3}$, containing 16 per cent of the flux at
rest-frame 4500\AA.

\onecolumn
\begin{table}
\caption{Percentages of the stellar components at (rest-frame) 4500\AA\ in
the best fit to the BCG spectra. The BCG spectra were extracted from the
aperture shown in column 2, and corrected for
intrinsic reddening by the amount shown in the column marked E(B-V). The monochromatic 4500\AA\ luminosity of the elliptical galaxy
component to the fit shown in the final column.  \label{tab:starcpts}}
\begin{tabular}{lrcccccccccc}
  Name       &  aperture & E(B-V) & \% O5    & \% B5       & \% A5      & \% F5 & \% Elliptical  & L(Elliptical) \\
             &  (kpc)    &        &       &           &        &       &   &        (10$^{40}$\ergpspA)    \\
A115          & 8.8 & 0.00   &  4 $\pm$ 1  & ---  &  4 $\pm$ 1   & ---  & 92 $\pm$ 1  & 0.61 $\pm$ 0.01 \\
A291          & $<$25.1 & 0.00   & 15 $\pm$ 1  & ---  & ---       & ---  & 85 $\pm$ 1 &  1.46 $\pm$ 0.01 \\
RXJ0751.3+5012b & 10.2 & 0.00 & --- &  3 $\pm$  2 & --- & --- & 97 $\pm$ 3 & 0.26 $\pm$ 0.01 \\
A646          & 10.5 & 0.06   &  8 $\pm$  1   & ---   & ---   & 9 $\pm$  3  & 83 $\pm$  3& 0.48 $\pm$ 0.02 \\
RXJ1000.5+4409 & $<$21.0 & 0.00 & --- & --- & --- &  22 $\pm$ 10 & 78 $\pm$ 11 & 0.43 $\pm$ 0.06 \\
A795          & 8.8 & 0.00   &  5 $\pm$  1   & ---   & ---   & ---  & 95 $\pm$ 1&  0.68 $\pm$ 0.01\\
Z3146         & $<$32.0 & 0.20   & 58 $\pm$   1   & ---   & ---   & ---  & 42 $\pm$  2 & 2.93 $\pm$ 0.12\\
A1068         & $<$19.3 & 0.39   & 9 $\pm$  2  & 27 $\pm$  1  & ---  & 34  $\pm$ 1  & 30  $\pm$  3 & 2.61 $\pm$ 0.27 \\
A1361         & $<$16.8 & 0.00   & 5 $\pm$   1 & --- & --- & --- &   95 $\pm$  1 &  0.58 $\pm$ 0.01 \\
RXJ1206.5+2810 & 9.9 & 0.00 & --- & --- & --- & 4 $\pm$  1 & 96 $\pm$   1 & 0.35 $\pm$ 0.01  \\
RXJ1223.0+1037 & 10.1 & 0.00 & --- & --- & ---- & 2 $\pm$ 2 & 98 $\pm$ 2 & 0.25 $\pm$ 0.01 \\
RXJ1230.7+1220 & 9.9 & 0.00 &  7 $\pm$ 1 & --- & --- & --- & 93 $\pm$ 1 &0.08 $\pm$ 0.01  \\
A1664         & $<$18.0 & 0.46   & 46 $\pm$  1   & ---  & 43 $\pm$  4   & ---  & 11 $\pm$  3 &  0.73 $\pm$ 0.22\\
A1795         & 10.3 & 0.15  & ---  &  9 $\pm$  2  & 19 $\pm$  4  & 10 $\pm$ 7 & 62 $\pm$   3 & 0.60 $\pm$ 0.03  \\
A1835         & 10.4 & 0.40   & 24 $\pm$  3  & 18 $\pm$  6   &29 $\pm$  4  & 23 $\pm$  5  &  6 $\pm$  2 &1.02 $\pm$ 0.38 \\
& 10.4 & 0.55    & 23 $\pm$  4  & 31 $\pm$  8  & 27 $\pm$  6  & 19 $\pm$  2  & --- & --- \\
& $<$29.7 & 0.38   & 10 $\pm$  3  & 35 $\pm$  8  & 12  $\pm$ 5  & 16 $\pm$  6  & 27 $\pm$  3& 6.55 $\pm$ 0.73 \\
RXJ1442.2+2218 & 10.1 & 0.00 & 1 $\pm$  1 & --- & --- & -- &  99 $\pm$ 2 & 0.44 $\pm$ 0.01 \\ 
A1991         & 9.9 & 0.00 & --- & --- & --- &  4 $\pm$ 2 & 96 $\pm$ 2 & 0.49 $\pm$ 0.01 \\
Z7160         & $<$30.1 & 0.24   &  2 $\pm$  1  & ---  &  6 $\pm$  3  & 22  $\pm$ 6  & 70 $\pm$  3& 5.04 $\pm$ 0.18 \\
A2052         & 9.4 & 0.22    & ---   & 3 $\pm$  1   & 8  $\pm$ 3  & 17 $\pm$  7 & 72 $\pm$  3 & 0.54 $\pm$ 0.02  \\
A2072         & 10.5 & 0.00 & 2 $\pm$ 1 & --- & --- & --- &  98 $\pm$ 1 & 0.61 $\pm$ 0.01 \\
RXJ1532.9+3021 & 8.6 & 0.21   & 25 $\pm$   1  &---  & 46 $\pm$  2  & --- &29 $\pm$   1&  1.34 $\pm$ 0.06 \\
A2146         & $<$28.3 & 0.20   & 10 $\pm$  1  &---   & ---  & 24 $\pm$  6  & 66 $\pm$  5& 2.66 $\pm$ 0.18 \\
A2199         & 10.2 & 0.10   & ---   & ---   & ---  & 11 $\pm$  1  & 89 $\pm$   1 & 0.39 $\pm$ 0.01 \\
A2204         & 9.6 & 0.00   & 26 $\pm$ 1   & ---   & ---  &  ---  & 74  $\pm$ 1 &  0.92 $\pm$ 0.01 \\
RXJ1715.3+5725 & 9.3 & 0.00 &  5 $\pm$ 1 & --- & --- & --- &  95 $\pm$  1 & 0.61 $\pm$ 0.01 \\
Z8197         & 9.6 & 0.33    & ---  & 21 $\pm$ 1    & ---  & 16 $\pm$   4 &63 $\pm$  2 & 1.70 $\pm$ 0.07  \\
Z8276         & 9.4 & 0.24    & 8 $\pm$ 1  & ---   & ---   & ---  & 92  $\pm$ 1 &0.82 $\pm$ 0.01 \\
 A2390         & 9.9 & 0.22    &15 $\pm$ 1   & ---   & 9  $\pm$ 5  & 32 $\pm$  9  & 44 $\pm$  4 & 1.06 $\pm$ 0.10\\
A2495         & 10.1 & 0.00 &  4 $\pm$ 1 & --- & --- & --- &  96 $\pm$ 1 & 0.41 $\pm$ 0.01 \\
              &         &              &       &             &              &      \\

\end{tabular}

\noindent The central galaxies in Z235, A262, Z808, Z2701, Z3179, A1668,
RXJ1320.1+3308, A1930, RXJ1733.0+4345,
A2626b and A2634 did not require any excess stellar population above that of
the template elliptical. RXJ0107.4+3227  only required an 0.4 per cent
excess of O5 stars  at 4500\AA.  \\
\end{table}

\begin{figure}
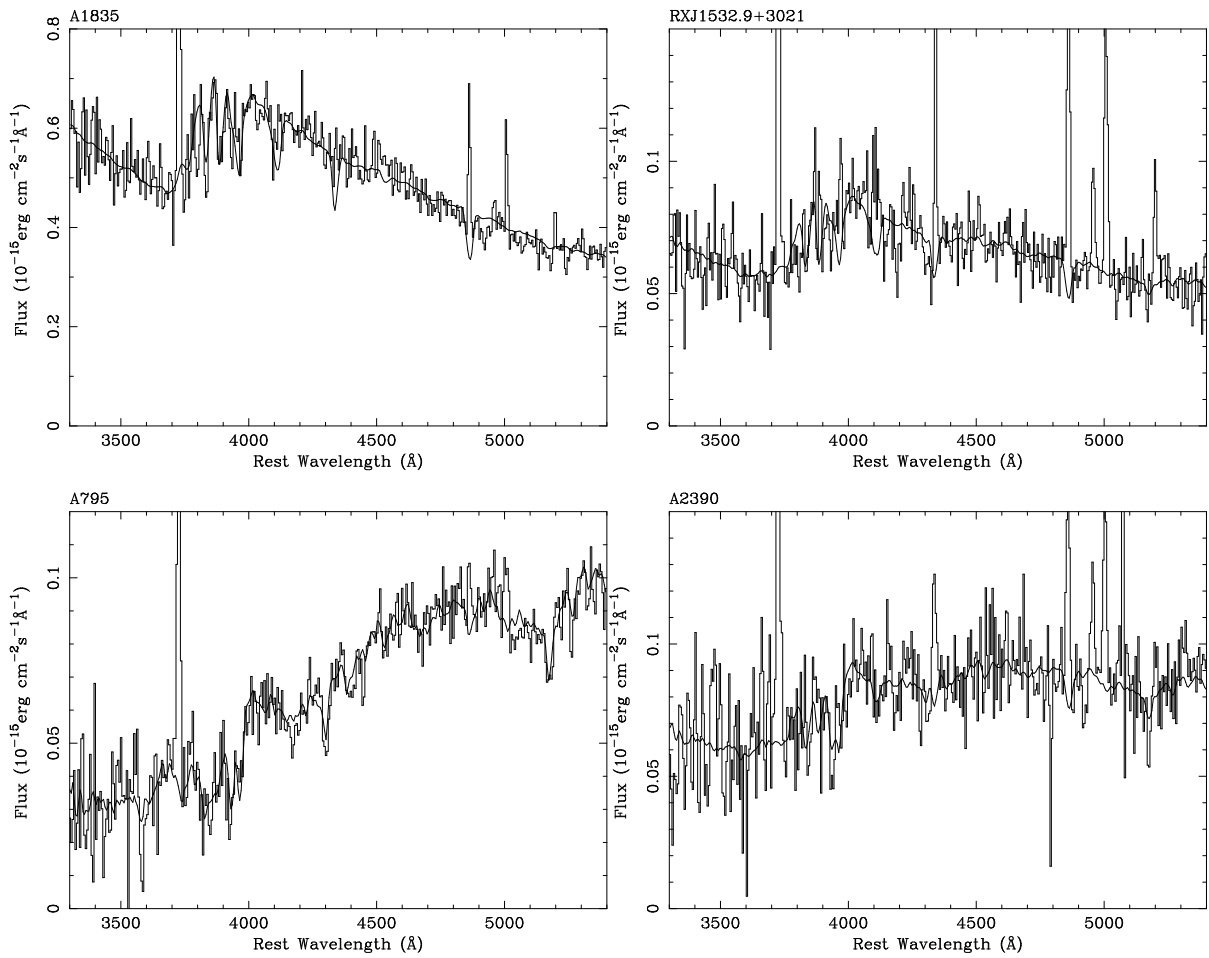

\hbox{\vbox{
\psfig{figure=fig17a.ps,width=0.45\textwidth,angle=270}
\vspace{0.25cm}
\psfig{figure=fig17b.ps,width=0.45\textwidth,angle=270}
}\vbox{
\psfig{figure=fig17c.ps,width=0.45\textwidth,angle=270}
\vspace{0.25cm}
\psfig{figure=fig17d.ps,width=0.45\textwidth,angle=270}
}}\caption{\label{fig:starfits} Four central cluster galaxy spectra
(corrected for intrinsic reddening) each
with their best-fit stellar  model (smooth line) as in
Table~\ref{tab:starcpts}. The emission-line
regions of the spectrum were not used in the fit.}
\end{figure}

\begin{table}
\caption{Numbers of O5 stars and number ratios of massive stars from the fits given in 
Table~\ref{tab:starcpts}. The rate of visible star formation (for the whole
aperture as given in Table~\ref{tab:starcpts}) inferred from
these fits is given in the last column. \label{tab:starnos}}
\begin{tabular}{lccccr}
  Name       & No. of O5 stars & B5/O5 & A5/O5 & F5/O5 & Visible SFR  \\
             & &      &       &       &     (\Msunpyr) \\
A115         &  $9.3\pm1.2\times10^3$ & --- &  2257     $\pm$ 864 & --- & 0.20  \\
A291          & $9.1\pm0.3\times10^4$ & --- & --- & --- &  0.99 \\
RXJ0751.3+5012b & --- & --- & --- & --- & 0.09 \\
A646          & $1.7\pm0.2\times10^4$ & --- & --- &   19254    $\pm$ 7158 & 0.40 \\
RXJ1000.5+4409 & --- & --- & --- & --- & 0.52 \\
A795          & $1.3\pm0.1\times10^4$ & --- & --- & --- & 0.14 \\
Z3146         & $1.51\pm0.03\times10^6$ & --- & --- & --- & 16.43 \\
A1068         & $5.0\pm1.2\times10^5$ & 86 $\pm$     61 &   1744 $\pm$ 875&
29840 $\pm$  10093 &30.59 \\
A1361         & $1.1\pm0.2\times10^4$ &  --- & --- & --- & 0.12 \\
RXJ1206.5+2810 & --- & --- & --- & --- & 0.06 \\
RXJ1223.0+1037 & --- & --- & --- & --- & 0.03 \\
RXJ1230.7+1220 & $2.3\pm0.2\times10^2$ & --- & --- & --- & 0.02 \\
A1664         & $1.17\pm0.03\times10^6$ & ---&  1927 $\pm$     173 &---& 23.06\\
A1795 & --- & --- & --- & --- & 1.94 \\
A1835         & $1.35\pm0.16\times10^6$ & 86 $\pm$     31  &  2599 $\pm$
505&   17149 $\pm$ 4249 &76.98 \\
              & $1.79\pm0.28\times10^6$ & 144 $\pm$  43  &  2382 $\pm$ 636&
13852 $\pm$   2814 &124.37 \\
              & $9.2\pm3.1\times10^5$ & 360  $\pm$   145&    2427 $\pm$1369&
25877 $\pm$ 13472&125.13 \\
RXJ1442.2+2218 & $2.1\pm1.4\times10^2$ & --- & --- & --- & 0.02\\
A1991         & --- & --- & --- & --- & 0.09 \\
Z7160         & $4.2\pm1.4\times10^4$ & --- &    7565 $\pm$   4562&  232815
$\pm$  94533 & 8.55 \\
A2052  & --- & --- & --- & --- & 0.96 \\
A2072 & $3.5\pm1.6\times10^3$ & --- & --- & --- & 0.04 \\
RXJ1532.9+3021 & $4.4\pm0.9\times10^5$ & ---&    3724  $\pm$   154  &---&12.26\\
A2146         & $1.4\pm0.2\times10^5$ &---&---& 41956 $\pm$ 11194&5.62 \\
A2199 & --- & --- & --- & --- & 0.21 \\
A2204         & $1.2\pm0.1\times10^5$ & --- & --- & --- &  1.29 \\
RXJ1715.3+5725 & $1.2\pm0.1\times10^4$ & --- & --- & --- & 0.13 \\
Z8197 & --- & --- & --- & --- & 7.86 \\
Z8276         & $2.8\pm0.2\times10^4$ & --- & --- & ---& 0.30 \\
A2390         & $1.3\pm0.1\times10^5$ & ---&  1218 $\pm$    676   & 36993
$\pm$  10350  & 5.40\\
A2495          & $5.7\pm0.9\times10^3$ & --- & --- & --- & 0.06 \\
&             &         &       &           &                  \\
\end{tabular}
\end{table}
\twocolumn

\begin{figure}
\vbox{
\psfig{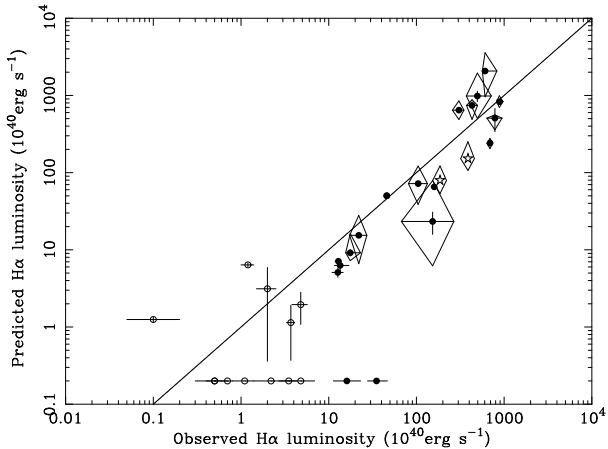}
}\caption{\label{fig:oha} The observed \Ha luminosity (corrected for
reddening) plotted against the \Ha luminosity predicted from the O
star component of the best-fit stellar model, assuming a covering fraction
of unity.   Error bars
on the observed luminosity are from the errors to the fit to the  \Ha intensity,  and on
the predicted luminosity from the errors on the O star component in the
stellar model.
 The error diamonds around some points show the errors in the
predicted luminosity whem using the full range of E(B-V) given by the errors
in Table~\ref{tab:stelind}. High \Han-luminosity systems
(L(\Han)$>10^{41}$\ergps) are shown by solid
circles, those at lower  \Han-luminosity by open circles, and the two
high-ionization systems A1068 and A2146 by open stars. 
The points  plotted at an arbitrary predicted luminosity of
$2\times10^{39}$\ergps are those galaxies for which no O stars were required
in the best-fit stellar population. The solid diagonal line gives the direct
one-to-one relationship between the predicted and observed fluxes. 
}
\end{figure}

\section{Discussion}

\subsection{Do brightest cluster galaxies differ between Abell and non-Abell clusters?}

The X-ray flux selection used to compile the BCS has produced a sample of
 the most luminous clusters within  $z<0.3$. Although the bulk of the
 sample is formed from clusters selected through cross-correlation against
 optical catalogues, the BCS also includes clusters selected from their
 X-ray properties alone. In fact, one of the key aims of the BCS is to
 examine the optical differences between clusters of similar X-ray
 luminosity and redshift. Whilst the determination of a robust galaxy
 density (ie cluster richness) is beyond the scope of this paper, we can use
 the brightest cluster galaxy to make a preliminary comparison between
 optically- and X-ray selected clusters. From Table~\ref{tab:bcslog} we can
 state that 48 of the 51 BCS clusters within a redshift of 0.05 have a BCG
 that is optically-catalogued (irrespective of its cluster environment) in
 either the MCG, NGC or IC optical catalogs, and they are all (where
 indicated) elliptical galaxies. The three exceptions are RXJ0338.7+0958,
 RXJ0341.3+1524, RXJ1205.1+3920, each of which is centred on a galaxy on the
 Palomar Sky Survey. The first source is the well-studied poor group
 2A~0335+096 (eg Sarazin, Baum \& O'Dea 1995), the second is centred on an
 anonymous Zwicky galaxy (III Zw 054) and the third, while at much higher
 galactic latitude than the other two sources, is missing from all available
 optical catalogues. There are a number of optically-catalogued central
 cluster galaxies beyond 0.05 but the fraction in both Abell and non-Abell
 clusters drops to essentially zero in these catalogues beyond a redshift of
 0.075. These numbers clearly indicate that the brightest member galaxies of
 X-ray selected clusters are almost always easily identified at $z<0.05$,
 even if the cluster has itself been missed in optical cluster/group
 searches. This implies that none of the BCS clusters are intrinsically
 `dark' and the brightest galaxies within them show a limited scatter in
 optical luminosity.

\subsection{Cooling flow nebulae}

Around one-third of the BCG in this paper show optical line emission of some
degree -- from those with low-level [NII]-only emission, to some of the
brightest non-AGN line emitters known (Z3146, RXJ1532.9+3021). The
\Han-emitting dominant galaxies  show properties very similar to those at the centre of
 known cooling flow clusters (eg JFN, Heckman \etal 1989). We defer detailed
discussion of whether individual clusters contain cooling flows until the
X-ray compilation paper (Crawford \etal 1999), but for the moment assume
that the majority of these systems are described as cooling flow nebulae
(the few exceptions are discussed further in section~\ref{sec:noncf}).

Central galaxies of cooling flow clusters have long been known to show
luminous line emission, spatially extended in filaments spread over the central
5-60\kpc\ of the cluster core (eg Lynds 1970; Heckman 1981; Hu \etal 1985; JFN; Heckman
\etal 1989; A92; C95). Such nebulae are too bright to be simply the
recombination phase of the cooling gas in the intracluster medium (JFN); an
additional local heat source is required -- distributed throughout the
optically emitting region of the core -- to re-ionize the atoms many times
in order to produce the observed luminosity (JFN; Johnstone \& Fabian 1988;
Heckman \etal 1989; 
Donahue \& Voit 1991; Crawford \& Fabian 1992). Despite this, the
presence of such line emission is dependent on the cluster properties: large
line luminosity extended 
emission-line regions are found in clusters with high cooling flow rates
(JFN; Allen 1998), and so far such systems have {\em not} been found around BCG in
non-cooling flow clusters. Line emission is not, however,  ubiquitous in BCGs contained
in strong cooling flows (eg A2029, A2063), and occurs at around 32 per cent
throughout our cluster sample which includes non cooling-flow clusters.
There is a propensity for the emission-line systems to be
associated with those BCG which contain a central radio source (Heckman \etal
1989; C95), though again not exclusively (eg A1060). The
probability of the BCG containing a central radio source will, however, be again
dependent on the general cluster properties. 
 Further discussion of the relation between the optical properties of
the BCG to the X-ray properties of the host cluster and any central radio
source are postponed to the next papers in this series where the X-ray data
and radio data are presented (Crawford \etal 1999, Edge \etal 1999b).

\subsubsection{High \Ha luminosity cooling flow  systems}

Only the cooling flow nebulae with a high line luminosity
(L(\Han)$>10^{41}$\ergps) are accompanied by an excess ultraviolet/blue
continuum (Figs~\ref{fig:mg2vsdcent} to ~\ref{fig:dvslha}; JFN). Both the
line emission  and the blue light are extended on scales of
5-60\kpc, and are seen to be spatially coincident from the flattening of the
radial gradient of the Mg$_2$ spectral index within the emission-line
regions (Cardiel \etal 1998). The morphology of the blue light also seems
(in the few cases imaged) spatially related to that of the contained radio
source (McNamara \& O'Connell 1993; Sarazin
\etal 1995b; McNamara 1995; McNamara \etal 1996a; McNamara 1997).
The ultraviolet continuum has been successfully ascribed to massive young
star formation, from both the spectra (once intrinsic reddening has been
corrected; Allen 1995; section~\ref{sec:stellarpop}) and from high-resolution
images (McNamara \etal 1996a,b; Pinkney \etal 1996). The line ratios
observed in these systems have strong Balmer lines relative to forbidden
lines such as [OIII], [NII] and [SII] (see Fig~\ref{fig:ratios}; mean values
around [NII]/\Han$\sim$0.8, [SII]\Han$\sim$0.4, [OI]/\Han$\sim0.2$ and
[OIII]/\Hbn$\sim$0.6). A good
relation is found between the observed \Ha luminosity (again corrected for
intrinsic reddening) and that predicted
from photoionization by the O stars predicted to be present from the  stellar fit to the
dereddened continuum (Allen 1995; Fig~\ref{fig:oha}). 
Detailed modelling of the BCG spectra (Johnstone \& Allen
1999) indicates that the ionizing fluxes required to produce the
emission-line ratios and luminosities observed in these 
systems can be fully accounted for by the observed O star
populations (assuming very hot stars, with $T\approxgt40000$\K, and
a low  mean ionization parameter, $U \sim -4$) in a 
 high metallicity environment (about Solar; see also Filippenko \& Terlevich 1992; Allen
1995). Together with the spatial correspondence of the blue light and line
emission (Cardiel \etal 1998) it seems that the most line-luminous cooling flow
nebulae are powered by a population of massive hot stars. 

The additional stellar population is too centrally concentrated for the
stars to be the immediate depository of gas cooling from the hot
intracluster medium, as X-ray profiles of cooling flow clusters show the
mass deposition to have a more distributed, flat profile (Fabian 1994 and
references therein). Large masses of X-ray absorbing material are detected
in the spectra of cooling flow clusters (White \etal 1991; Allen \& Fabian
1997), indicating that cold clouds can be an end-product of the cooling
process. Such clouds might coagulate and condense to form a central
reservoir of cold gas. As suggested by the double-lobed blue light
morphology of A1795 and A2597, and as hypothesized by several authors
(McNamara \& O'Connell 1993; Pinkney \etal 1996; McNamara 1997), the
interaction between the outflowing radio plasma and this gas reservoir could
induce a massive starburst. The distribution of stellar types inferred is
weighted to more lower-mass  stars (section~\ref{sec:stellarpop} and Allen
1995), suggesting that such a starburst is triggered at irregular
timescales. Those galaxies requiring an excess population of just O stars
(eg those in A291, Z3146, Z8276) may only have one very recent such
starburst, whilst others may have aged such that few O stars remain. This
scenario predicts that the very luminous systems are created by the {\em
combination} of a central radio source and a surrounding mass of cold clouds
accumulated from the cooling flow -- both a common consequence of a rich
(but non-binary) cluster environment. Thus it is possible a central radio
source will not induce star formation and luminous line emission
unless the cooling flow has accumulated a reservoir of cold gas. Hence any
BCG in a cooling flow cluster that contain  a strong radio source yet no
optical anomalies  should also lack evidence for intrinsic absorption in the
cluster X-ray spectrum. A full examination
of the correlations between cooling flow properties and both the type of
line emission and presence of a central radio source will be addressed in
the subsequent papers in this series.

The ionization state of the nebula in high \Han-luminosity systems are seen
to change away from the galaxy, but not in a consistent manner
(Fig~\ref{fig:extratios} and \ref{fig:extratiosall}; see also S1101 in
Crawford \& Fabian 1992). As shown in section~\ref{sec:spatext}, the
variation is not simply caused by stellar \Ha absorption in the galaxy
continuum increasing the central value of [NII]/\Ha derived, as the line
ratios both drop and rise off-centre in different galaxies
(Fig~\ref{fig:extratiosall}). This change in line ratios is not easily
ascribed to a radial metallicity gradient either, as any overabundance of
nitrogen and sulphur would likely be concentrated toward the centre of the galaxy,
producing a drop in the line ratios at larger radius.

\subsubsection{Low luminosity cooling flow  systems}

The variation in the properties of the cooling flow nebulae form a
continuous trend with line luminosity (eg Figs~\ref{fig:lhavsn2oha},
\ref{fig:ratios}), implying a gradual change in the dominance of ionization
mechanism, rather than two discrete populations (cf. Heckman \etal 1989).
The cooling flow galaxies with a lower luminosity emission-line region have
stronger forbidden line emission to Balmer line emission
([NII]/\Han$\sim$2.4, [SII]/\Han$\sim$1 but [OIII]/\Hb showing a large
spread). Such line intensity ratios are traditionally difficult to explain,
and require a steeper ionizing spectrum over 13.6-100eV than can be produced
by the O~stars present in the higher luminosity systems. There is no
evidence for an appreciable excess blue continuum in these galaxies
(Figs~\ref{fig:mg2vsdcent} to \ref{fig:dvslha}), suggesting that massive
stars cannot make such an important contribution to the ionization.

The line ratios suggest the introduction of an additional, harder ionizing
source, and the nebulae are also physically smaller than the high luminosity
systems, suggesting that the dominant ionization source is not as spatially
extended. Several processes have been proposed, such as self-irradiation
from X-rays in the cooling flow (Voit \& Donahue 1990; Donahue \& Voit 1991;
which is, however, in contradiction with the observed X-ray flux
distribution) to self-absorbed mixing layers tapping both the kinetic and
thermal energy of the gas (Crawford \& Fabian 1992). Another realistic
possibility is that as the amount of visible star formation decreases down
the \Han-luminosity `sequence', the relative importance of any low-level
nuclear activity as an ionization source increases (Johnstone \& Allen
1999). Determining the source of photo-ionization will require the radial
mapping of the ionization parameter across the nebula in such systems.

At the very low-luminosity extreme of the \Ha sequence we find a 
small fraction of [NII]-only line emitters in our sample, which
are most  common in the lower X-ray luminosity clusters ($<10^{44}$\ergps;
Fig~\ref{fig:lxfreq}). The line emission in these galaxies closely
resembles that found in around half of ordinary elliptical galaxies (Phillips
\etal 1986)  ascribed to low-level LINER activity.

\subsection{The non cooling-flow line emitters}
\label{sec:noncf}

We have marked four obvious exceptions from the general behaviour of
line-emitters by different symbols in the correlations presented in section
3. Specifically, there are two BCG (in A1068 and A2146) with a combination
of line ratios and line luminosity (high [NII]/\Ha and [OIII]/\Hb for their
\Ha luminosity) that suggests the dominant form of ionization is due to an
AGN (these are marked by open star symbols in Figs~\ref{fig:lhavsn2oha} to
\ref{fig:dvslha}; note that the dominant galaxy in A2089 also shows a
similar ionization state). A2146 also has  prominent HeII$\lambda$4686 line emission
 (Fig~\ref{fig:emspectra}; Allen 1995) in agreement with the suggestions of
 a harder ionizing continuum than is usually seen in
 cooling flow nebulae. Follow-up X-ray observations confirm the presence of
 a point-like component of X-ray emission at the BCG in A2146 (at a small
fraction of the overall cluster luminosity; Allen 1995). A similar
 situation may exist for A1068, where the presence of an AGN cannot be ruled
 out, although the {\em IRAS} fluxes and Wolf-Rayet features in this galaxy 
 argue in favour of a massive starburst (Allen 1995). Two other systems
 (A2204 and RXJ0439.0+0520) also have high [NII]/\Ha for their \Ha
 luminosity, but line ratios more consistent with the majority trend. 
Both these clusters contain strong central radio sources (Edge \etal 1999b) which might 
dilute the cooling flow properties. 

Similarly, there are two galaxies offset from the general trend in the
opposite direction (A2294, RXJ0821.0+0752). These two galaxies have stronger
Balmer lines relative to the forbidden line emission, placing them more into
the regime of classical HII regions (these are marked by open triangles in
the Figs~\ref{fig:lhavsn2oha} to \ref{fig:dvslha}). 

\section{Summary and conclusions}
\label{sec:summary}

We have presented spectral observations of 256 dominant galaxies in 217
clusters, including 87 per cent of the 203 clusters that comprise the BCS.
This sample has been supplemented with basic information about the
brightest galaxy in a further 21 BCS clusters, as available from the
literature. We have listed new redshifts for 18 clusters, mostly  systems
discovered through X-ray-selection. One such cluster, RXJ1532.9+3021, is at
a redshift of $z$=0.36 and is thus the second most distant object in the
BCS.

27 per cent of the central dominant galaxies show narrow low-ionization
emission lines in their spectra, all but five with line intensity ratios
typical of nebulae associated with galaxies at the centres of cooling flow
clusters. A further 6 per cent of the clusters have central galaxies showing
only [NII] line emission with \Ha in absorption. We find no evidence for an
increase in the frequency of line emission with X-ray luminosity, the
distribution being consistent with a constant fraction.  We also find that at low redshift 
a purely X-ray-selected cluster  has a higher probability of
containing an emission-line system around its dominant galaxy than an
optically-selected cluster. We find similar numbers of dominant galaxies in both
optically-selected and purely X-ray selected clusters at low redshift
included in optical catalogues, suggesting that  the two techniques
select clusters with similar properties. The projected separation between
the optical positions of the BCG and the X-ray cluster centroid determined
from RASS is smaller for the line-emitting BCG than for those without line
emission (at mean values of $\sim30\kpc$ and $\sim90$\kpc\ respectively).
Assuming that the line emission traces a cooling flow system, this is
consistent with the idea that there is a smaller offset between the
gravitational centre and the central dominant galaxy in cooling flow
clusters.

We have measured three stellar indices ($\delta BR$, Mg$_2$ and \Dn) from a
spectrum extracted from the central 10\kpc\ of each galaxy in our sample; we
have also fitted the emission lines in spectra extracted from both this
region and the total span of the emission-line nebula. The total slit \Ha
ranges over four decades in luminosity ($10^{39}-10^{43}$\ergps), and the
properties of the emission-line system follow continuous trends of behaviour
across this full luminosity range.  The more \Han-luminous systems
(L(\Han)$\approxgt10^{41}$\ergps) have the more spatially extended
line-emitting regions (diameter $>20$\kpc), and show a higher ratio of Balmer
to forbidden line emission (eg [NII]/\Han$\approxlt$1). Some of the more
luminous systems are sufficiently extended that we can plot spatially
resolved line ratios, and find that the ratio of forbidden to Balmer lines
can either increase or decrease away from the centre of the galaxy. We find
that this effect is too large to be simply ascribed to \Ha stellar
absorption affecting the line ratios where the emission is coincident with
the underlying galaxy continuum, but is rather a real ionization effect. The
Balmer decrement measured in the more luminous nebulae often implies the
presence of internal dust, with reddening typically of E(B--V)$\sim0.3$ (but
with a wide range). These stronger line emitters also show a significantly
blue spectrum. 

After correction for intrinsic reddening (where appropriate), we have
characterized the excess blue light in the high-L(\Han) galaxies by fitting
simple template stellar spectra to their central continuum. Most of the
galaxies require an excess population of O stars in sufficient numbers to
provide the photoionization that can produce the observed \Ha luminosity.
Even though the distribution of stellar types is only crudely determined,
the greater number of lower mass stars relative to the O star population
does not favour predictions from either continuous star formation or a
single starburst. These findings are compatible with models that suggest the
massive star formation is triggered from the interaction between outflowing
plasma from the central radio source, and cold clouds deposited from, and
embedded in a cluster cooling flow. 

The lower \Ha luminosity emission-line systems show line ratios (eg
[NII]/\Han$\approxgt2$) requiring the introduction of a much harder source
of ionization, and are much less spatially extended. The galaxy continuum shows
no evidence for an appreciable amount of massive stars, with stellar indices
little different from those expected from the stellar population of non
line-emitting dominant galaxies. Possibilities for the dominant ionization
mechanism include turbulent mixing layers or low-level nuclear activity. The
small fraction of [NII]-only emitters closely resemble the low-level LINER
activity found in many ordinary elliptical galaxies.

There are only five obvious exceptions from the general trends shown by the
line-emitting galaxies. At least one (A2146) shows evidence for
contaminating photo-ionization by a central active nucleus. Two other
galaxies (A2294 and  RXJ0821.0+0752) have line intensity ratios suggestive of
classical HII regions.

This paper is the first in a series detailing  the observed properties of the
BCS clusters in the radio, optical and X-ray wavebands. Later papers will 
investigate the dependence of the optical spectral characteristics of the
dominant galaxies presented here with both the global cluster properties and
those of  any radio source contained in the BCG. 

\section{Acknowledgements}
CSC, ACE and ACF thank the Royal Society for financial support. Thanks are
also due to Stefano Ettori for his help in observing during the December
1996 run, and to Roderick Johnstone for anything and everything to do with
the computers. The INT is operated on the island of La Palma by the Isaac
Newton Group in the Spanish Observatorio del Roque de los Muchachos of the
Instituto de Astrofisica de Canarias. This research has made use of the
NASA/IPAC Extragalactic Database (NED).

{}

\end{document}